\documentclass[twocolumn,amssymb, bibnotes, aps,prd,superscriptaddress]{revtex4-1}

\usepackage{graphicx}
\usepackage{amsmath, amsthm, amssymb}
\usepackage{mathdots}
\usepackage{float}
\usepackage{pgfplots} 
\usepackage{tikz-3dplot}
\usepgfplotslibrary{patchplots}
\usepgfplotslibrary{groupplots}
\pgfplotsset{compat=1.6,ylabsh/.style={every axis y label/.style={at={(0,0.5)}, xshift=#1, rotate=90}}}
\usetikzlibrary{positioning}
\usetikzlibrary{shapes}
\usetikzlibrary{shapes.geometric}
\usetikzlibrary{decorations.text}
\usetikzlibrary{decorations.pathreplacing,shapes.misc}
\usetikzlibrary{matrix}
\usepackage{caption,subcaption}
\usepackage[ruled,lined]{algorithm2e}
\usepackage{algorithmic}
\renewcommand{\algorithmiccomment}[1]{\bgroup\hfill//~#1\egroup}
\newcommand{\etal}{\textit{et al}.}

\usepackage{xcolor}

\begin{document}

	\title{Hyaluronic acid over-expression drives tissue architecture in both healthy and diseased states}
	\author{Kyungeun Kim}
	\email{kkim@math.ubc.edu}
	\affiliation{Department of Mathematics, University of British Columbia, Vancouver, BC V6T 1Z2, Canada}
	\author{Brittany E. Schutrum}
	\email{bs773@cornell.edu}
	\affiliation{Nancy E. and Peter C. Meinig School of Biomedical Engineering, Cornell University, Ithaca, New York, 14853, USA}
	
	\author{Claudia Fischbach}
	\email{cf99@cornell.edu}
	\affiliation{Nancy E. and Peter C. Meinig School of Biomedical Engineering, Cornell University, Ithaca, New York, 14853, USA}
	\affiliation{Kavli Institute at Cornell for Nanoscale Science, Cornell University, Ithaca, New York, 14853, USA}
	\author{J. M. Schwarz}
	\email{jmschw02@syr.edu}
	\affiliation{Physics Department, Syracuse University, Syracuse, NY 13244 USA}
	\affiliation{Indian Creek Farm, Ithaca, NY 14850 USA}

	\date{\today}
	\begin{abstract}
		Tissue architecture is shaped by a variety of biomechanical and biochemical mechanisms, including morphogen patterning, differential growth, and spatial variations in cellular contractility. These mechanisms typically emphasize forces generated within cells. Here, we explore a complementary mechanism in which intercellular space itself exerts forces that shape tissue architecture. More specifically, we focus on hyaluronic acid (HA), a glycosaminoglycan of the extracellular matrix, that draws fluid into the spaces between cells through osmotic pressure, allowing these spaces to expand and mechanically deform the surrounding tissue. We propose that spatial patterning of HA-mediated extracellular forces, acting together with tissue geometry and junctional mechanics, provides a common physical framework capable of generating distinct tissue architectures. We test this hypothesis using a computational framework coupling a cell-based extended vertex model for biological tissues to a flow network representing HA and extracellular fluid in the extracellular space. This coupled model demonstrates a range of architectural responses spanning changes in tissue topology and morphology: global HA overexpression drives fluid accumulation and cell-cell junction rupture to produce multiple lumens in nontrivial patterns; a localized HA-rich region generates pressure gradients that drive budding; and extended HA-rich patterns can drive tubular structures. Our results have direct experimental implications and establish the spatial patterning of HA-mediated intercellular forces as a mechanical morphogen  helping shape tissue architecture.

	\end{abstract}
	\maketitle
	
	\section{Introduction}
	Tissue architecture emerges from the coordinated action of biochemical and biomechanical processes across multiple length scales and is essential to normal development. During development, spatially patterned morphogen signaling, differential cell growth and proliferation, cell rearrangements, and variations in actomyosin contractility generate forces that drive folding, budding, branching, tubulation, and tissue-scale bending~\cite{collinet2021programmed, anlacs2018tissue,briscoe2015morphogen}. Related mechanisms are altered in disease, where dysregulated growth, adhesion, contractility, and extracellular matrix remodeling can disrupt normal tissue organization and generate pathological structures~\cite{wirtz2011physics}. Much of our understanding of tissue morphogenesis has therefore focused on how forces generated by cells, or changes in cellular material properties in response to biochemical signals, sculpt tissues. And yet, groups of cells do not act in isolation: they are separated and surrounded by extracellular spaces containing fluid, polymers, and other matrix components that can themselves generate mechanical stresses~\cite{humphrey2014mechanotransduction,stramer2024extracellular}. This multicomponent property of tissues raises a broader question: {\it to what extent can forces generated within the intercellular space determine tissue architecture in both healthy and diseased states?}
	
	If the initiation location and size such intercellular spaces can be patterned by the cells themselves, then the resulting forces are patterned to lead to nontrivial tissue architectures. As for how cells can pattern intercellular space, we turn to hyaluronic acid.  Hyaluronic Acid (HA), a key glycosaminoglycan component of the extracellular matrix, is known to play a critical role in tissues across multiple developmental and disease contexts~\cite{toole2004hyaluronan, spicer2004hyaluronan,garantziotis2019hyaluronan, rabelink2024roles}. For example, HA generates osmotic gradients that drive fluid accumulation and hydrostatic pressure, processes essential for lumen expansion during embryogenesis—including the early blastocel and more specifically otic vesicle formation in the inner ear~\cite{MUNJAL2021}. In pathological contexts, HA upregulation is associated with cyst formation  \cite{Kato2012}  \cite{Kato2016} and, in the case of breast tissue, correlates with malignant transformation~\cite{Zoltan-Jones2003,chanmee2016hyaluronan}.
	
	Given the plethora of possible tissue architectures, we will focus on several motifs that can be driven by the forces in intercellular space, namely, lumen formation, tissue asymmetry, tubular formation, and budding/branching formation. See Fig. 1. Lumen formation, the emergence of fluid-filled cavities within cellular tissues, is a fundamental morphogenetic process critical to both development and disease. During embryonic development, lumen formation drives organogenesis in structures ranging from the inner ear to mammary glands, neural tubes, and vascular systems~\cite{FUJI2022,TORRESSANCHEZ2021}. Dysregulated lumen formation in mature tissues, however, contributes to pathological conditions including cyst formation in the breast and kidney~\cite{ATHANASIOU2014,wilson2007cystic}.  Tissue asymmetry and bending provide another common architectural motif and underlie processes including left-right asymmetric organ positioning and gut looping~\cite{Kurpios2008,Sivakumar2018}. Tubular architectures are ubiquitous in organs responsible for transport, secretion, and exchange can arise through several morphogenetic routes, including folding and invagination of epithelial sheets~\cite{lubarsky2003tube}. Budding and branching generate new tissue protrusions that can subsequently elongate, divide, and differentiate into complex organ architectures. These processes are central to the formation of organs such as the lung, mammary gland, kidney, and salivary glands~\cite{affolter2009tissue}.
	
	To examine how tissue architecture emerges in response to HA-driven forces within intercellular spaces, we choose a cell-as-deformable-polygons/polyhedrons model for tissues known as the vertex model~\cite{honda1980much,Honda1982,fletcher2014vertex,honda2022vertex}. Vertex models where neighboring cells share interfaces have provided insights into tissue mechanics, including predictions of rigidity transitions in disordered tissues \cite{Bi2015,zhang2022topologically}, such as those observed during tail growth in zebrafish embryos \cite{Mongera2018,Lenne2022}. They have also shown that how fluid-like embedded spheroids  can remodel the extracellular matrix more effectively than solid spheroids given their nontrivial dynamics~\cite{zhang2025,ho2026vimentin}.
	
\begin{figure}
		\captionsetup{singlelinecheck = false, justification=raggedright}
		\centering
		\includegraphics[width=0.95\linewidth]{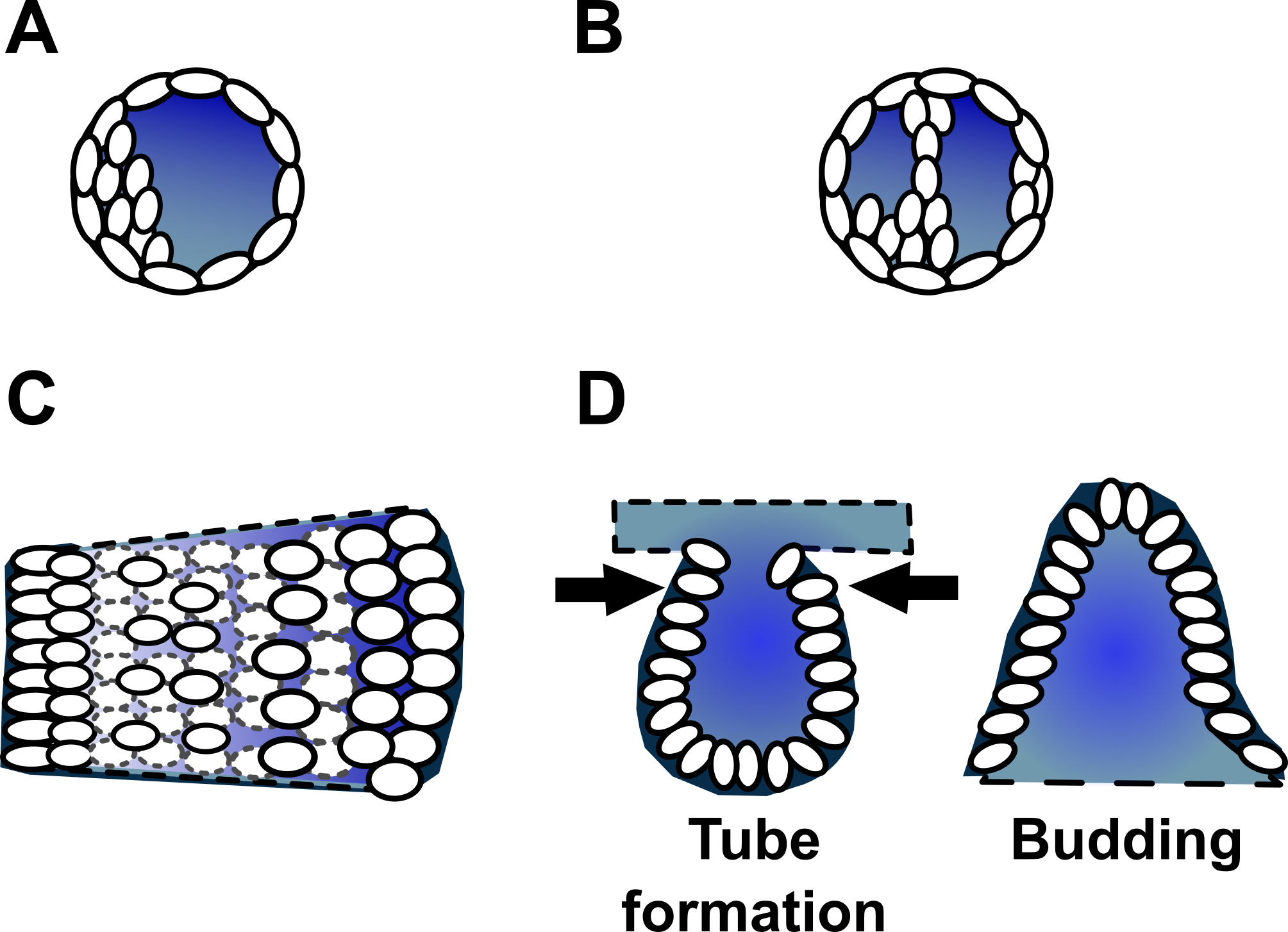}
		\caption{{\it Schematic of lumen architectures in development and disease.} A: A single large fluid-filled region, as in the blastocyst. B: Multiple discrete lumens, as in cystic tissue. C: Developmental asymmetry, with color denoting the pressure gradient across the tissue, as in gut looping. D:~Budding and tube formation, as in development of the inner ear.}
		\label{fig-drawings}
	\end{figure}

	Here, we expand upon the conventional two-dimensional vertex model by incorporating an extracellular fluid flow network, providing a framework that explicitly links cell and tissue deformation to fluid accumulation in the intercellular spaces. The flow is driven by HA-generated osmotic pressure. To accommodate the flow network, we implement an extended vertex model in which neighboring cells no longer share a common interface, but instead possess independent boundaries separated by intercellular compartments. The flow network couples to the cells through forces generated by the fluid.  In total, we arrive at a cellular-based model that captures the interplay between intracellular and intercellular forces, with the latter mediated by flow network.  Our models differ from prior lumen formation modeling by using pressure gradients and modified cell target area/perimeter to establish an indirect mechanistic link between lumen formation and buckling instabilities. Second, we simplify the complex osmotic processes into a flow network framework that allows local paracellular fluid dynamics to be interpreted as emergent global pressure gradients. This approach enables us to systematically investigate how the spatial distribution of HA-overexpressing cells physically shapes tissue architecture to create a variety of experimentally observed structures, as shown in Fig.~\ref{fig-drawings}.
	
	\section{Computational Model for cells and intercellular fluids}
	We introduce the theoretical framework used to describe the coupling between HA-mediated fluid accumulation and tissue mechanics. We first formulate the cellular mechanics using an extended vertex model and then introduce a network representation of the intercellular spaces through which HA and extracellular fluid are distributed. We next describe how HA generates osmotic pressure within these spaces and how this pressure couples back to cell shape and tissue deformation. We then discuss the numerical implementation and incorporate experimental data obtained under HA-rich conditions to inform our model and refine key coefficients based on empirical observations. Dimensionless parameters and modeling assumptions informed by relevant references are used in the simulations. Dimensionless quantities are denoted by hats throughout.
	
	\begin{figure*}
		\captionsetup{singlelinecheck = false, justification=raggedright}
		\includegraphics[width=\textwidth]{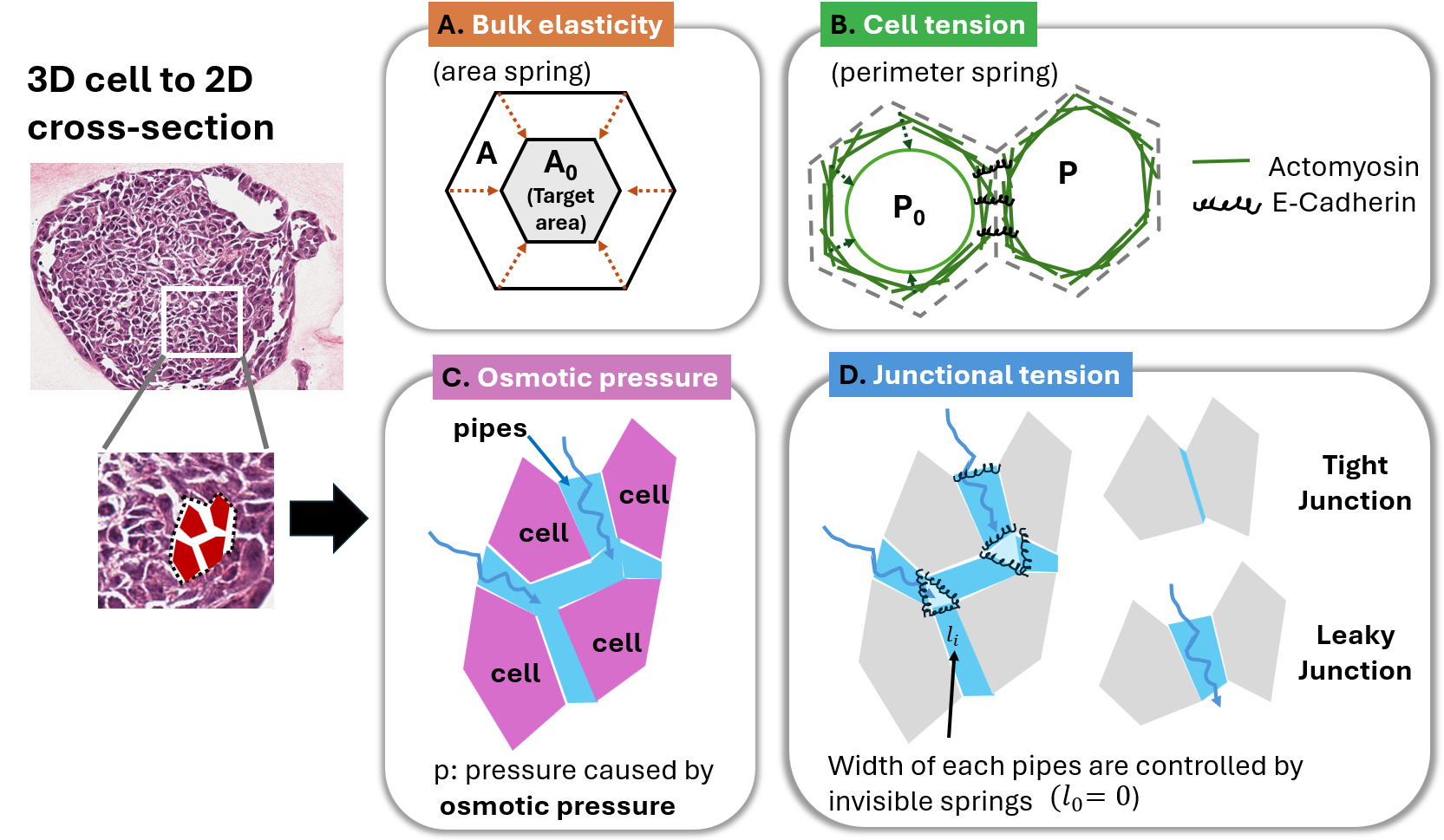}
		\caption{{\it Visualization of the vertex model energy function.} The first two terms
			are from the traditional vertex model: (A) area spring ($A_0$ is a target area), (B) perimeter spring (originating from the tension of the actomyosin cortex and $P_0$ represents target perimeter). (C)--(D) Additional energy terms for the extended vertex model, where (C) represents the pressure energy term modeled by a pipe network and (D) shows the junctional tension representing leaky and tight junctions,
			based on the experimental observations of Ref.~\cite{Mukenhirn2024}.}
		\label{fig-vertex_model}
	\end{figure*}
	
	\subsection{Tissues}
	To model the mechanics and morphology of a cross-section of a three-dimensional tissue, we begin with the energy function for the mechanical energy $E_\alpha$ for the $\alpha$th cell in a cell packing system~\cite{Farhadifar2007,Staple2010,Bi2015}, where the cross-sectional area and perimeter are $A_\alpha$ and $P_\alpha$, respectively. We define: 
	\begin{equation} 
		\label{single_cell_energy} 
		\quad E_\alpha =\frac{1}{2} K_{A} (A_\alpha - A_{0})^2 + \frac{1}{2}K_P P_\alpha^2 + \gamma P_\alpha. 
	\end{equation} 
	The first term reflects the cell's bulk elasticity \cite{hufnagel2007mechanism,zehnder2015cell}, where $K_{A}$ represents area stiffness and $A_0$ is the preferred area. The second term captures the contribution of the cell's perimeter, modeling the intracellular contractile forces from the actomyosin cortex with elastic constant $K_P$~\cite{Bi2015}. The third term accounts for interfacial tension $\gamma$, which arises from the balance between cortical tension and cell-cell adhesion energies~\cite{Manning2010} at the interface of adjacent cells. For isolated cells, $\gamma$ is interpreted as surface tension. For simplicity, we assume uniform values for all parameters, $K_A$, $A_0$, $K_P$, and $\gamma$, across all cells. Adding a constant to the mechanical energy, which leaves the forces unchanged, allows us to complete the square, yielding: 
	\begin{equation} 
		E_\alpha = \frac{1}{2}K_{A} (A_\alpha - A_{0})^2 + \frac{1}{2}K_P (P_\alpha - P_{0})^2,\label{eq-e_tot} 
	\end{equation} 
	where $P_0 = -\gamma / (K_P)$ is the effective target perimeter. 
	
	The energy can be further non-dimensionalized as: 
	\begin{equation} 
		\label{scaled_e_tot} 
		\hat{E}_\alpha= \frac{1}{K_A A_0^2} E_\alpha = \frac{1}{2}(\hat{A}_\alpha - 1)^2 +\frac{1}{2} \frac{K_P}{(K_A A_0)}(\hat{P}_\alpha - \hat{P}_0)^2, 
	\end{equation} 
	where $\hat{A}_\alpha = A_\alpha / A_0$ and $\hat{P}_\alpha = P_\alpha / \sqrt{A_0}$ are the dimensionless shape parameters for area and perimeter. Additionally, $\hat{p}_0 = P_0 / \sqrt{A_0}$ is termed the target shape index, a key quantity that governs the mechanical behavior of tissues in two dimensions. For example, the shape index for a regular hexagon is $\hat{p}_0^{hex}=2 \sqrt{2} \sqrt[4]{3}\approx3.72$, while for a regular pentagon it is $\hat{p}_0^{pent}=2 \sqrt{5} (5-2 \sqrt{5})^{1/4}\approx3.81$. Cells with fewer edges tend to have shape indices larger than that of a hexagon~\cite{Kim2024}.
	
	\subsection{Fluid flow model assumptions}
	
	Extracellular HA accumulation creates an osmotic gradient that drives fluid flow into intercellular spaces, termed here as compartments. The resulting hydrostatic pressure from fluid accumulation deforms cells and drives, for example, lumen expansion. We model this process accordingly for our four scenarios, with specific formulations detailed in the respective sections below. The morphologies reported here arise from the coupling of hyaluronan transport in the extracellular space to the mechanics of the surrounding epithelium. Before specifying the fluid energy functional and the transport equations, we state the assumptions on which the model rests.
	\begin{enumerate}
		\item \textbf{Osmotic influx.} Fluid is drawn into an extracellular compartment by the osmotic pressure of the hyaluronan it contains, which varies with the local HA concentration. Because hyaluronan is a polyanion, this pressure originates in Donnan partitioning of mobile ions rather than in the polymer concentration itself~\cite{Lai1991,Buschmann1995}, and is, therefore, nonlinear in the fixed charge density of the network.
		
		\item \textbf{Hydrostatic pressure.} The hydrostatic pressure in each intercellular compartment is set by its hyaluronan content. Compartments carrying high HA are driven to expand against the elasticity of the surrounding cells; those where the field has decayed are not. It is this spatial differential, rather than any uniform pressure, that deforms the tissue.
		
		\item \textbf{Incremental steady flow.} The full transport dynamics is decomposed into a sequence of iterations, each representing an incremental steady state. Within an iteration the geometry is fixed, the concentration field is solved at steady state on the current configuration, and the mechanical state is relaxed; between iterations the field and the stress-free configuration are updated. Fast microscale processes are thus coarse-grained into the update, while the slow accumulation of hyaluronan carries the dynamics.
		
		\item \textbf{Junction state.} Fluid accumulation can separate neighboring cells, allowing fluid to enter or redistribute through intercellular spaces. In most of the scenarios considered here, these openings are assumed to reseal rapidly compared with the timescale of HA accumulation, so that cell–cell connectivity is preserved. However, sufficiently large separation causes irreversible junction rupture, leaving a persistent fluid-filled cavity and thereby changing tissue topology. Thus, we distinguish transient junction opening, which facilitates fluid redistribution without altering tissue connectivity, from permanent junction failure, which generates stable luminal structures.
		
		\item \textbf{Interfacial flow.} Flow through cell interfaces is taken to be governed principally by the hydrostatic pressure difference across them.
		
	\end{enumerate}
	
	For item 1, in its standard form, the Donnan equilibrium determines the swelling of the charged network self-consistently with the ion partitioning~\cite{Overbeek1956}. Here the two are separated: the osmotic pressure enters the mechanical energy and drives flow through the extracellular space, while the growth of the cells is treated as an independent contribution. Including the cell areas in the pipe network produced no appreciable difference in the resulting morphology (see Fig.~\ref{fig-result_by_asymmetry2}(c)--(d)).
	
	Related to the fourth item, to test the sensitivity of the outcome to this choice, we compared a simulation maintaining cell--cell adhesion at every iteration with one in which adhesion in the higher-concentration region was removed on alternate iterations, mimicking repeated fracture and healing. The two gave no appreciable difference in the resulting morphology, consistent with the fracturing--healing cycle acting on a timescale shorter than a single iteration and therefore being absorbed into it.
	
	These five assumptions allow the model to toggle between two established fluid descriptions: it retains the osmotic and hydrostatic terms used in compartment-level treatments of lumen formation, while resolving the extracellular space as a spatially extended network rather than as one or a few lumped reservoirs.
	
	\subsection{Hyaluronan concentration field and osmotic coupling}
	Hyaluronan is synthesized on the surface of the cell membrane by HAS-expressing cells where it either remains in the pericellular coat or is released into the extracellular space, where it spreads by diffusion and is cleared by degradation \cite{toole2004hyaluronan} . Because HA is a large polyanion localized outside the cells, we solve for its transport within the extracellular network alone; this same confinement is what renders it osmotically active across the membrane. To do so, we denote $\hat{c}_\alpha$ for the dimensionless concentration in the extracellular compartment $\alpha$. We adopt the linear reaction--diffusion description typically used for signaling species in multicellular lumens~\cite{Yu2024},
	\begin{equation}
		\frac{\mathrm{d} \hat{c}_\alpha}{\mathrm{d}t} = \hat{D} \sum_{\beta \in \mathcal{N}(\alpha)} \left( \hat{c}_\beta- \hat{c}_\alpha \right) + \hat{s}_\alpha - \hat{k}\, \hat{c}_\alpha,
		\label{eq:ha_rd}
	\end{equation}
	where $\mathcal{N}(\alpha)$ are the compartments sharing an edge with $\alpha$, $\hat{D}$ is the diffusivity, $\hat{k}$ the dimensionless degradation rate, and $\hat{s}_\alpha$ is nonzero only adjacent to HAS-expressing cells. We retain the diffusion and linear kinetics, but omit the autocatalytic reaction terms of the Schnakenberg scheme, since hyaluronan is synthesized enzymatically rather than through a self-activating network~\cite{Yu2024}. The accompanying reactant equation is therefore not required. Since the geometry is fixed within an iteration, Eq.~\eqref{eq:ha_rd} is solved at steady state for the current configuration. For localized sources, the superposition of screened kernels yields
	\begin{equation}
		\hat{c}_\alpha = \hat{c}_0 \sum_{\mu \in \mathcal{S}} \exp\!\left( - \frac{\left| \mathbf{\hat{r}}_\alpha - \mathbf{\hat{r}}_\mu \right|}{\hat{\lambda}} \right), \qquad \hat{\lambda} = \sqrt{\hat{D}/\hat{k}},
		\label{eq:ha_superposition}
	\end{equation}
	where $\mathcal{S}$ is the set of HAS-expressing cells and $\mathbf{\hat{r}}_\mu$ their centroids. The single dimensionless length $\hat{\lambda}$, measured in compartment spacings, controls the profile; we normalize $\mathbf{\hat{c}}$ by its maximum so that the coupling strength is set separately.
	
	The field enters the mechanics through the osmotic pressure it exerts on the flow network. Following the van 't Hoff relation used for osmotically driven lumen growth~\cite{SunFeng2022}, the pressure in compartment $\alpha$ is $\Pi_\alpha = k_\mathrm{B}T\,(c_\alpha - c_\mathrm{ref})$, and the corresponding contribution to the energy is the pressure--volume work
	\begin{equation}
		\hat{E}_\mathrm{fluid}
		= -\sum_{\alpha}\hat{\Pi}_\alpha \hat{A}^f_\alpha,
		\label{eq:pA}
	\end{equation}
	where $\hat{A}^f_\alpha$ is the area of fluid compartment $\alpha$ in the current configuration. Compartments carrying high HA therefore expand against the elasticity of the surrounding cells, while those where the field has decayed do not, and the resulting differential is what deforms the tissue. Crucially, $\hat{A}^f_\alpha$ is not prescribed: the compartment areas follow from minimization of the total energy, so the flow network adopts whatever configuration balances the osmotic drive against the cell mechanics.
	
	We adopt two elements of the framework of Sun and Feng: the coupling of osmotic pressure to tissue mechanics through the pressure–area term in Eq.~\eqref{eq:pA}, and the decomposition of deformation into changes in the stress-free configuration and an elastic response~\cite{SunFeng2022}. The treatment of fluid transport, however, differs substantially. Sun and Feng describe transport into and out of a single lumen using spherically symmetric solute pumping, Darcy influx, and Poiseuille outflow~\cite{SunFeng2022}. Here, we instead resolve the extracellular space as a network of interconnected compartments, with HA transport described by Eq.~\eqref{eq:ha_rd}. At each iteration, the HA field is solved on the current deformed network and the resulting osmotic pressure enters the mechanical problem through Eq.~\eqref{eq:pA}. Thus, the HA concentration field and tissue geometry evolve self-consistently, rather than assuming a prescribed spatial concentration profile.
	
	The four scenarios described here differ in the membership of $\mathcal{S}$ and in the strength and distribution of cell junctional tension, with the latter being described below. Specifically, the cellular HA sources ($\mathcal{S}$) are uniformly distributed for multi-lumen formation. In contrast, for budding, the HA sources are localized to predefined regions. For asymmetric development, the HA sources confined to one ``side" of the tissues. 
	More precisely, compartments close to the sources carry high HA and swell osmotically, while those on the far side, where the field has decayed resulting in low HA concentration, remain near their reference size and are placed in relative contraction by their expanding neighbors. A tissue layer whose two faces change length by different amounts cannot remain straight, and the mismatch is accommodated by total layer curvature. Finally, tube formation implements cell proliferation in addition to HA accumulation, in the presence of a single large compartment on the apical side of all cells in the monolayer. 
	
	\subsection{Cell junctional tension and the four scenarios}
	
	Now that we have described the energetic contributions of the intracellular mechanics and the fluid, we turn to junctional tension, the component that distinguishes the four models summarized in Table~\ref{tab:scenarios}. It enters through two distinct contributions, which act on different parts of the geometry. The first is cell--cell adhesion, which acts across the extracellular gaps separating neighboring cells. Where the intercellular space is resolved as a fluid compartment in the form of a pipe network, adhesion is represented by junction springs connecting the corresponding vertices of the two cells bounding that compartment. The stiffness assigned to each junction therefore controls how readily the gap opens under the osmotic pressure of the fluid it contains, and hence whether hyaluronan accumulation is accommodated by separation of the cells or resisted by their adhesion.
	
	The second is surface tension, which acts on the boundaries between tissue and bulk fluid, where cells are exposed to fluid rather than to a neighboring cell. This term maintains the outer boundary of the spheroid, and where it differs between two such boundaries it produces a mechanical asymmetry. In the asymmetric scenario the tension differs between the left and right boundaries of the tissue; in the tube formation scenario it differs between the apical and basal surfaces of the epithelial sheet. In the budding scenario, by contrast, no apical--basal tension asymmetry is imposed. In contrast, Munjal \etal~ used direct stiffness differences between apical and basal tensions as input to their vertex model~\cite{MUNJAL2021}. Since our aim is to induce the asymmetry through fluid energy alone, we do not prescribe these terms.

	The two contributions are therefore complementary: adhesion governs the internal cohesion of the tissue and the opening of intercellular space, while surface tension governs the shape of its exposed boundaries. For Scenario 1, to maintain spheroid integrity, we include:
	\begin{align}
		\hat{E}_{\text{spheroid}} &= \hat{K}_{PS} \cdot \hat{P}_{\text{spheroid}},\\
		\hat{E}_{\text{boundary}} &= \sum_{i,j \in \partial\Omega}
		\hat{K}_B \, d(\hat{v}_i,\hat{v}_j)^2,\\
		\hat{E}_R &= \sum_{i\in\Omega} \hat{K}_R
		\left( (\hat{x}_i \cdot (\hat{S}_R)^n-\hat{x}_i(t))^2 \right.
		\nonumber\\
		&\qquad \left. {}+
		(\hat{y}_i \cdot (\hat{S}_R)^n-\hat{y}_i(t))^2 \right).
	\end{align}
	where $\hat{E}_{\text{spheroid}}$ represents the two-dimensional analog of surface tension and $\hat{E}_{\text{boundary}}$ maintains outer cell connectivity through springs connecting boundary vertices. Furthermore, we employ a resistance function, $\hat{E}_R$, to represent the target coordinates after swelling, where $\hat{S}_R$ represents the swelling target at the $n$-th iteration~\cite{Kim2024}. 
	For Scenarios 2 and 3, we implement surface tension with the following: 
	\begin{equation}
		\hat{E}_{\text{tension}} = \hat{K}_{t}\hat{L}_t 
	\end{equation}
	where $L_t$ refers to arc length of specific HAS-treated boundary cells for Scenario 2 and left/right boundary cells for Scenario 3.  Thus, for Scenario 1 (Multi-lumen), $\hat{E}_{\text{total}} = \hat{E}_{\text{cell}} + \hat{E}_{\text{junction}} + \hat{E}_{\text{fluid}} + \hat{E}_{\text{spheroid}} + \hat{E}_{\text{boundary}}+\hat{E}_R$, where $\hat{E}_{\text{junction}}$ includes variable junction stiffness. For Scenarios 2 and 3, $\hat{E}_{\text{total}} = \hat{E}_{\text{cell}} + \hat{E}_{\text{junction}} + \hat{E}_{\text{fluid}} + \hat{E}_{\text{tension}}$. Scenario 4 is a strip of cells without intercellular compartments, and junctional tension is therefore included in the surface tension term.
	
	Cell rearrangement through T1 transitions is a mechanism by which an epithelium relieves accumulated junctional tension: when a shared junction contracts below a threshold length, the four cells surrounding it exchange neighbors, and the tension that drove the contraction is redistributed. In a conventional vertex model the energy of that junction is $\hat{\Lambda}_{ij}\hat{\ell}_{ij}$ for a single edge shared by cells $i$ and $j$. In the present model the shared junction is replaced by two edges bounding a fluid compartment. Since these edges belong to different cells and have distinct vertices, we denote by $\hat{\ell}_{ij}$ the edge of cell $i$ facing cell $j$, and by $\hat{\ell}_{ji}$ the edge of cell $j$ facing cell $i$, so that the corresponding energy is
	\begin{equation}
		\hat{E}_{ij} = \hat{\Lambda}_i \hat{\ell}_{ij} + \hat{\Lambda}_j \hat{\ell}_{ji} - \hat{p}_{ij} \hat{A}_{ij} + \hat{K}_S \hat{d}_{ij}^2 ,
	\end{equation}
	where $\hat{A}_{ij}$ is the dimensionless area of the compartment between junctions, $\hat{d}_{ij}$ the separation of the two cells, and $\hat{K}_S$ the adhesion stiffness. Contracting the junction to zero now requires collapsing the compartment as well, which the pressure term opposes. Therefore, an explicit representation of fluid-filled intercellular compartments also raises questions about cell rearrangements that are not addressed in the present model. In conventional vertex models, T1 transitions occur through the shrinkage and disappearance of a cell–cell edge followed by the formation of a new edge~\cite{Farhadifar2007,Bi2015}. Here, such a transition would additionally require displacement of fluid from the intervening compartment and may therefore depend on hydrostatic pressure and fluid transport. 
	
	T1 transitions are, thus, not implemented in the present simulations, as we do not determine how intercellular fluid modifies their energetic barriers. Moreover, in Scenario 1, where swelling and the fluid flow act continuously through Donnan osmotic uptake as well as the external hydrostatic pressure, neighboring cells are in any case more likely to separate than to converge, despite being densely packed. In Scenarios 2 and 4 the cells remain adhered, but the deformation is driven by the fluid pressure acting on the compartments rather than by junctional tension, so a rearrangement would not relieve the forces producing it.
	Cell rearrangement has also been associated with high cell aspect ratio~\cite{Bi2015,Claussen2024}. In Scenario 1 most cells are bounded by fluid on at least one side rather than by neighboring cells on all sides, so the aspect ratio of individual cells is not the relevant quantity, and it is maximized at the tissue periphery or for isolated cells. In Scenarios 2 and 4 the four-cell configuration required for an exchange is rarely or never realized. The purpose of Scenario 3 is in any case to demonstrate the feasibility of asymmetric structure formation under an asymmetric fluid gradient, rather than to compare numerical data. 
	
	\subsection{Parameters}
	Rather than calibrating each scenario to a single tissue, we select parameters lying within the ranges reported for hyaluronan-containing epithelia, and check that the resulting osmotic pressures are mutually consistent. Taking $E = 10^4$~Pa, a lumen radius $R_0 = 46.1\,\mu$m and a membrane permeability $\alpha = 7.35\times 10^{-10}$~m\,s$^{-1}$Pa$^{-1}$, as reported for the hyaluronan-rich otic vesicle by Wang \textit{et al.}~\cite{Wang2026}, gives a hydrostatic relaxation time $\tau = R_0/(\alpha E) = 6.3$~s. Since a single iteration advances the stress-free configuration by one to five percent, and growth of epithelial tissue proceeds over hours, the iteration timestep exceeds $\tau$ by three to four orders of magnitude. This separation is what the incremental steady-state scheme rests on: mechanical and hydrostatic relaxation are complete long before the stress-free configuration has changed appreciably.
	
	The strength of the osmotic coupling is not an independent amplitude but follows from the fixed charge density of the hyaluronan network. Because hyaluronan is a polyanion confined to the extracellular space, its fixed negative charge density $c_F$ retains a counterion cloud, so that the concentration of mobile ions inside the gel exceeds that outside by
	\begin{equation}
		c_1 - c_0 \simeq \frac{c_F^2}{4 c_0} ,
		\label{eq:donnan}
	\end{equation}
	where $c_0$ is the external mobile ion concentration~\cite{Lai1991}. Combining Eq.~\eqref{eq:donnan} with the van 't Hoff relation $\Pi = RTc$ gives the dimensionless osmotic ratio
	\begin{equation}
		\hat{\Pi}_0 = \frac{RT}{E}\,\frac{c_F^2}{4 c_0}, \qquad
		\hat{\Pi}_\alpha = \hat{\Pi}_0 \, \hat{c}_\alpha
		\label{eq:pihat}
	\end{equation}
	where $c_0$ is the external mobile ion concentration. Because $\hat{\Pi}_0$ scales as the square of $c_F$, a modest difference in hyaluronan content produces a much larger difference in osmotic pressure. Two independent estimates bracket the relevant range. Hydrostatic pressures inferred for intercellular flow in epithelial tissues are of order tens of pascals, which for $c_0 = 323$~mM corresponds to $c_F \approx 4$~mM; the fixed charge density adopted from the literature by Wang \etal\ for the hyaluronan-rich pillar of the otic vesicle is $10$~mM, corresponding to $\Pi_0 \approx 190$~Pa. In the simulations reported here $\hat{\Pi}_0 = 0.05$, and since the concentration field is not normalized the osmotic ratio varies across the tissue with the local hyaluronan content, reaching this value where $\hat{c}_\alpha = 1$ and reaching a peak of $\hat{\Pi}_\alpha = 0.18$, corresponding to fixed charge densities of $16$ and $31$~mM respectively. These lie between the two estimates above and within the range measured for hyaluronan-containing tissues. As for the cells, lengths in the simulation are measured in units of the cell diameter, so that the reference cell area is unity, and energies in units of $K_A A_0^2$ as suggested in Eq. 3. All energy terms are therefore expressed in dimensionless form. In the budding scenario the target shape index is $\hat{p}_0 = 4$, which corresponds to a square. In the asymmetric scenario the target is advanced with the iteration index as $\hat{p}_0^{(n)} = 3.72\,(1.01)^n$, reaching $4.11$ after ten iterations, so that the cells begin at the hexagonal minimum and are progressively softened over the course of the simulation.
	
	The growth of the stress-free configuration is imposed through the cell targets, which differ between scenarios and are stated with the corresponding results. Writing $n$ for the iteration index, targets are advanced geometrically as $\hat{A}_0^{(n)} = \hat{g}_A^{\,n} \hat{A}_0$ and $\hat{P}_0^{(n)} = \hat{g}_P^{\,n} \hat{P}_0$. Isotropic swelling requires $\hat{g}_P = \sqrt{\hat{g}_A}$, since area scales as the square of a length while perimeter scales linearly, so the ratio $\hat{g}_P/\sqrt{\hat{g}_A}$ measures the departure from self-similar growth: values above unity elongate the cells, values below make them more compact. In the budding scenario the HAS-treated cells are given $\hat{g}_A = 1$ and $\hat{g}_P = 1.05$, so that their area is unchanged while their target perimeter increases; the volume increase resides entirely in the extracellular compartments. In the asymmetric scenario the two sides are advanced oppositely, with $\hat{g}_A = 0.99$ on one side and $\hat{g}_A = 1.02$ on the other, and with $\hat{g}_P/\sqrt{\hat{g}_A}$ equal to $0.995$ and $1.005$ respectively.
	
	Junction stiffnesses are assigned per junction from a small set of discrete levels; the levels used span a factor of five to ten between the most and least adherent groups. Surface tensions on the bounding arcs are equal in the budding scenario and differ by an order of magnitude between the two sides in the asymmetric one. The remaining parameter is the dimensionless HA decay length $\hat{\lambda} = \sqrt{\hat{D}/\hat{k}}$, set by the ratio of the hyaluronan diffusivity in the matrix to its turnover rate. The values used here, of order three and thirty cell diameters, are chosen to produce differentials on the two length scales relevant to the respective morphologies.
	
	\subsection{Quasi-static Approach}
	We use energy minimization to relax the configuration at each iteration: the Nelder--Mead method via NMinimize in Mathematica 13.3 for Scenario 1, and a quasi-Newton method via FindMinimum in Mathematica 15.0,  
	limited to 100 iterations per relaxation, for Scenarios 2--4.  For Scenarios 2 and 3, the energy decreases only marginally beyond 100 iterations; for Scenario 4 the limit is used to compare the effect of each constraint. The sequence of relaxed configurations traces the progressive change in shape. To model progressive cell swelling over time with no cell division, we incrementally increase target dimensionless area and perimeter according to $\hat{A}_{0}(t) = \hat{A}_{0}(0) \cdot \hat{S}_A\cdot (\hat{S}_R)^n$ and $\hat{P}_{0}(t) = \hat{P}_{0}(0) \cdot \sqrt{\hat{S}_P\cdot (\hat{S}_R)^n}$, where $n$ is the iteration step and $\hat{S}_A$, $\hat{S}_P$ denote the swelling ratios for area and perimeter applied $n$ times. This promotes increasing shape index $\hat{P}_{0}(t)/\sqrt{\hat{A}_{0}(t)}$   for $\hat{S}_A<\hat{S}_P$ while allowing cell area expansion.
	
	Our model addresses four experimentally motivated scenarios that differ in HA spatial distribution and resulting mechanical response. Table \ref{tab:scenarios} summarizes the key differences between widespread HA distribution leading to multi-lumen formation versus localized HA inducing buckling instabilities. For computational convenience and geometric simplicity, Scenarios 2 and 3 are run on a subset of cells taken to represent a region of a larger tissue mass, and are intended to demonstrate the capabilities of the model rather than to reproduce a specific experiment.
	
	Initial model development for Scenario 1 (multiple lumen formation) was inspired by a unique \textit{in vitro} experimental system in which a pre-formed multicellular spheroid, comprising thousands of cells, was pharmacologically stimulated to induce excess hyaluronic acid synthesis following the establishment of the initial aggregate~\cite{Schutrum2026}. \textit{In vitro} data suggests heterogenetity in cell-cell adhesions formed by HA producing cell  and consequently in this scenario, the lumen patterning is driven by differences in cell-cell adhesion. Moreover, we adopt the two-stage model proposed by Indana \etal, who highlighted the dual mechanisms involved in lumen formation: first, apical actin polymerization drives initial lumen expansion, and once the apical domain matures, tight junctions form, allowing the lumen to expand further due to osmotic pressure~\cite{Indana2024,GUO2024}. By simplifying the complexities through this approach, we can capture key biological insights without excessive detail. We employ a simpler spring and pipe network to handle multi-cell structures consisting of approximately 150 cells in our 2D vertex model, where junction stiffness varies for cell heterogeneity, flow network energy accounts for distributed gaps, and progressive junction breaking occurs when distances exceed a threshold. More details for the two-stage simulation model can be found in the Appendix. The implementation identifies different cell-cell adhesion distributions based on position, allows fluid accumulation in gaps between cells as all cells produced HA, and models lumen expansion through a flow network and cell swelling. This leads to multiple discrete lumens forming at locations with leaky junctions.
	
	For Scenario 2 (budding by local HA overproduction), our hypothesis is that buckling is driven by local pressure gradients, inspired by the experimental observations of HA-driven tissue deformation, in which cells remain cohesive around a pre-existing single lumen without dissemination~\cite{MUNJAL2021}. We model HA-induced lumen budding as a single-stage process, by modeling simple, localized HA with no cell swelling, rather than using the two-stage model. In this scenario, HA production restricted to this specific cell region creates a localized osmotic pressure gradient that drives mechanical buckling. The implementation starts with approximately two straight rows of cells and uses fixed target area $A_0$ and perimeter $P_0$ for HA-treated cells, while non-HA cells retain their original targets $A_0$ and $P_0$.
	
	Scenario 3 (asymmetric swelling) is inspired by the work of Sivakumar \textit{et al.}~\cite{Sivakumar2018} on midgut asymmetry. For example, in gut laterality during embryonic morphogenesis, enzymatic modification of HA by Tsg6 specifically on the right side of the dorsal mesentery creates an expanding HA matrix that initiates midgut rotation. In this context, right-sided HA accumulation generates mechanical expansion and vascular exclusion, breaking left-right symmetry to enable proper gut looping~\cite{Kurpios2008}. We model this as an asymmetric HA concentration field between the left and right cell populations, with different target swelling ratios and surface tension strengths imposed on the two sides.
	
	Scenario 4 (tube formation) is inspired by the work of \cite{Sai2008,Sai2015} on otocyst formation during morphogenesis of the inner ear. The otic placode is a thickened patch of ectoderm in the head of the embryo, one of several such patches that go on to form sensory organs and cranial nerves. Once it has separated from the surrounding tissue, it begins to bend inward, forming a pit. Basal expansion dominates at this stage. Invagination is thereafter driven principally by apical constriction, and the pit deepens progressively into a cup; closure of the cup produces the enclosed structure. The essential sequence is therefore that a sheet of cells first thickens, then undergoes basal expansion and apical constriction, and finally closes. Depending on the tension gradient and on whether the rims meet, the same sequence can yield either a closed vesicle or an open groove that extends into a tube. We model this as a one-dimensional array of cells subject to a tension gradient concentrated on a subset of them, with different surface tension strengths imposed on the apical and basal surfaces, and with the surrounding fluid represented by a simple triangulated region.
	\begin{table*}[t]
		\centering
		\caption{Comparison of the four HA-driven morphological scenarios.}
		\label{tab:scenarios}
		\renewcommand{\arraystretch}{1.3}
		\begin{tabular}{|p{0.13\textwidth}|p{0.19\textwidth}|p{0.19\textwidth}|p{0.19\textwidth}|p{0.19\textwidth}|}
			\hline
			\textbf{Feature} & \textbf{Scenario 1: Multi-lumen} & \textbf{Scenario 2: Budding} & \textbf{Scenario 3: Asymmetry} & \textbf{Scenario 4: Tube formation} \\
			\hline
			HA distribution & Widespread (all cells produce HA)& Localized (only central cells)& Localized (one lateral side)& Localized (central region of the sheet)\\
			\hline
			Primary mechanism & Junction ruptures& Pressure gradient & Lateral cell swelling differential& Basal expansion and apical constriction\\
			\hline
			Junction state & Gradient of adhesion strengths with leaky junctions spatially defined& Leaky in HAS-overexpressing cells only& Leaky in HAS-overexpressing cells only& Intact throughout\\
			\hline
			Junction breaking & Yes, above threshold distance & No & No & No \\
			\hline
			Fluid energy & Flow network ($\hat{E}_{\text{fluid}}$) & Flow network & Flow network & Triangulated external compartment \\
			\hline
			Cell targets (HA producing)& Progressive swelling over phases & $\hat{A}_0$, $\hat{P}_0 \approx 4$ (floppy) & Progressive swelling over iterations & Reference $\hat{A}_0$, apical and basal tension gradient \\
			\hline
			Cell targets (non-HA producing)& N/A & Reference $\hat{A}_0$, $\hat{P}_0$ &Reference $\hat{A}_0$, $\hat{P}_0$ & Reference $\hat{A}_0$, $\hat{P}_0$ \\
			\hline
			Pressure response & Linear & Nonlinear & Nonlinear & Nonlinear \\
			\hline
			Base geometry & Spherical shape (disk) & Two lines of cells with fluid network & Packed cells with intercellular spaces & A line of cells with fluid network \\
			\hline
			Outcome & Multiple discrete lumens & Local morphological change (bulging)& Lateral asymmetry & Invagination and closure into a vesicle \\
			\hline
		\end{tabular}
	\end{table*}

	\section{Results}
	\subsection{Multi-lumen formation via global HA production}
\begin{figure*}
		\captionsetup{singlelinecheck = false, justification=raggedright}
		\begin{center}
			\begin{tabular}{c c c}
				\includegraphics[height=0.22\textwidth]{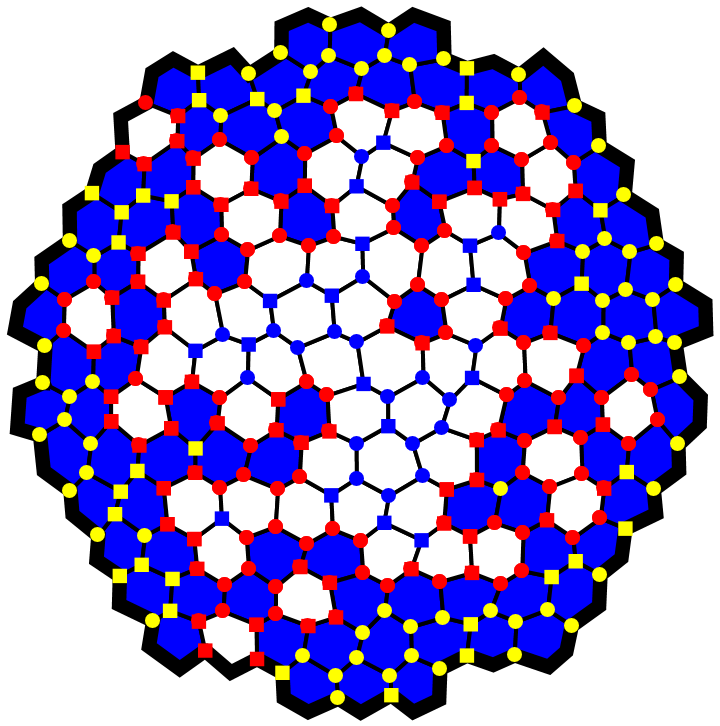}&
				\includegraphics[height=0.22\textwidth]{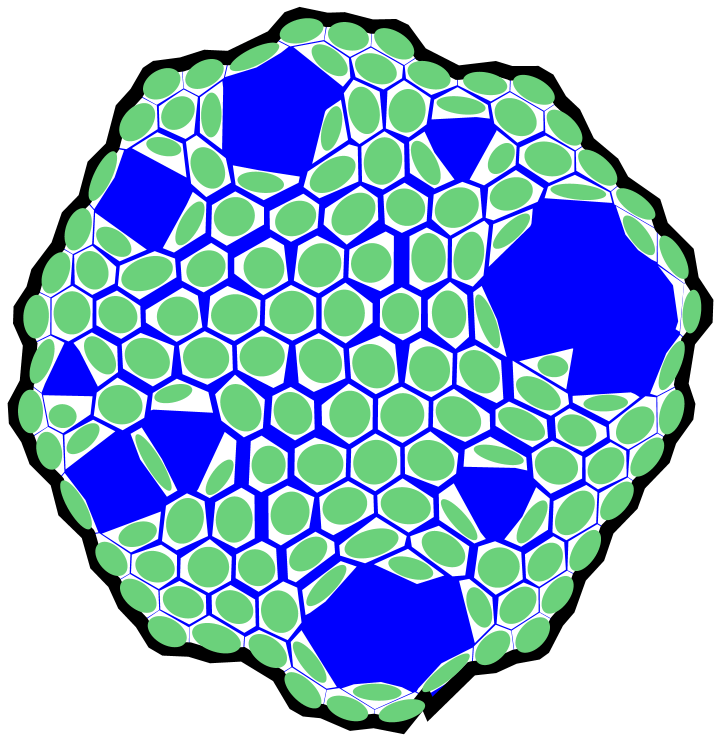}&
				\includegraphics[height=0.22\textwidth]{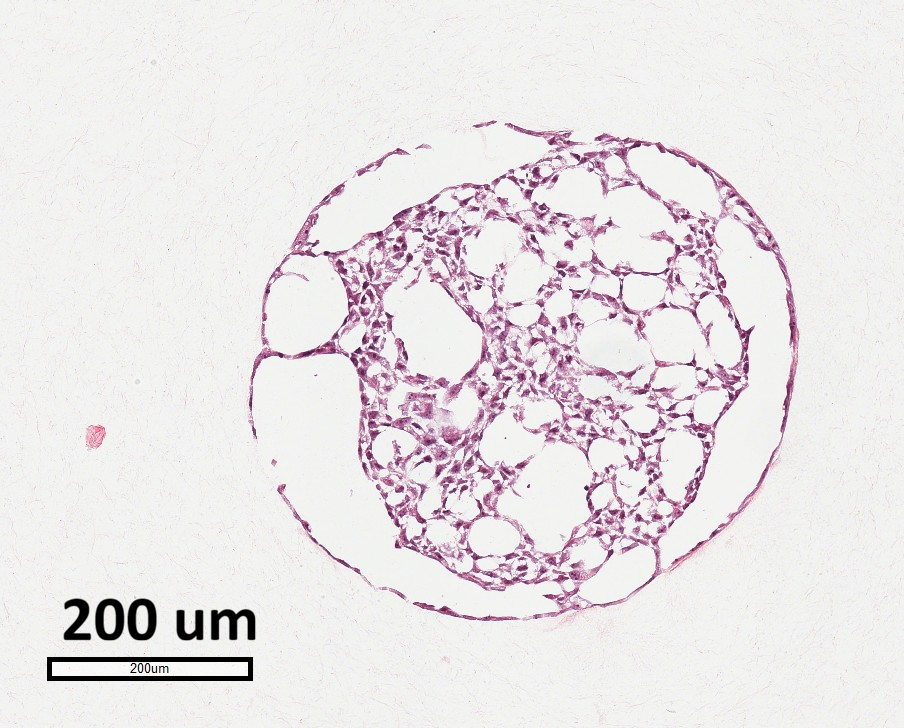}\\
				(a) Initial Distribution  &(b) Result&
				(c) Experimental Image\\
			\end{tabular}
			\begin{tabular}{c c}
				\includegraphics[height=0.22\textwidth]{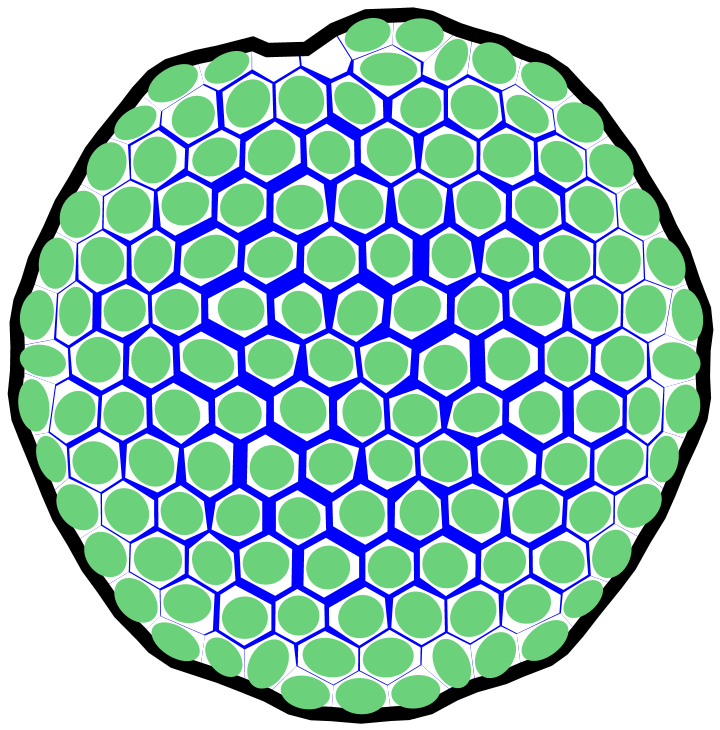}&
				\includegraphics[height=0.22\textwidth]{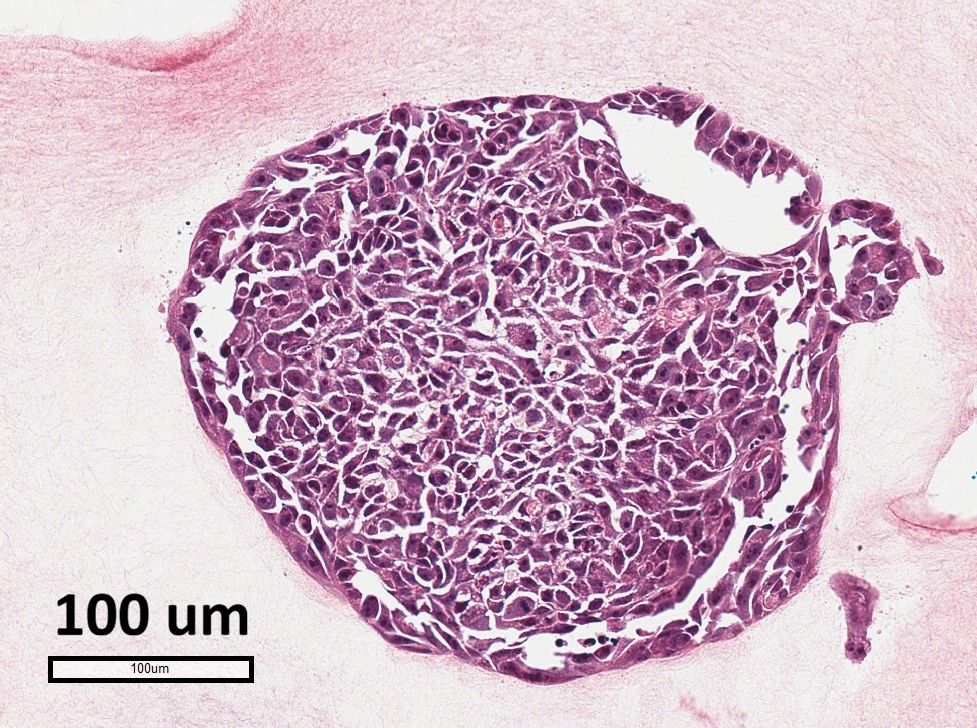}\\
				(d) Control group &
				(e) Experimental image\\
			\end{tabular}
			\caption{{\it Scenario 1 simulation results of HA producing spheroid cross-sections compared to experimental  \textit{in vitro} collagen embedded spheroids:} (a) Initial patterning of the multi-luminal cyst phenotype: Cells near the boundary have weaker cell-cell junctions (blue cells, yellow junctions = weakest) due to matrix confinement induced cell sorting  based on junction strength with some assumed level of natural heterogeneity and imperfection. (b) Simulation results shows lumen formation (blue regions) in high-concentration areas of low cell-cell adhesive regions at the periphery of the spheroid. Green ellipses were fit to each cell to visualize orientation and aspect ratio(c) Histology image of a cross-section of a an H\&E stained HAS overexpressing collagen embedded spheroid. White void spaces in the spheroid are lumen-like structures that accumulated HA-rich fluid during the 2 day culture period.  (d) Control group: No leaky cells all junctions were strong and non-breakable. (e) H\&E stained cross-section of a day 2 collagen embedded spheroid that was not induced to overexpress HAS showing minimal lumen formation.}
			\label{fig-result_by_groups}
		\end{center}
	\end{figure*}

	We begin with two groups of cells classified by adhesion strength organized into the initial pattern as shown in Fig. \ref{fig-result_by_groups} (a) in accordance to experimentally observed cell organization based on adhesion strength\cite{Cai2024}.  Blue cells represent low adhesion cells, while white cells represent high adhesion cells. We categorize the junction springs connecting the cells into three groups: the first group connects blue cells to other blue cells (low-low adhesion, yellow), the second connects blue cells to white cells (low-high adhesion, red), and the third connects white cells to white cells (high-high adhesion, blue) Fig. \ref{fig-result_by_groups} (a).  The lumen formation process follows a two-phase dynamic where Phase 1 involves leaky junction formation with fluid accumulating in intercellular gaps, followed by Phase 2 where junctions exceed critical distance, rupture, and permanent lumens establish.
	
	Figure~\ref{fig-result_by_groups} shows the simulation results after the 2-phase lumenogenesis process:  (b) simulation result (green= ellpise fit to each cell , blue= modeled accumulated fluid), and (c) experimental end-point image for comparison. The model is able to replicate the combination of peripherally located and more centrally located lumens as observed in an in vitro model of a breast cyst with a HA overproducing spheroid embedded into a collagen hydrogel. Importantly, lumen formation is diminished when the cells are not pharmacologically stimulated to produce HA and develop the weaker cell-cell adhesion that accompany the phenotype. These results are mirrored in the computational model when junction breaking is not permitted as Fig.~\ref{fig-result_by_groups} (d)--(e) represents a control group (no HA overproduction) and shows no lumen formation. These simulation results suggest that the density and initial distribution of cell-cell adhesion influence both the size and positioning of lumens within the spheroid. 
	
	We examined the sensitivity of the simulated lumen patterns to selected parameters and initial conditions, including spatial variations in cell
	swelling increments, alternative energy formulations for fluid flow through the pipe network, and connected versus discrete lumen configurations. In the cases examined, changing the initial distribution of leaky junctions had a more pronounced effect on lumen number and position than the parameter variations tested. Surface tension also affected lumen geometry: higher values favored rounded, convex lumens, whereas lower values permitted star-shaped or irregular configurations (Appendix A). These comparisons characterize sensitivity only for the parameter ranges and initial conditions tested, and only within the present quasi-static framework, in which mechanical and hydrostatic relaxation are assumed to occur on timescales much shorter than the time interval represented by each simulation iterationn (see Appendix B.4 for details).
	
\begin{figure}
		\captionsetup{singlelinecheck = false, justification=raggedright}
		\begin{center}
			\begin{tabular}{c c c c}
				\includegraphics[width=0.11\textwidth]{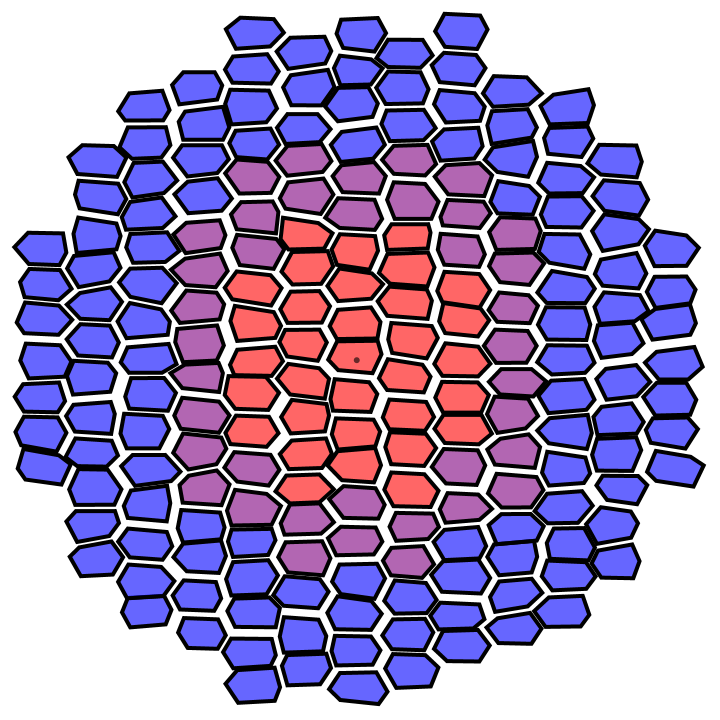}&     \includegraphics[width=0.11\textwidth]{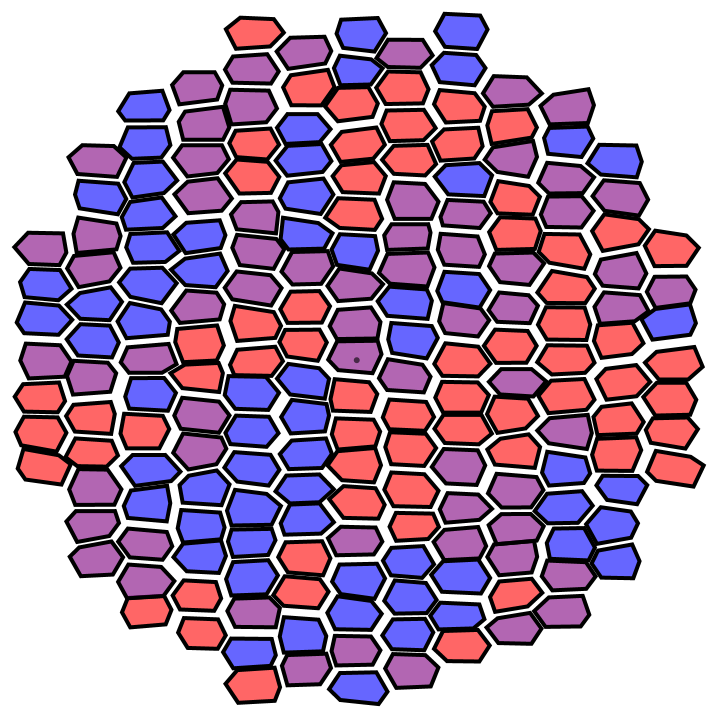}&     \includegraphics[width=0.11\textwidth]{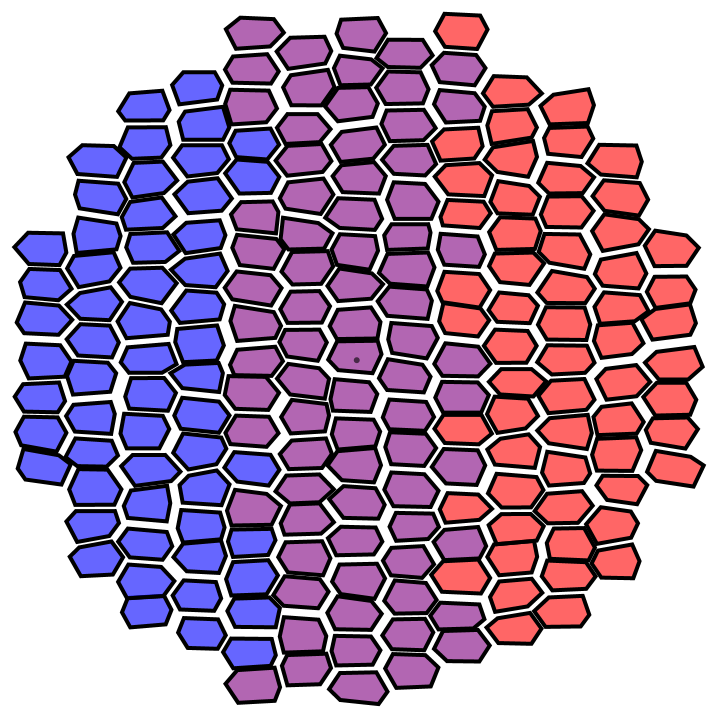}&     \includegraphics[width=0.11\textwidth]{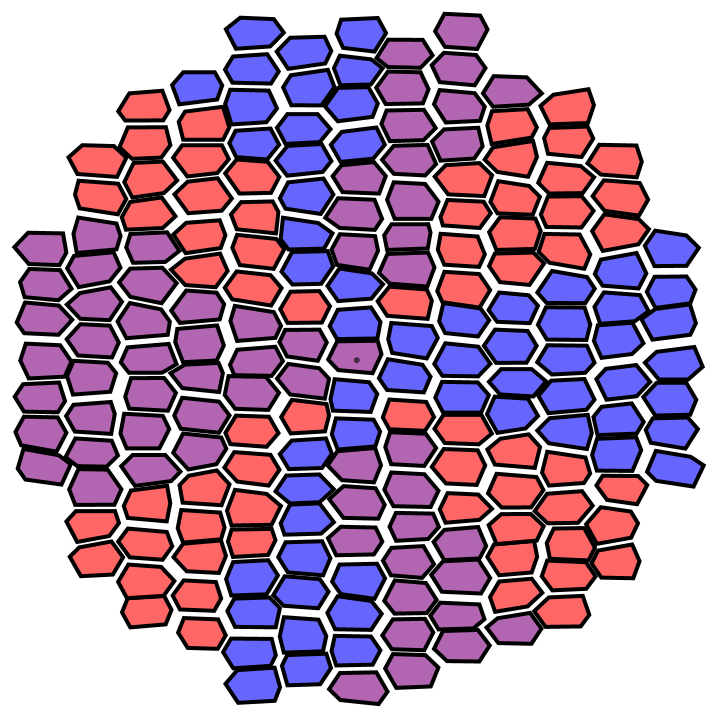}  \\
				
				\includegraphics[width=0.11\textwidth]{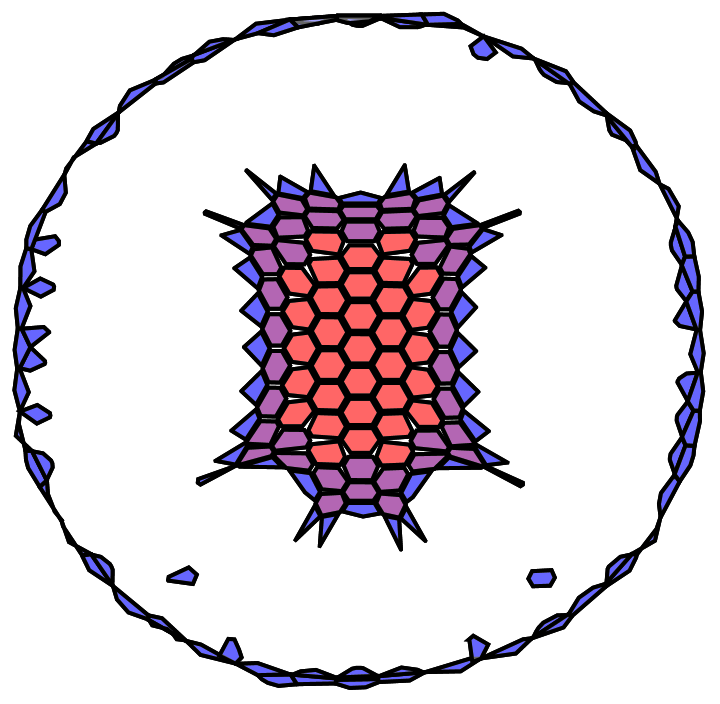}&     \includegraphics[width=0.11\textwidth]{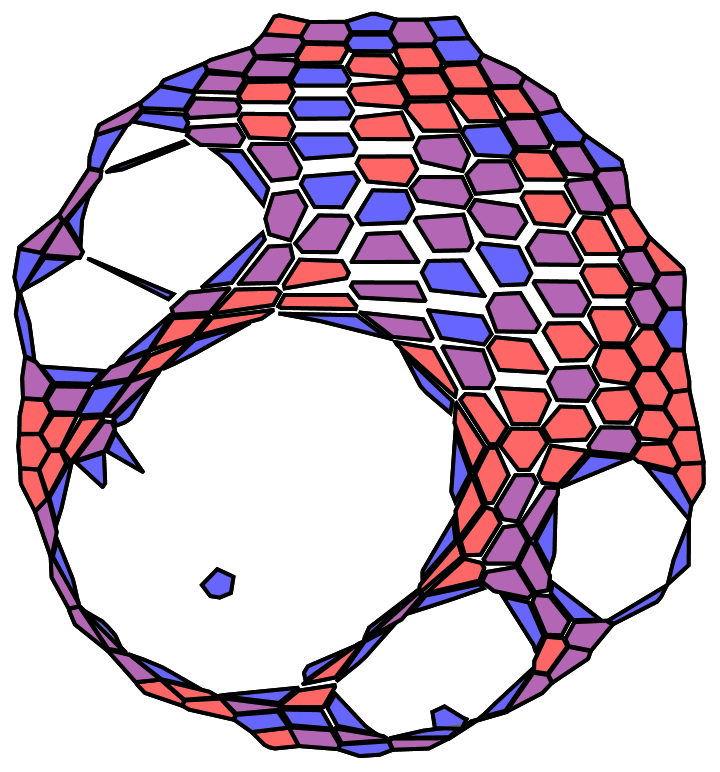}&     \includegraphics[width=0.11\textwidth]{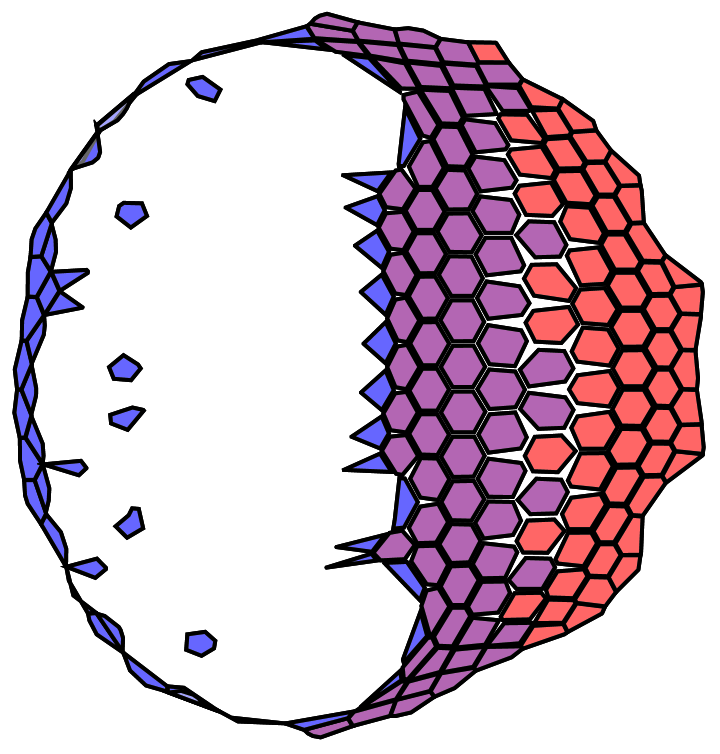}&     \includegraphics[width=0.11\textwidth]{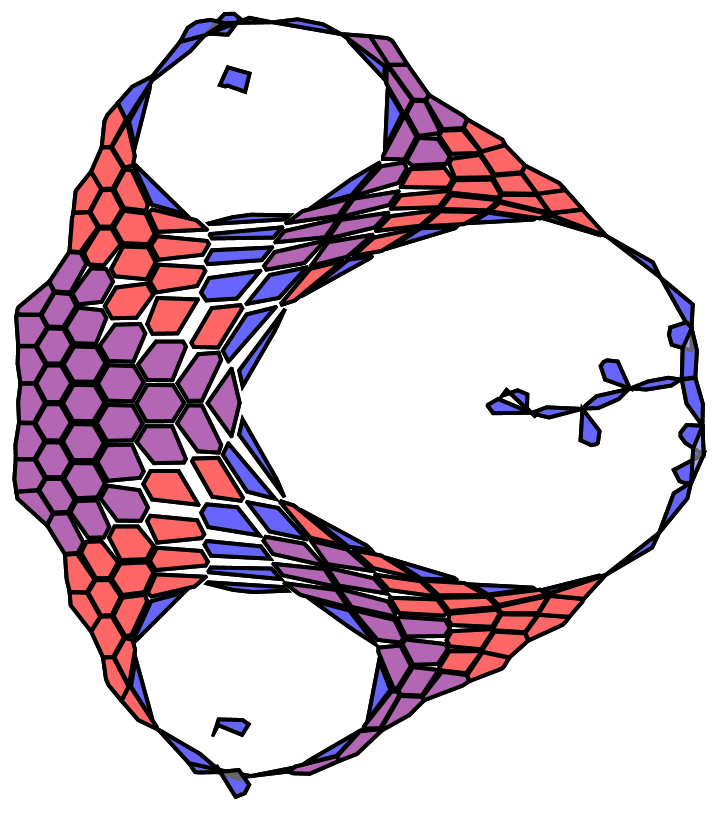}  \\\end{tabular}\caption{{\it Scenario 1 simulation results (bottom row) for different initial patterning (top row) of adhesion strengths within the spheroid.} Blue: low cell--cell adhesion; red: high cell--cell adhesion; purple: intermediate strength, fluid filled lumens displayed here as white void space.}
			\label{fig-result_by_initial}
		\end{center}
	\end{figure}

	To explore how junction strength  patterning alone may regulate lumen initiation and cohesion~\cite{Julien2019}, we tested 4 alternative initial cell--cell adhesion patterns displayed in Fig.~\ref{fig-result_by_initial}.  Critically, whether neighboring lumens fuse is not prescribed here, in contrast to Lu \etal: fusion occurs when the intervening junctions open under the local fluid pressure, and is otherwise absent~\cite{Lu2019}. In panels (c) and (d) the large lumens merge over the simulated time course, reflecting the proximity of the leaky junctions in the initial adhesion pattern. Specifically, the lumen shape induced by the vertical cell--cell arrangement in Fig.~\ref{fig-result_by_initial}(c) resembles that of the blastocyst~\cite{Xue2025}, showing that the model can span multi-lumen and single-lumen outcomes without imposing additional constraints. Thus, under conditions of uniform HA secretion by all cells into the extracellular space, lumens preferentially initiate at sites of weak cell-cell adhesion in our model.  Rupture at these locations creates cavities that accumulate HA and expand into lumens to recapitulate not only the morphology observed in \textit{in vitro sytems}, but also structures seen blastocyst formation~\cite{Blastocystimage}. Our model builds on an experimentally supported mechanism in which actin remodeling contributes to lumen initiation, while subsequent growth is driven by osmotic pressure from accumulated ions and accompanied by spatial remodeling of cell--cell adhesion, as observed in collagen-embedded spheroids~\cite{Schutrum2026} Within this framework, we adopt the hypothesis that prescribed junction-breaking behavior constrains where lumens initiate. Our results therefore describe lumen patterning through the implemented junction-breaking mechanism and do not establish a general requirement for weak adhesion during subsequent lumen growth, which may be related to Scenario 4.
	\subsection{Single lumen budding through localized HA overexpression}
	
	We performed simulations with localized HA treatment to investigate whether HA pressure gradients alone are sufficient to induce budding, without incorporating cell-generated forces through separately prescribed basal and apical junctional tensions~\cite{MUNJAL2021}.  To model this scenario we begin with two straight rows of cells embedded in the extracellular network, with HA producing cells assigned a fixed target area and an elevated target perimeter, and untreated cells retaining their reference targets. In contrast to Scenario 1, all junctions remain intact, so that no cell--cell separation occurs. The hyaluronan field is obtained from the steady-state solution of Eq.~\eqref{eq:ha_steady} for the current configuration, with sources in the compartments adjacent to the treated cells (mimicking cell membrane embedded HAS secreting HA into the extracellular space), and enters the mechanics through the pressure--volume work of Eq.~\eqref{eq:pA}.
	
\begin{figure}
		\captionsetup{singlelinecheck = false, justification=raggedright}
		\begin{center}
			\begin{tabular}{c c}
				\includegraphics[width=0.25\textwidth]{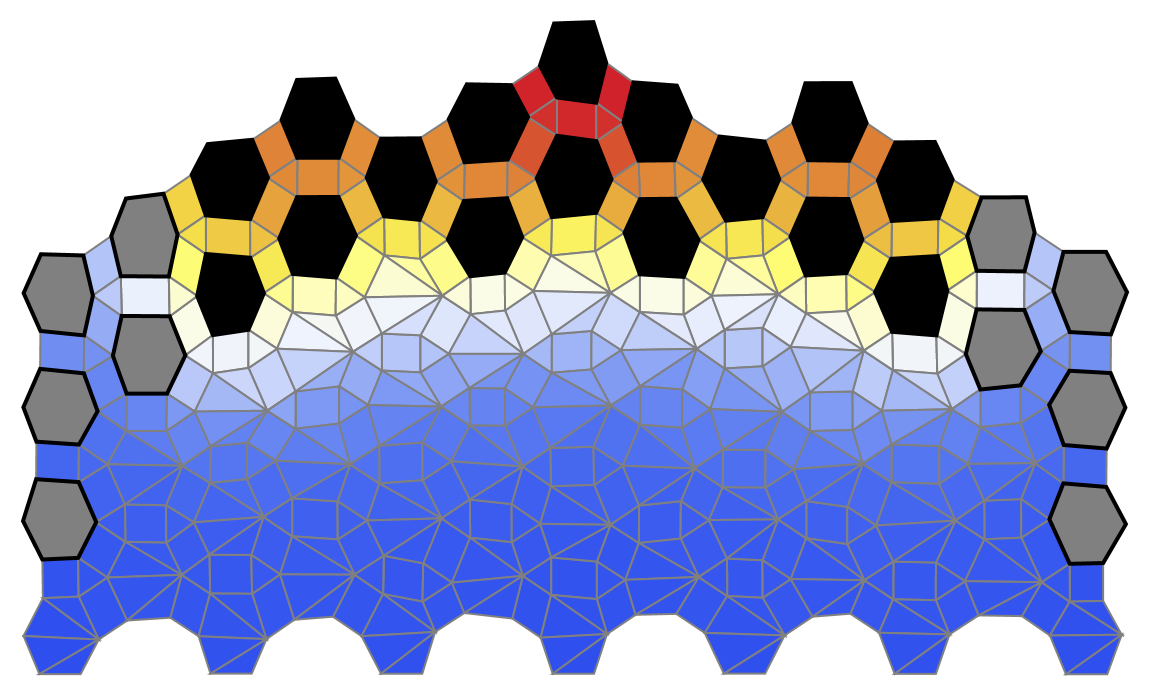}
				& \includegraphics[width=0.25\textwidth]{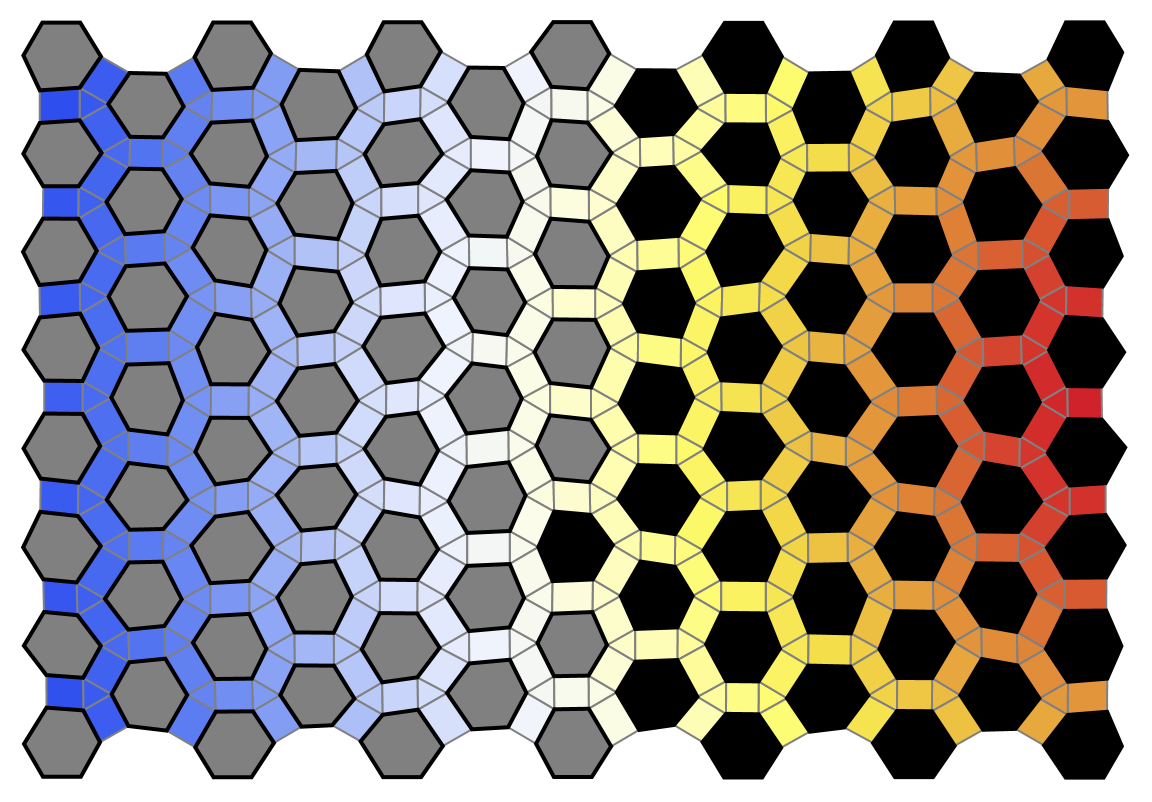} \\
				(a)  &  (b)
			\end{tabular}
			\caption{{\it Steady-state hyaluronan concentration in the extracellular
					network for Scenarios 2 and 3.} (a) Scenario 2 Budding. The HA concentration field is obtained by solving
				Eq.~\eqref{eq:ha_steady} for the current cell configuration, with sources located
				in the compartments adjacent to HAS-expressing cells (black). Color denotes $\hat{c}_\alpha$
				normalized by its maximum $1$, shown on a temperature scale from low $0$ (blue) to high $1$ (red); the interiors of the black HAS-expressing cells  are excluded from the
				transport network (gray: other cells). The concentration decays over
				$\hat{\lambda} = \sqrt{\hat{D}/\hat{k}} \approx 3$ cell diameters, and no functional form for the
				gradient is imposed. (b) Scenario 3 Asymmetric development. Sources of HA are confined to one
				side of the tissue (the right), so that the HA field decays laterally and establishes both
				horizontal and vertical gradients (minimum $\sim 2.691$ blue, maximum $\sim 3.726$ red).}
			\label{fig-HAconc}
		\end{center}
	\end{figure}

\begin{figure}
		\captionsetup{singlelinecheck = false, justification=raggedright}
		\begin{center}
			\begin{tabular}{c c}
				\includegraphics[width=0.23\textwidth]{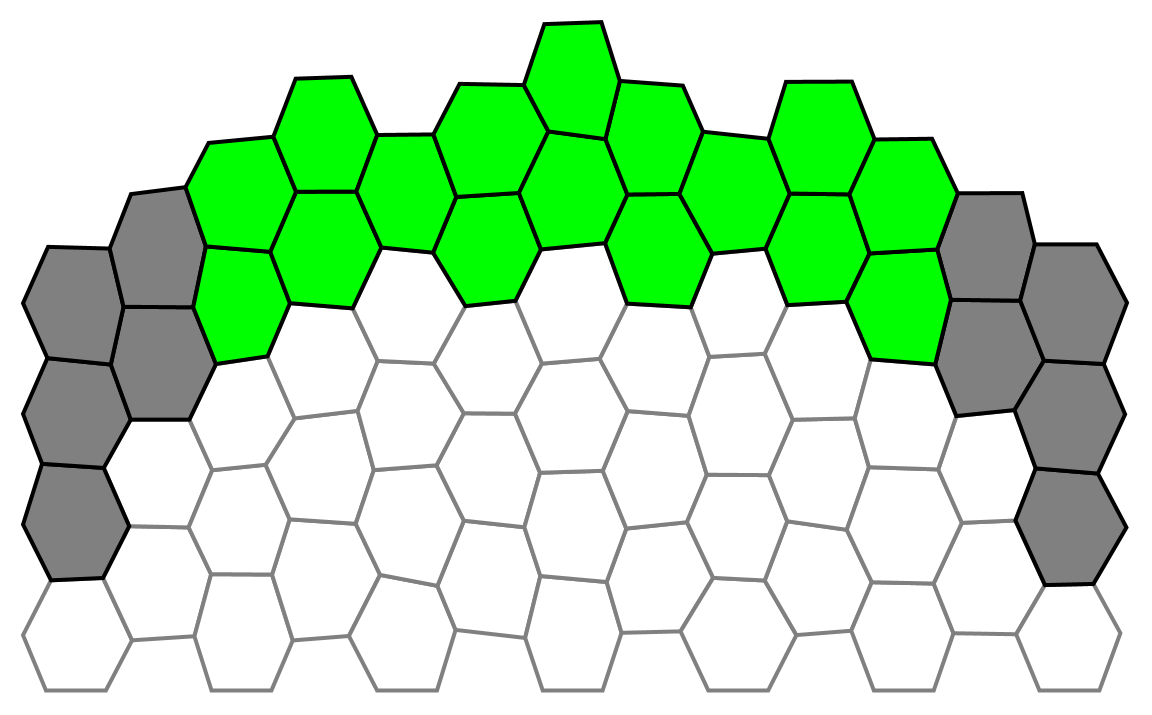}
				& \includegraphics[width=0.23\textwidth]{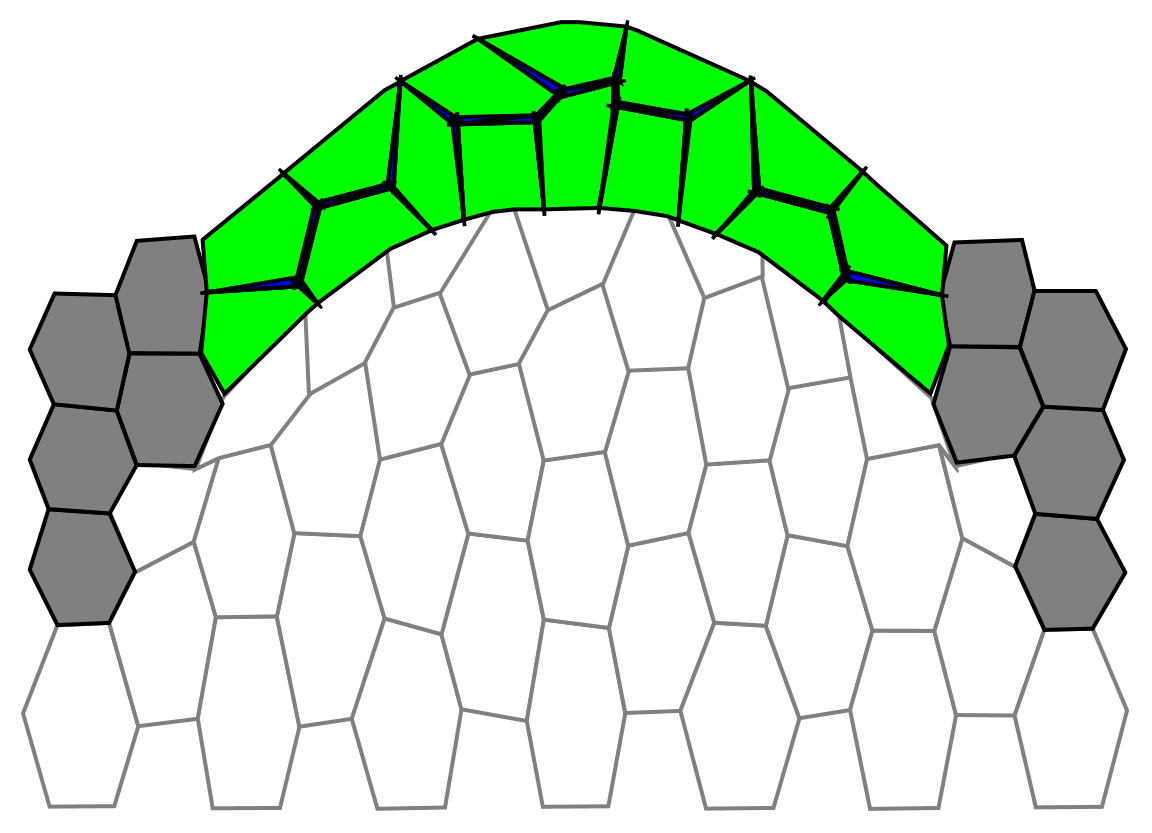} \\
				(a)  & (b)\\
				\includegraphics[width=0.23\textwidth]{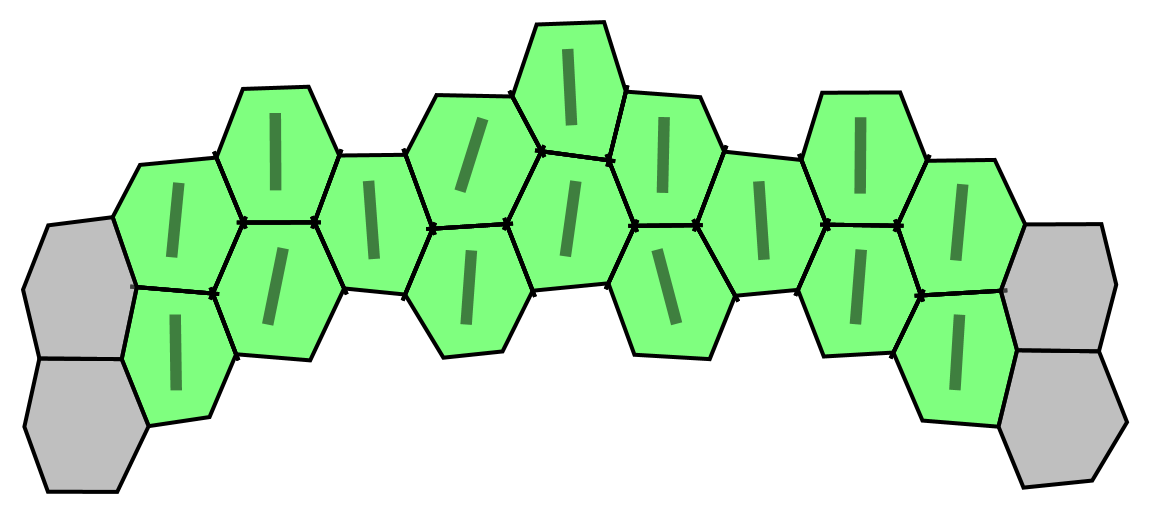}
				& \includegraphics[width=0.23\textwidth]{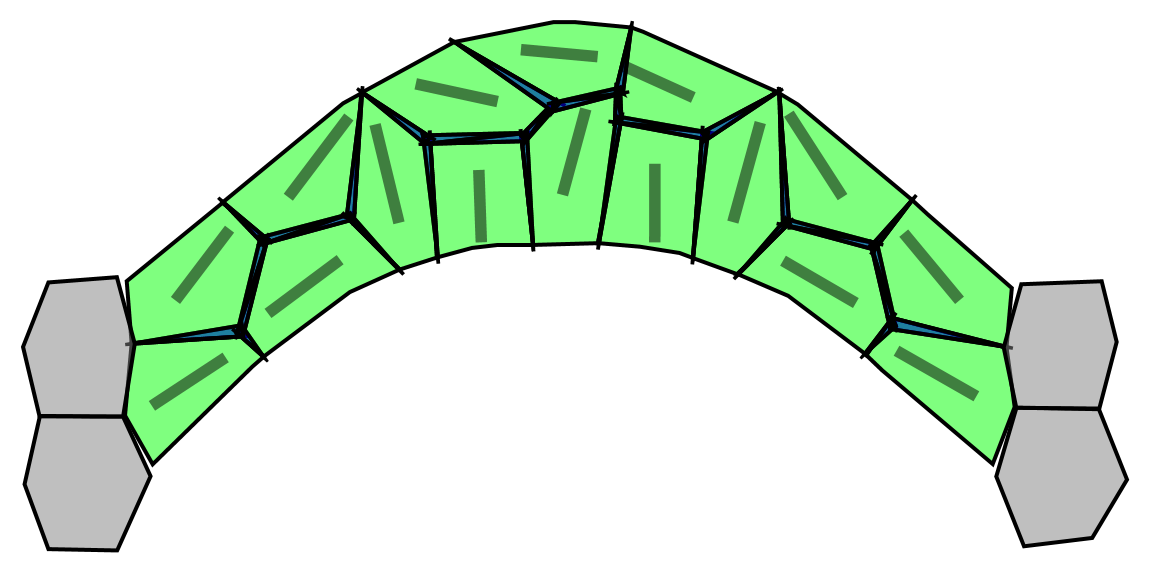} \\
				(c)  & (d)
			\end{tabular}
			\caption{{\it Budding driven by hyaluronan concentration gradients.} (a)~Initial configuration: two rows of epithelial cells (gray, with the HA producing subset in green) embedded in the extracellular fluid network. (b)~Configuration after $10$ iterations of the coupled scheme, in which the HA field is solved given the deformed network and the mechanical energy is then minimized. Osmotic swelling of the compartments adjacent to the HAS-treated cells is greater on the upper surface than on the lower one, and the resulting mismatch drives the layer out of plane. The direction of budding is not prescribed; it follows from the asymmetry of the concentration field about the treated region. (c)~Cell orientation in the initial configuration, where the
				long axes (black lines) are predominantly aligned with the vertical. (d)~Cell orientation in (b), where the long axes in the outermost layer of the bud have reoriented toward the horizontal.}
			\label{fig-result_by_budding}
		\end{center}
		
	\end{figure}

	The result is that budding is reproduced without adjusting individual junctional tensions: the deformation follows from the fluid energy acting on the extracellular compartments together with the change in target shape index of the treated cells. Intercellular compartments where hyaluronan has accumulated expand against the elasticity of the surrounding tissue, while those where the field has decayed do not, and since the non-HAS cells of the surrounding oval boundary remain unchanged and so act as an effective boundary, and the resulting differential across the layer drives it out of plane. The direction of the bud is not imposed; it follows from the asymmetry of the concentration field about the treated region. The accompanying change in cell orientation is shown in Fig.~\ref{fig-result_by_budding}. In the initial configuration the long axes of the cells are predominantly vertical, panel~(c); after the bud has formed, the cells in the outermost layer have reoriented toward the horizontal (perpendicular to the direction of the deformation), panel~(d). This reorientation is consistent with the experiments~\cite{MUNJAL2021}, and is notable because no apical--basal tension asymmetry is prescribed in the model: the anisotropy emerges from the hyaluronan field and the mechanical response to it. Budding therefore results from the localized concentration of hyaluronan in the intercellular compartments alone, without requiring a programmed cell phenotype. Budding requires a spatial difference in the HA field: the HAS-treated region accumulates more HA and expands relative to the surrounding tissue. Fig.~\ref{fig-HAconc} (a) shows the HA concentration field for the budding case, which spans nine cells horizontally. Budding without explicit apical--basal tension changes requires both a sufficiently long decay length, approximately three cells here, and a sufficiently wide treated region: we expect that reducing the decay length to one cell and the treated width to less than three cells would not produce buckling.
	
	
	\subsection{Lateral asymmetry from one-sided hyaluronan exposure}
	
\begin{figure}
		\captionsetup{singlelinecheck = false, justification=raggedright}
		\begin{center}
			\begin{tabular}{c c}
				\includegraphics[width=0.23\textwidth]{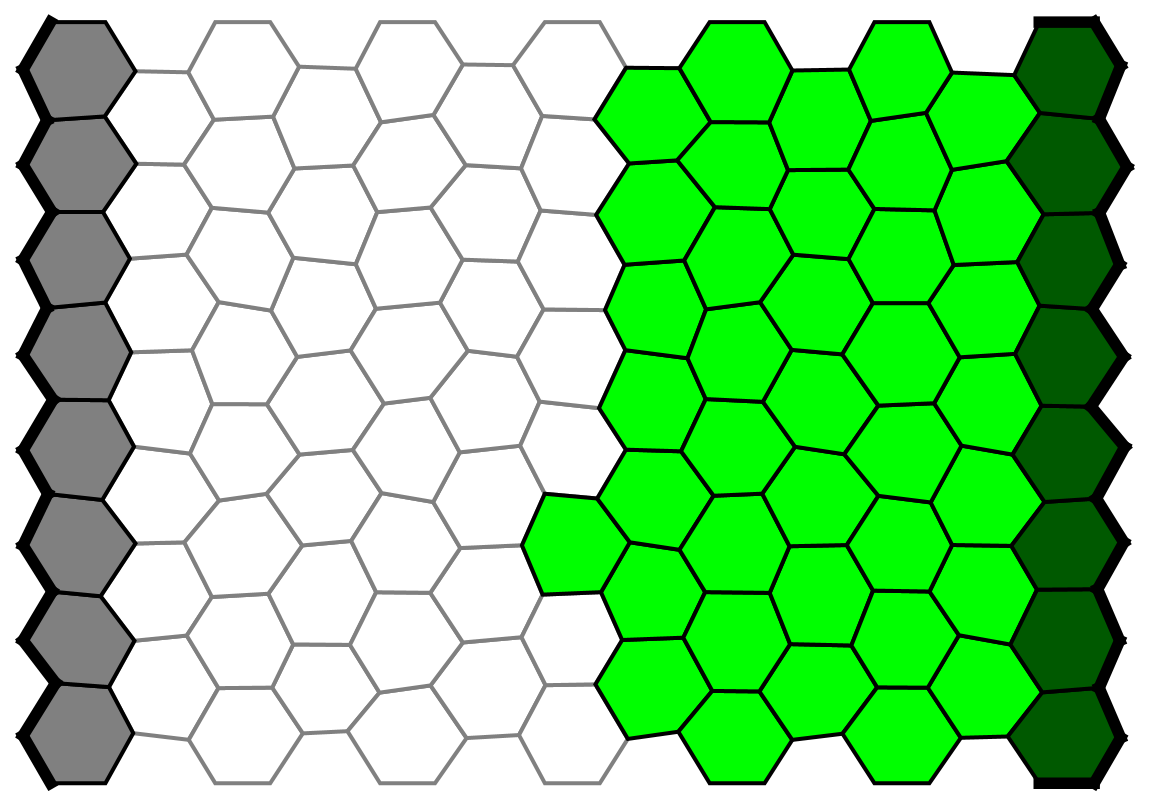}&
				\includegraphics[width=0.23\textwidth]{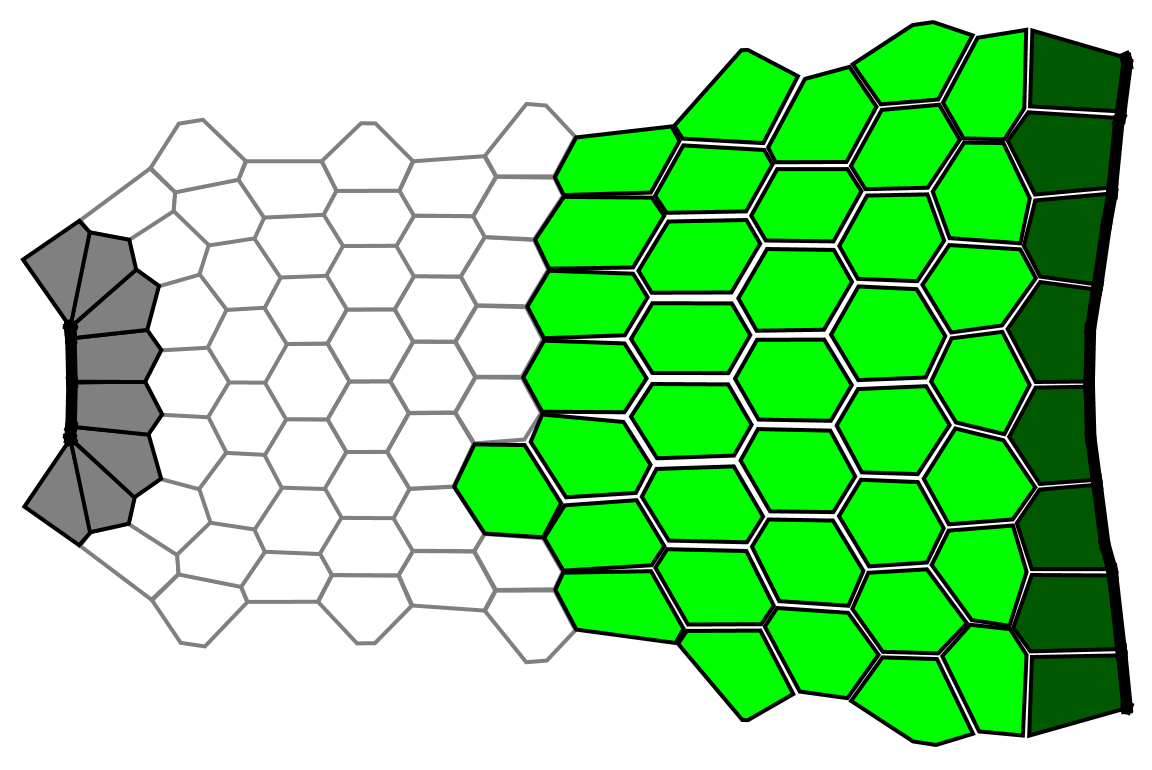} \\
				(a) & (b)
			\end{tabular}
			\caption{{\it Asymmetry driven by a hyaluronan concentration gradient.}
				(a)~Initial configuration: two rows of epithelial cells (white and gray, with the HAS-treated subset in green) embedded in the extracellular fluid network. (b)~Configuration after $10$ iterations of the coupled scheme, in which the HA field is solved on the deformed network and the mechanical energy is then minimized. Compartments on the HAS-treated side swell osmotically while those on the far side remain near their reference size, and the resulting lateral mismatch tilts the tissue away from its initial symmetry.}
			\label{fig-result_by_asymmetry}
		\end{center}
	\end{figure}

	In the third scenario the HAS-expressing cells are confined to one lateral side of the tissue, in contrast to a localized region used for budding. Two differences between the configurations in Figs.~\ref{fig-result_by_budding}(a) and~\ref{fig-result_by_asymmetry}(a) should be noted. The first is the orientation of the HAS-expressing cells relative to the untreated cells, which lie horizontally in the budding case and along the lateral edge in the asymmetric case. The second, and the more consequential, is the treatment of the neighboring tissue. In the budding configuration the treated region is flanked on both sides by cells held fixed since the tissue adjacent to a bud is not observed to deform with it. These cells act as a boundary condition that confines the deformation to the treated region. In the asymmetric configuration no such cells are present and the treated region extends to the edge of the tissue, so that the response is a bend of the tissue as a whole rather than a localized protrusion.
	
	The hyaluronan field calculated from Eq.~\eqref{eq:ha_superposition} is shown in Fig.~\ref{fig-HAconc}(b). Although the concentration contribution from each source decays isotropically with distance, the sources are confined to one lateral side of the tissue. Their superposition therefore produces a spatially asymmetric concentration field. This asymmetry arises from the source distribution, without introducing directional dependence into the kernel.
	
	The mechanical response is shown in Fig.~\ref{fig-result_by_asymmetry}. Compartments on the HAS-exposed side carry high HA and expand osmotically, while those on the opposite side, where the field has decayed, do not. The two sides of the tissue therefore change length by different amounts, and the mismatch is accommodated by curvature rather than by uniform expansion. Three contributions act together: the fluid pressure gradient across the tissue, the difference in target swelling ratio between the two cell groups, and the difference in shape index, which sets the relative fluidity of the two sides. The pressure gradient is obtained from a two-dimensional exponential kernel; see the Appendix for details.
	
	Cell swelling and compartment swelling enter as independent inputs, and both contribute here: the cells are given different target areas on the two sides, and the pressure acts on the compartments. Because the cells remain adhesive, the two act together provided the junctions do not rupture, regardless of the difference in their strengths. Physically the two are linked, since the same hyaluronan would drive both through Donnan osmotic swelling of the cells as well as of the extracellular space. We prescribe the cell targets rather than deriving them from the field, since the two can be measured independently; this allows their contributions to be varied one at a time. The two could be merged into a single input provided the resulting model remains well constrained. The present simulations include both prescribed changes in cell target area and the hydrostatic pressure generated by Donnan swelling in the extracellular compartments, and we are not aware of studies that establish their relative dominance in this setting. The prescribed cell targets also help provide sufficient constraints for the energy minimization; weakening them in our implementation produces unphysical, distorted geometries, limiting a direct comparison. When strong adhesion restricts cell separation, swelling is accommodated through the cell target areas rather than through expansion of the intercellular compartments, which is a modeling prescription rather than an independently derived osmotic response of the cells. We tested two alternatives, assigning the cell targets from the pressure field and removing the fluid energy (Fig. A6 in Appendix), and these show that both the intracellular and intercellular compartments contribute to the final geometry. We therefore cannot identify either contribution as dominant from the present simulations, which show only changes in morphology.
	
	Two features of this result are worth noting. First, the deformation is not a transient of the minimization: the curvature reflects a change in the stress-free state rather than an incomplete relaxation (see Methods). Second, the same energy functional produces budding when the sources are centered and lateral asymmetry when they are displaced to one side, together with the corresponding change in the surrounding tissue, so that the two morphologies follow from one mechanism under different hyaluronan distributions and boundary conditions rather than from distinct physical processes. This asymmetrical swelling mechanism provides a candidate physical route to the left--right asymmetry observed during gut development, in which one side of the tube elongates relative to the other. In the present model the differential arises from osmotic swelling driven by asymmetric HA exposure, without recourse to differential proliferation.
	
	\subsection{Tube formation by closure, driven by tension and fluid pressure gradient}
\begin{figure}
		\centering
		\captionsetup{singlelinecheck = false, justification=raggedright}
		\includegraphics[width=0.7\linewidth]{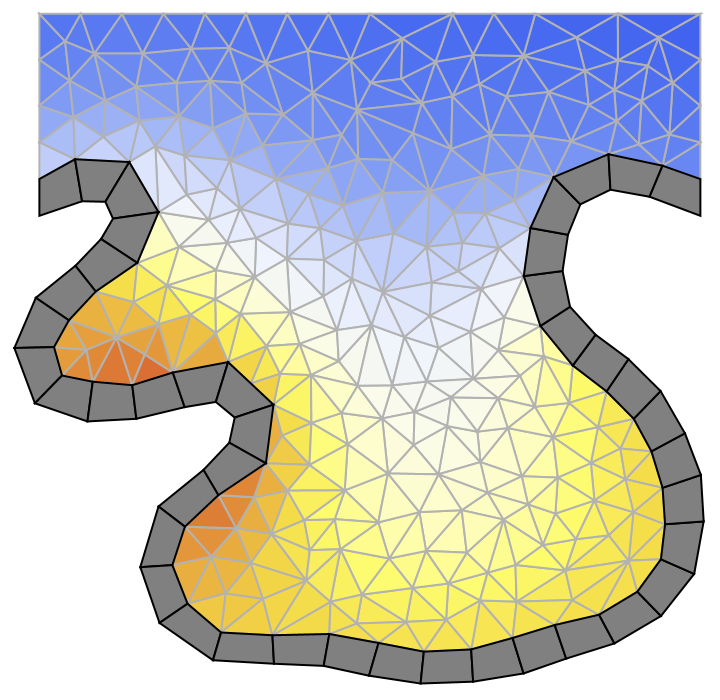}
		\caption{{\it Triangulated extracellular compartment.} The fluid region above the epithelial sheet is filled with triangles bounded below by the apical surface and above by a curve The lower vertices coincide with the apical vertices of the tissue, so that the two deform together, and the triangulation is regenerated at each iteration on the current configuration. Color denotes the local osmotic pressure, obtained from the hyaluronan field on the same geometry. Pressure is shown from a minimum of $0$ (blue) to a maximum of $0.035$ (red).}
		\label{fig-external_fluid}
	\end{figure}

	The imposed forces at the two ends of the row are motivated by the literature on otic vesicle formation. In this scenario, the imposed forces bring the tissue edges together to promote closure. Whether closure can occur without these forces has not been established by the present simulations. The final scenario unfolds over a longer developmental interval than the preceding scenarios, so cell division must be included. Each dividing cell is split so that its daughters lie side by side, extending the layer along its length. The outward forces generated during mitotic rounding can themselves aid buckling and thereby promote invagination~\cite{Kondo2013,Finegan2019,Gupta2022}, which is the aspect of division relevant to the present model. To investigate the simplest tubular formation, we consider a single row of columnar cells and remove the cell--cell gaps. In terms of the fluid, we retain only the interaction between the cells and the surrounding fluid, as shown in Fig.~\ref{fig-external_fluid}, which is represented by a triangulated compartment above the sheet and regenerated at each iteration for the current configuration.
	
	The tension and pressure fields are obtained from the same superposition used in the preceding scenarios, with the HAS-expressing cells as sources. Apical and basal tensions are assigned separately, so that the sheet first thickens under tension applied equally to both surfaces, then bends as the basal target is lengthened relative to the apical, and finally deepens as the apical tension is increased. (As noted earlier, we omitted the intercellular gaps to reduce computational complexity.) The fluid pressure acts on the triangulated compartment through the pressure--volume work of Eq.~\eqref{eq:pA}, and the compartment shares the apical vertices of the sheet, so that pressure and tension act on the same surface. 
	
	The simulation begins from a single row of $n=18$ cells with target area $\hat{A}_0 = 1.2$ and target shape index $\hat{q}_0 = 4.0$, so that $\hat{P}_0 = \hat{q}_0\sqrt{\hat{A}_0}$. The tension gradient is initially centered on cells $6$ through $12$. Both the tension and the pressure fields are obtained by superposing screened kernels centered on the cell centroids of this source set, so that a cell at distance $\hat{r}$ from a source receives a contribution $\hat{g}_0\exp(-\hat{r}/\hat{\lambda})$, summed over all sources. The apical and basal tensions, $\hat{\gamma}_a$ and $\hat{\gamma}_b$, and the fluid pressure $p$ are each constructed in this way, differing only in their amplitude and decay length. The energy minimized at each step is
	
	\begin{align}
		\hat{E} = &\sum_\alpha \left[ \tfrac{1}{2}\hat{\mu}_A (\hat{A}_\alpha - \hat{A}_{0\alpha})^2 + \tfrac{1}{2}\hat{\mu}_P (\hat{P}_\alpha - \hat{P}_{0\alpha})^2 \right]\notag\\
		+ &\sum_\alpha \left[ \hat{\gamma}_{a,\alpha}\, \hat{\ell}^a_\alpha + \hat{\gamma}_{b,\alpha}\, \hat{\ell}^b_\alpha \right]
		- \sum_k \hat{p}_k \hat{A}_k ,
	\end{align}
	where $\hat{\ell}^a_\alpha$ and $\hat{\ell}^b_\alpha$ are the apical and basal edge lengths of cell $\alpha$, and the final sum runs over the triangles of the fluid region, with $\hat{\mu}_A = 50$ and $\hat{\mu}_P = 2$.
	
\begin{figure*}
		\centering
		\captionsetup{singlelinecheck = false, justification=raggedright}
		\includegraphics[width=1\linewidth]{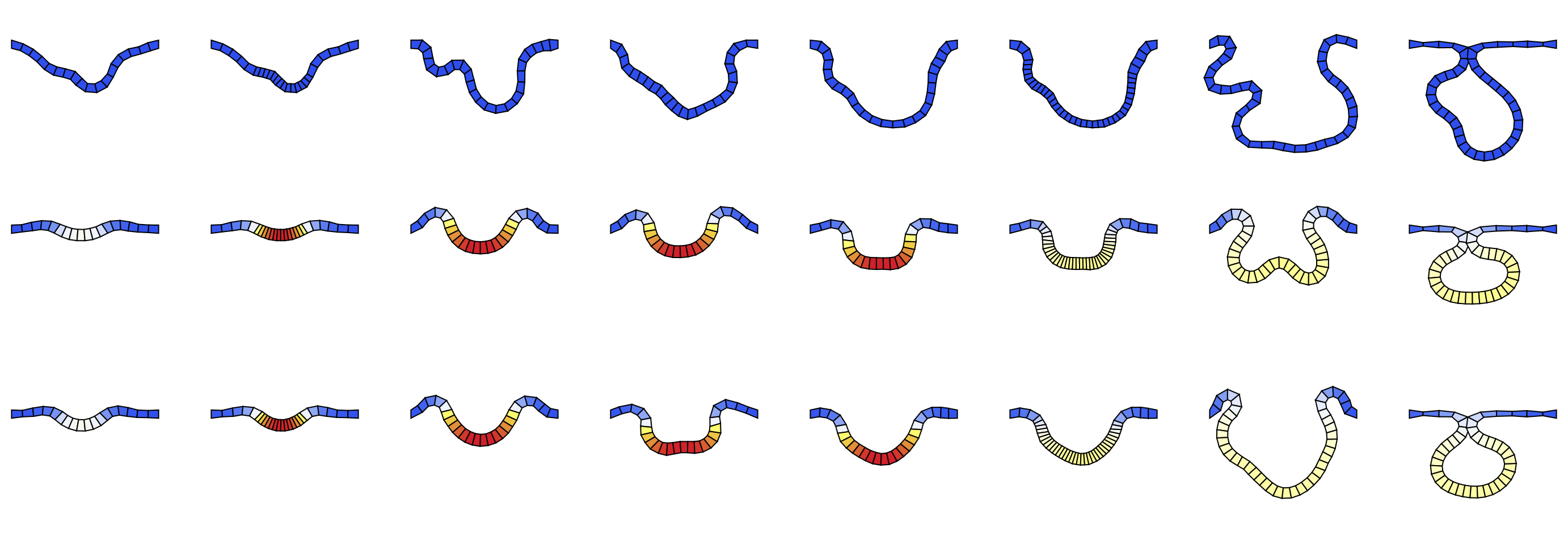}
		\caption{{\it Tube formation by closure under tension and fluid pressure.} Columns show successive stages of the simulation, from the initial sheet (a straight horizontal line) to the closed vesicle. Rows compare three cases obtained by setting one contribution to zero at a time, with each relaxation limited to 100 iterations: fluid pressure alone (top), tension alone (middle), and both together (bottom). Color denotes the local apical tension amplitude (basal tension is lower or set to zero), from blue ($0$) to red ($0.6$). All three cases close, and the final vesicles differ only in depth, but the intermediate morphologies differ: the fold is angular under fluid pressure alone, smooth under tension alone, and widest when both act together.}
		\label{fig-cyst-formation}
	\end{figure*}

	To separate the two contributions we compare three cases, obtained by running the same sequence with the same functions and setting one term to zero at a time: tension alone, with the pressure amplitude set to zero; fluid pressure alone, with the tension amplitudes set to zero; and both together. Table~\ref{tab:closure-steps} lists the parameters used at each step. The apical tension is taken positive and the basal tension negative, so that the apical surface contracts while the basal surface lengthens, mimicking basal expansion. Each cell is thereby driven toward a wedge shape, and the layer bends toward the apical side, corresponding to apical constriction. We assume that this activity decreases once the fold has formed, and the apical amplitude is therefore reduced in the later steps. The closure step differs from the others in that two pairs of cells on either side of the opening are pulled together over ten iterations, mimicking the mechanical forces that close the structure. Closure is represented by a line tension acting between the paired apical vertices,
	\begin{equation}
		\hat{E}_{\text{closure}} = \hat{k}_P \sum_{p} \left| \hat{\mathbf{r}}_{p_1} - \hat{\mathbf{r}}_{p_2} \right| ,
	\end{equation}
	which is a tension rather than a spring, being linear rather than quadratic in the separation. Because $k_P$ has units of energy per length, it is a force, and can therefore be placed on an absolute scale. Throughout, an asterisk denotes simulation units, and the calibration given in Appendix~\ref{app-units} sets $1$~Pa$^* = 1600$~Pa, with $1$~$\mu$m$^{*2}$ corresponding to $50$~$\mu$m$^2$ as measured from Fig.~\ref{fig-result_by_groups}(c),(e); one simulation force unit is then $80$~nN. Ramping $k_P$ from $0.1$ to $1.0$ over ten iterations therefore corresponds to a closure tension increasing from $8$ to $80$~nN, below the $100$--$120$~nN measured for cell--cell junction forces, so that closure proceeds by progressively raising the tension around the closing cells rather than by drawing them together mechanically. The same conversion places the apical and basal tensions at $8$ and $4$~nN and the fluid pressure at $8$~Pa, the latter consistent with the hydrostatic pressures inferred for intercellular flow. The results are shown in Fig.~\ref{fig-cyst-formation}.
	\begin{table}[t]
		\centering
		\caption{Parameters used at each step of the closure sequence. $\hat{\lambda}$ is the decay length of the tension ($\hat{\lambda}_\gamma$) and pressure kernels ($\hat{\lambda}_p$), $\hat{\gamma}_a$ and $\hat{\gamma}_b$ the apical and basal tension amplitudes, and $\hat{p}_0$ the pressure amplitude. Dashes indicate that the term is not applied at that step.}
		\label{tab:closure-steps}
		\renewcommand{\arraystretch}{1.3}
		\begin{tabular}{|l|l|c|c|c|c|c|}
			\hline
			\textbf{Step} & \textbf{Operation} & $\hat{\lambda}_\gamma$ & $\hat{\gamma}_a$ & $\hat{\gamma}_b$ & $\hat{\lambda}_p$ & $\hat{p}_0$ \\
			\hline
			1 & Relax & 1.5 & 0.1 & 0 & 3 & 0.005 \\
			\hline
			2 & Divide (cells 6--12) & -- & -- & -- & -- & -- \\
			\hline
			3 & Relax & 1.5 & 0.1 & $-0.01$ & 3 & 0.005 \\
			\hline
			4 & Relax & 1.5 & 0.1 & $-0.05$ & 3 & 0.005 \\
			\hline
			5 & Relax & 1.5 & 0.1 & $-0.05$ & 3 & 0.005 \\
			\hline
			6 & Divide (cells 4--20) & -- & -- & -- & -- & -- \\
			\hline
			7 & Relax & 1.5 & 0.05 & 0 & 3 & 0.005 \\
			\hline
			8 & Closure & 1.5 & 0.05 & 0 & 3 & 0.005 \\
			\hline
		\end{tabular}
	\end{table}
	The strip geometry is based on the model of Munjal \etal, in which tension alone drives anisotropic budding; here, however, we start from a flat geometry, incorporate the fluid network, adding cell division steps, and add a closure step. Note that the fluid-only case can grow without bound, since it has no tension term and no cells or fluid connected on the opposite, basal side; we therefore limited each relaxation to 100 minimizer iterations to compare the effect of each constraint, and as a result the fluid-only case shows some fluctuation in its profile. A fluid-only case with cells (or fluid) connected on the basal side would become similar to Scenario 3, and adding tension would change the budding profiles. All three cases eventually close, but their final shapes differ. The pressure-only case produces a vertically elongated vesicle because nothing restricts it (in a more detailed simulation, the structure underneath the tube could provide this limit), whereas the tension-only case produces a horizontally elongated vesicle. When both contributions are present, the vesicle is more nearly circular in cross-section. The difference in depth arises because the fluid pressure acts only downward, whereas the tension acts as a counteracting force that limits the deformation. The local undulations in the pressure-only case result from the limited number of optimization iterations in this underconstrained case; adding further constraints would make it equivalent to Scenario 3.

	\section{Discussion}
	
	Our extended vertex model coupled to an HA-driven extracellular fluid network demonstrates how spatially patterned osmotic forces can generate distinct tissue architectures. Rather than acting solely through biochemical signaling or changes in cell-cell adhesion, HA can directly generate mechanical forces within intercellular spaces that deform tissues and, under certain conditions, alter their topology. In this sense, HA can act as a mechanical morphogen: its spatial distribution encodes a pattern of extracellular forces that, together with tissue geometry and cellular mechanics, determines tissue architecture. Widespread HA overexpression, coupled with heterogeneous cell-cell adhesion, can generate multiple lumens through junction rupture, whereas localized HA accumulation can drive budding and tissue asymmetry through spatially varying osmotic forces. In tube formation, HA-mediated pressure acts together with apical–basal tension to influence the geometry of tissue folding.
	
	The analogy to a classical morphogen lies in the conversion of a spatial concentration field into spatially organized tissue architecture. Unlike conventional morphogens, which primarily act through biochemical signaling and cell-fate specification, HA can directly generate spatially patterned mechanical forces through osmotic pressure in the intercellular space. Although HA also engages signaling pathways through receptors such as CD44 and RHAMM~\cite{Toole2004}, these effects are not included in our model. More broadly, our results suggest that cells can pattern extracellular materials to generate mechanical forces that shape tissue architecture, providing a complementary mechanism to conventional biochemical morphogenesis. 
	
	These mechanisms emerge from the same biophysical framework, differing in the HA distribution pattern, the patterning and/or breaking potential of cell--cell junctions, and the resulting mechanical response. The multi-lumen phenotype resembles the complex cystic masses observed in mammary tissue pathology~\cite{ATHANASIOU2014}, suggesting that dysregulated HA production affecting many cells would create the mechanical conditions for multiple cyst formation. The budding mechanism provides a potential link to branching morphogenesis should there be two neighboring localized regions of HA overexpression. Moreover, the graded case suggests that HA gradients could create the asymmetric swelling forces and bending moments capable of initiating the leftward tilt characteristic of gut rotation.
	
	One useful distinction is between topological remodeling, in which cell–cell connectivity changes, and geometrical remodeling, in which tissue shape changes while connectivity is largely preserved. In the multi-lumen scenario, HA-driven fluid accumulation can rupture junctions and create persistent cavities, thereby altering tissue topology. By contrast, in the budding, asymmetric, and tube-formation scenarios, the tissue primarily deforms through curvature, bulging, or folding without widespread junction failure. Thus, HA-mediated forces can act in two qualitatively different regimes: by driving topological changes when junctions fail, or by driving geometrical changes when tissue cohesion is maintained. Measurements of junctional strength (in the form of different types of adhesion molecules) would predict whether or topological or geometrical changes dominate. 
	
	The localized-HA results suggest a particularly appealing patterning experiment: the number and spacing of HA-producing regions should control the number and spacing of buds or branches. One localized source should generate one bud; two sufficiently separated sources should generate two buds, while closely spaced sources may merge into a single broader protrusion. The distinct intermediate morphologies predicted during tube closure provide additional experimentally testable signatures of the underlying mechanical mechanism. Although fluid pressure and cellular tension can both ultimately produce closure in our model, they generate different shape trajectories: fluid pressure alone produces a sharper, more angular fold, whereas tension produces a smoother, progressively deepening cup, with the combination of pressure and tension generating a broader intermediate structure. Thus, the final closed morphology alone may be insufficient to identify the forces responsible for tubulation; instead, time-resolved measurements of tissue shape during closure can discriminate among competing mechanisms. In particular, observations of fold width, curvature, and their evolution during closure could be compared directly with the predicted trajectories, allowing pressure-dominated, tension-dominated, or coupled mechanisms to be supported or falsified.
	
	As for limitations of our model, the model does not contain mechanochemical feedback in the full sense, although such feedback is well documented in epithelia~\cite{Gudipaty2017,DiMeglio2022}. Models that retain such a feedback exhibit qualitatively richer behavior: deactivation of a signaling species by cellular deformation gives rise to Turing patterns and sustained oscillations~\cite{Yu2024} and intermittent hydrostatic bursting produces oscillations in lumen size~\cite{Ruiz-Herrero2017}. A related point concerns bistability. Xue and collaborators~\cite{Xue2025} show that mechanosensitive feedback between cytoskeletal tension and luminal pressure renders intestinal organoid morphogenesis bistable, so that the final state depends on the history of volume changes rather than on the instantaneous parameters alone. Our model contains no such feedback: the hyaluronan field sets the cell targets but is unaffected by the mechanical state, and each configuration is therefore determined by the current parameters rather than by the path taken to reach them. Introducing a strain dependence into hyaluronan synthesis would close this loop, and represents the most direct extension of the present work. 
	
	Moreover, extensions to three dimensions with cells as deformable polyhedrons and cell interfaces as areas such that the flow network becomes three-dimensional pipes, would benefit all of the scenarios considered here, and would be essential for fully characterizing the chirality of gut looping, which depends on the complex patterning that a two-dimensional cross-section cannot capture. Together, these results establish a physical framework in which HA-driven osmotic pressure in the intercellular space, interacting with cell-cell junction mechanics, is sufficient to generate diverse tissue architectures. By varying only the spatial distribution of HA, our extended vertex model produced distinct morphologies in four tissue geometries, indicating that the spatial localization of fluid accumulation is a key determinant of tissue structure. More broadly, because the model depends on osmotic pressure rather than HA specifically, it can be adapted to gradients generated by other sources, such as $\text{Na}^+/\text{K}^+$ ion pumps, and applied to organoids and tissues in which fluid-filled spaces shape morphology, including the breast, lung, pancreas, gut and kidney~\cite{Bagnat2007,Chan2019,Chugh2022}.

	\bibliography{main}
	\section*{Funding}
	
	JMS acknowledges funding support from the National Science Foundation via DMR-2204312. CF acknowledges financial support from NCI R01CA259195, R01CA276392, and R01CA297166.

	\section*{Acknowledgements}
	JMS and KK acknowledge discussion with Tao Zhang during early stages of project.

	\appendix
	
	\section{Scenario 1}
	In this section, we provide a step-by-step simulation procedure with a simpler structure to demonstrate how Scenario 1 is implemented. We briefly present simulation results using different swelling mechanisms and energy functions. Under the constraints tested, we observed no significant effects from applying different energy functions or swelling methods. Our tests identified surface tension, the location of leaky junctions, and the stiffness of the spring network for each group as key factors determining lumen shapes and locations.
	\subsection{More on the methodology}
	We begin by reviewing previous simulations of lumen morphogenesis to identify key approaches that inform our study. Indana \etal~developed two-stage modeling approaches for lumen growth in the epiblast, where early lumen formation is driven by mechanical forces from extensive actin remodeling, while later stages are governed by osmotic pressure from accumulated ions. Their model, although informative, was limited in complexity when scaled to larger cell systems. Lu \etal~ adopted a similar biphasic model, applying it to the generation of small lumens and to their subsequent fusion.
	
	Sabyasachi \etal~\cite{Sabyasachi2018} quantitatively studied the balance between paracellular leaks and osmotic pressure build-up in lumen development. Similarly, McEvoy \etal~\cite{McEvoy2020} explored how gap junctions amplify spatial variations in cell volume within multicellular spheroids, noting the dependence on proliferation-induced solid stress. However, both studies focused primarily on liquid flow dynamics between two cells, making them less applicable to large-scale multicellular systems. Moreover, these lumen models were not designed to capture multi-lumen formation near spheroid boundaries. Therefore, we adopt a cellular-based framework similar to Vasquez \etal~\cite{Vasquez2021} but scaled to larger cell numbers to investigate multi-lumen structural organization.
	
	For simulation, we assume simple hypotheses to compare with collagen-embedded samples:
	\begin{itemize}
		\item We assume a locational distribution of high adhesion cells at the center~\cite{GUO2024}.
		\item We assume a linear pressure gradient from the spheroid boundary to the spheroid center.
		\item The Differential Adhesion Hypothesis~\cite{Foty2005} states that aggregates of cells will self-sort and segregate based on relative adhesion strength.
	\end{itemize}
	
	For the second assumption, we tested both a Laplacian pipe network and a linear pressure gradient. The linear pressure model showed gradual changes, while the Laplacian pipe network shows more accumulation in low adhesion regions. However, this does not change the location of lumens. Additionally, experimental measurements by Lee \etal~\cite{Lee2025} indicate pressure differences between the spheroid center and the outside of the spheroid.
	\renewcommand{\thefigure}{A\arabic{figure}}
	\setcounter{figure}{0}
	
	\begin{figure}
		\captionsetup{singlelinecheck = false, justification=raggedright}
		\begin{center}
			\begin{tabular}{c c c}
				\includegraphics[width=0.15\textwidth]{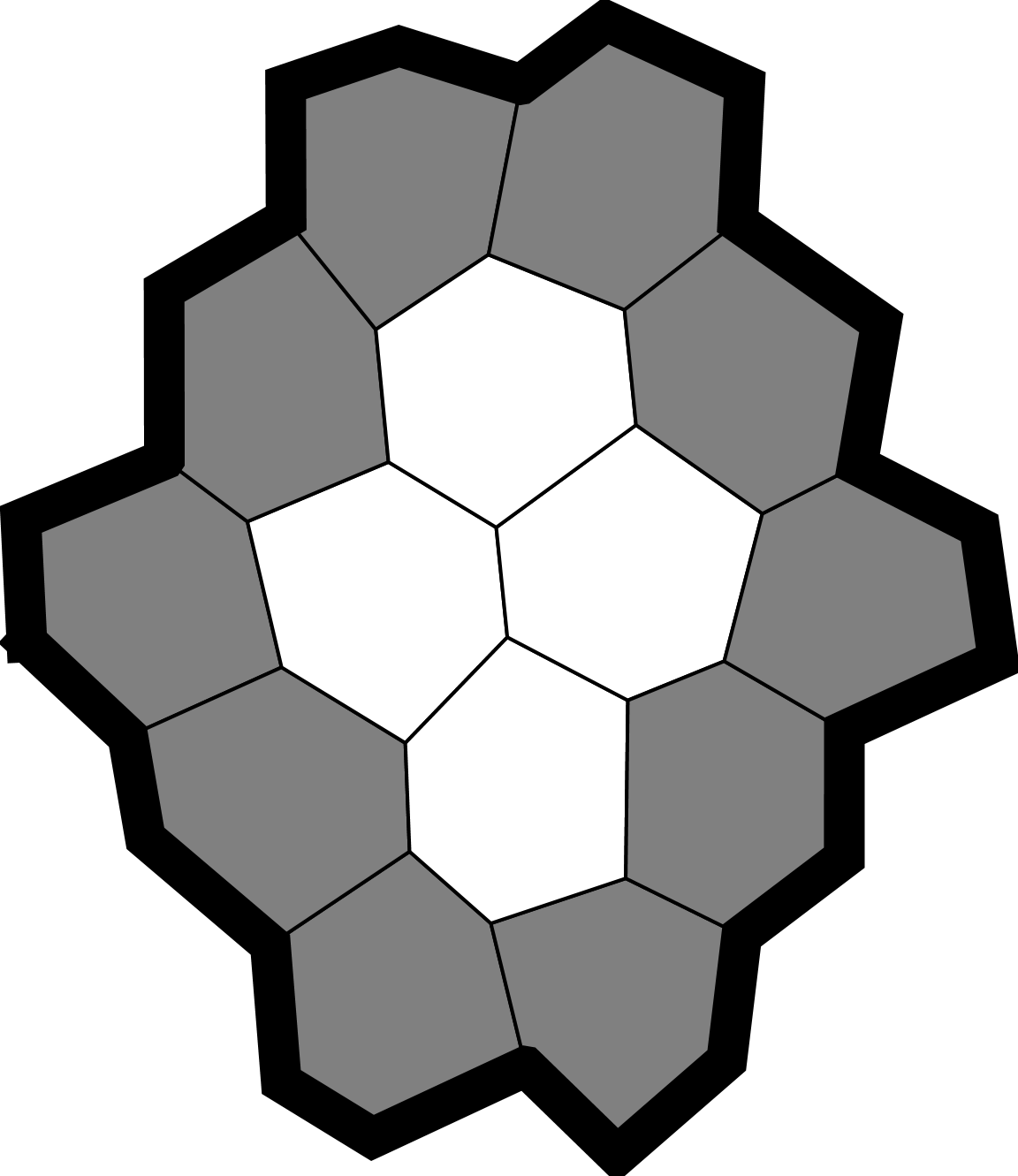}&
				\includegraphics[width=0.15\textwidth]{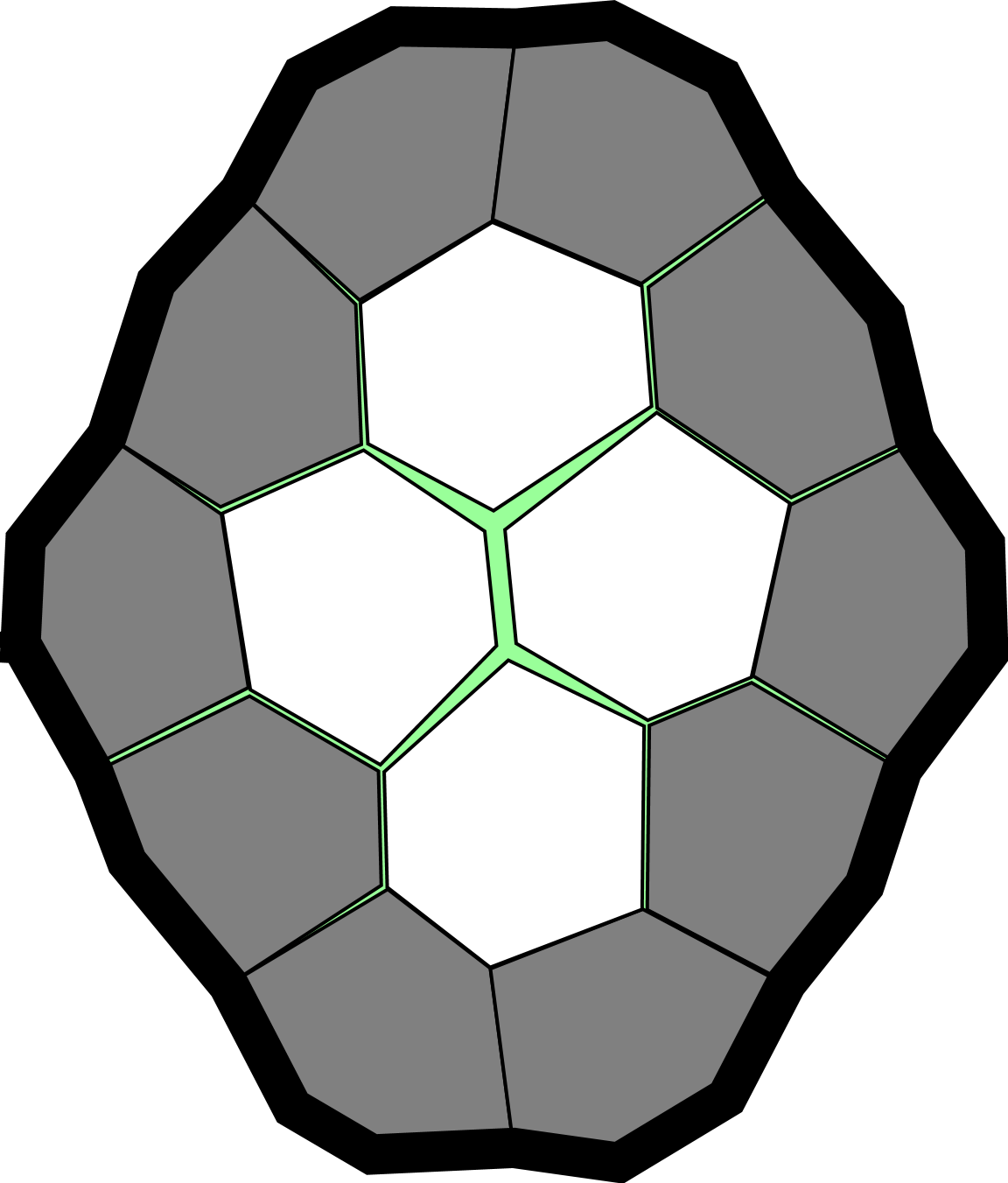}&
				\includegraphics[width=0.15\textwidth]{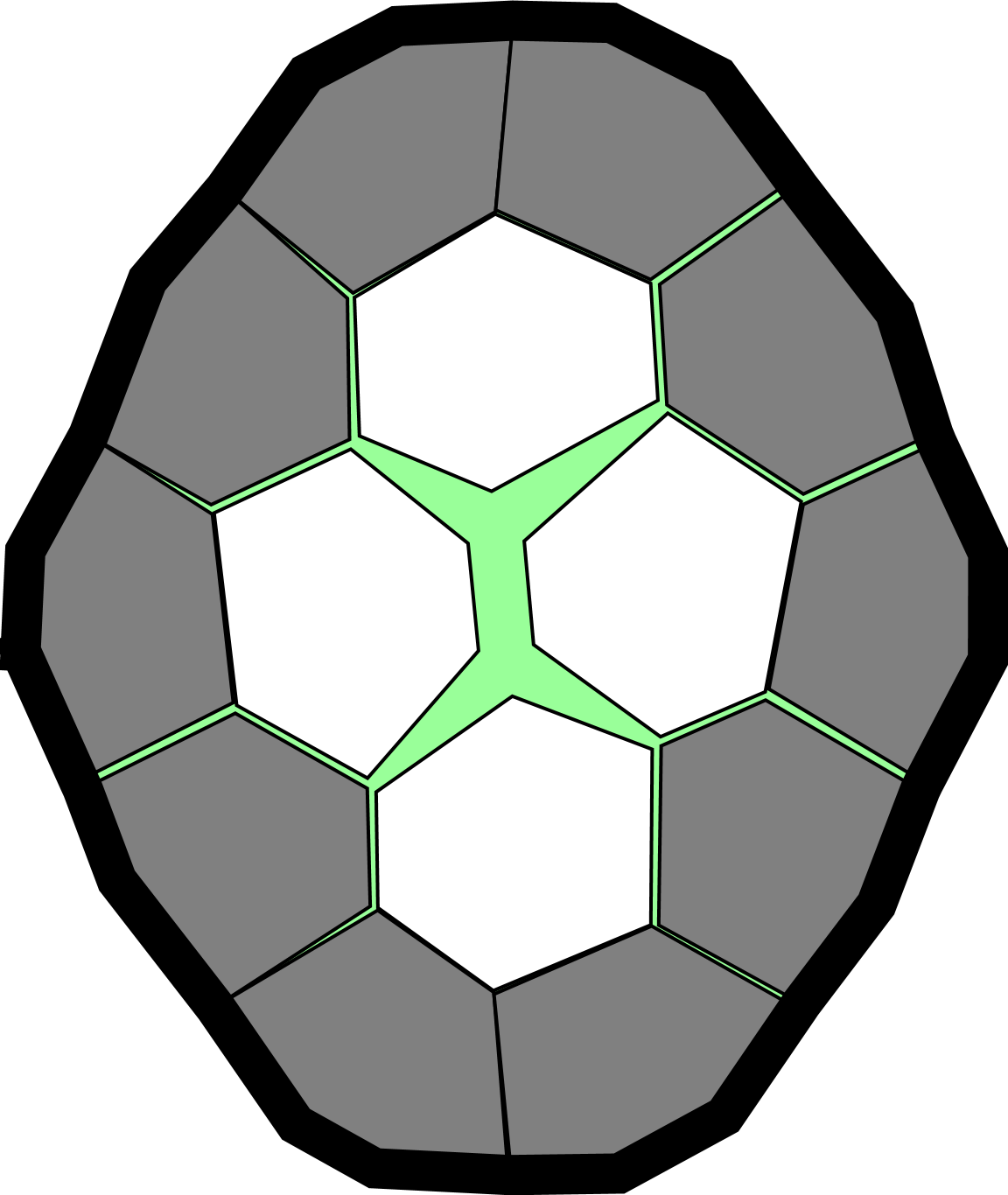}\\
				(a) Spheroid &(b) Pre-phase &(c) During Phase 1\\
				\includegraphics[width=0.15\textwidth]{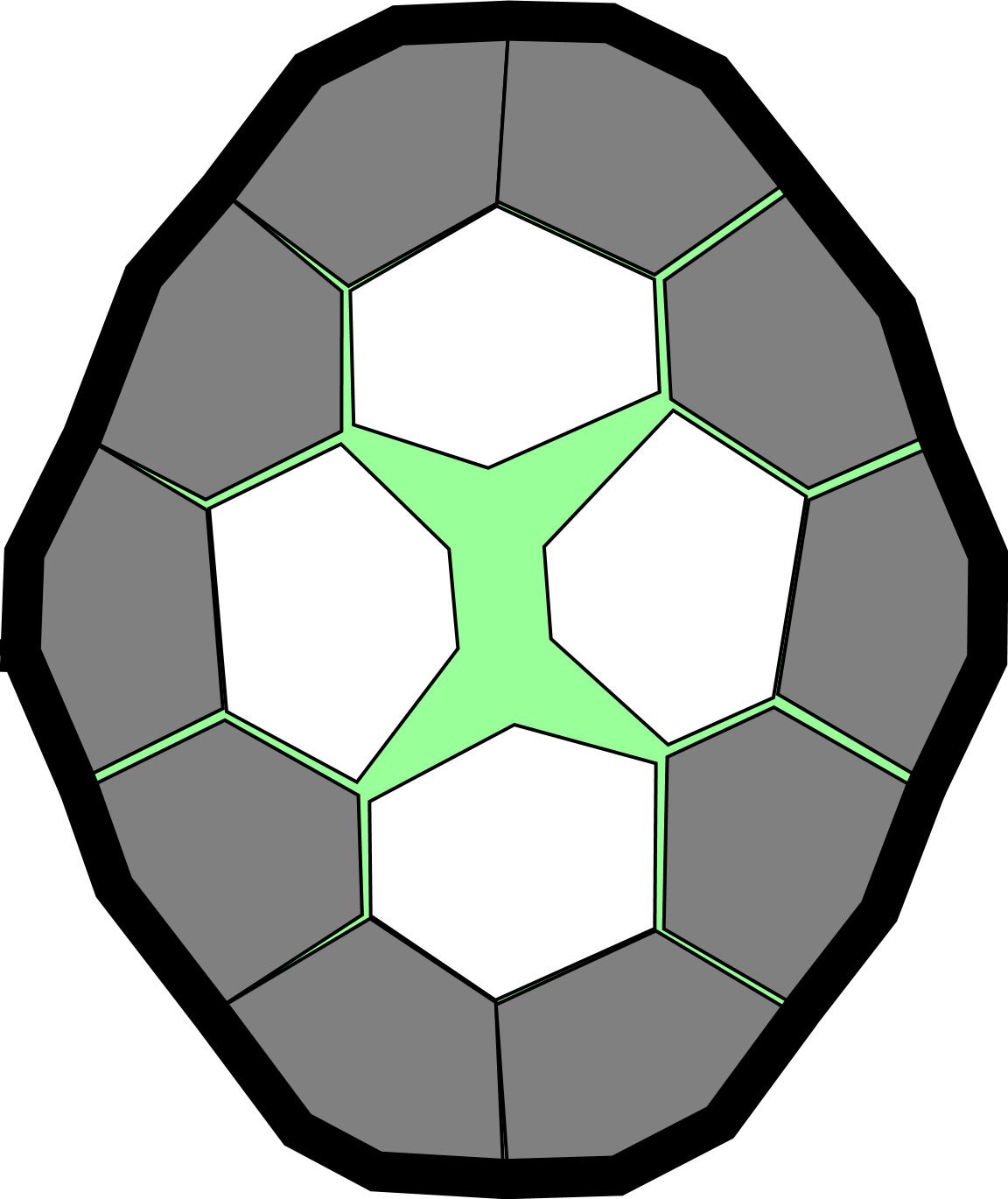}&
				\includegraphics[width=0.15\textwidth]{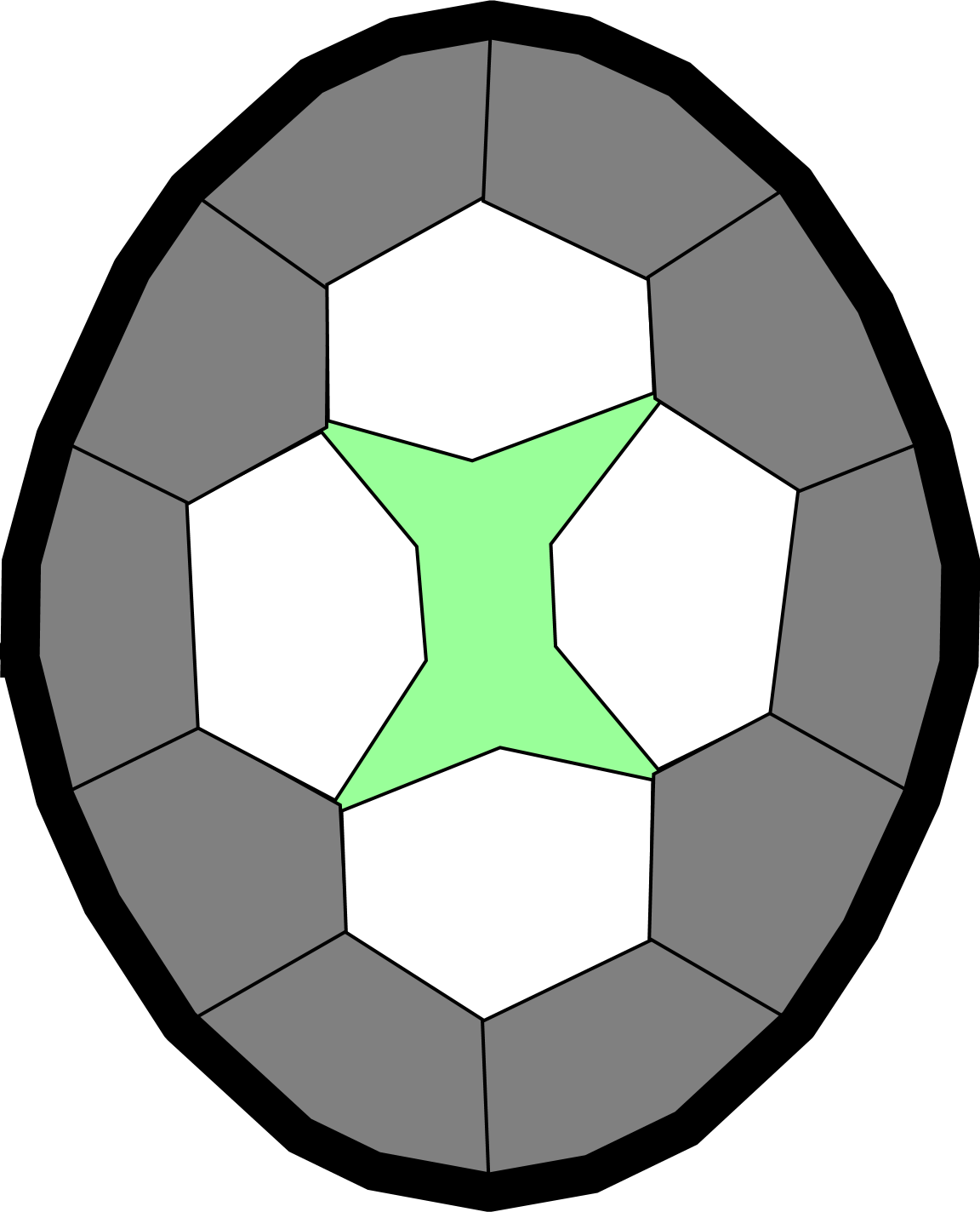}&
				\includegraphics[width=0.15\textwidth]{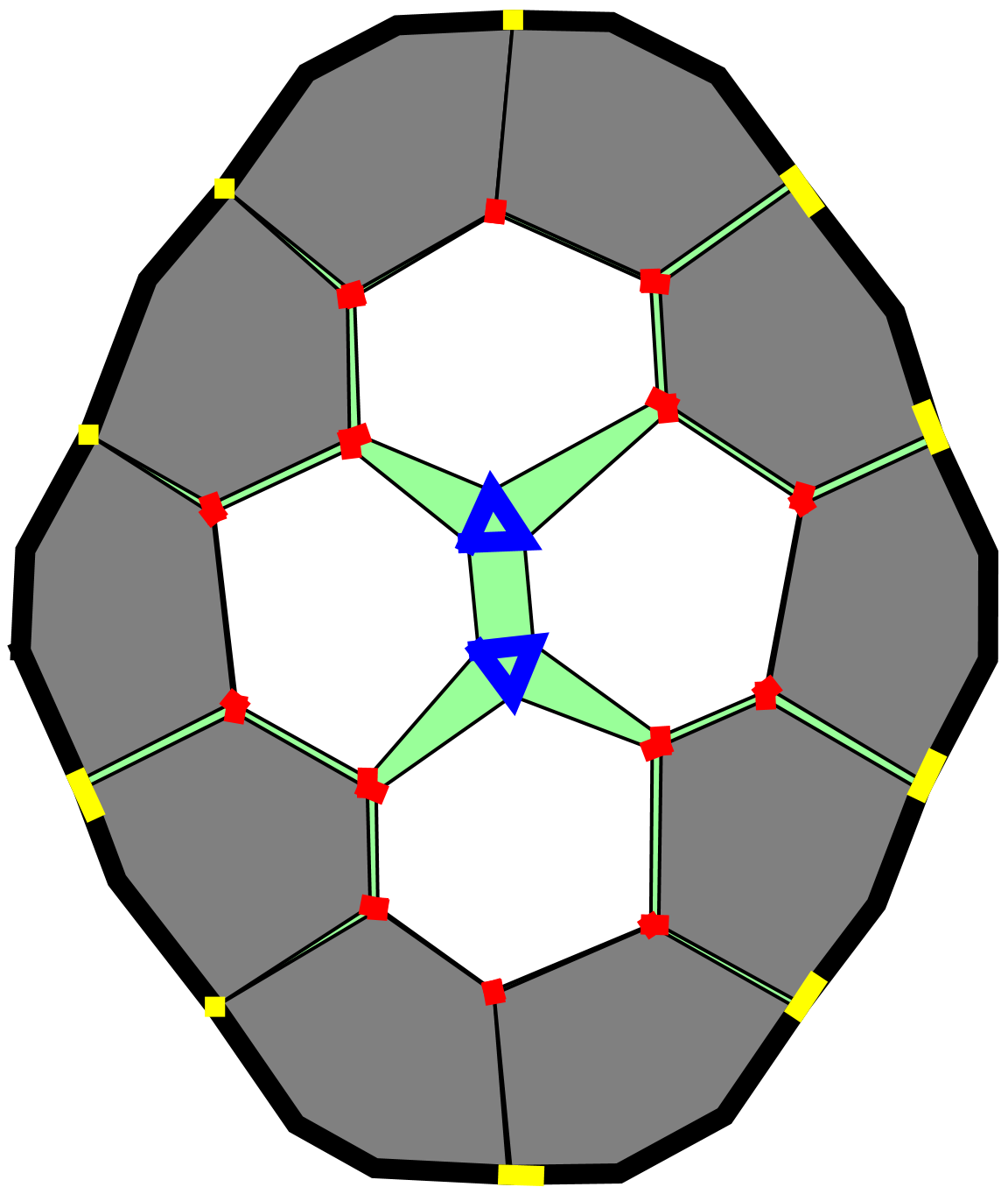}\\
				(d) After Phase 1&(e) Phase 2 &(f) Spring networks\\
			\end{tabular}
			\caption{{\it Small spheroid simulation for centered lumen:} (a) Initial state of the spheroid, (b) Before Phase 1, (c) During Phase 1, (d) After Phase 1, (e) Phase 2, (f) Spring networks: blue lines represent springs between white cells, red lines represent springs between white and gray cells, and yellow lines represent springs between gray cells. These springs depict cell-to-cell adhesion and illustrate how the connections are leaky.}
			\label{fig-mini_spheroid}
		\end{center}
	\end{figure}
	
	We divided cells into two groups: $\alpha$ for leaky junctions (lower cell-cell adhesion) and $\beta$ for higher cell-cell adhesion. The total energy functions used in the minimization process were as follows:
	\begin{itemize} 
		\item Pre-phase: $\hat{E}_{\text{cell}} + \hat{E}_{\text{junction}} + \hat{E}_R$ 
		\item Phase 1: $\hat{E}_{\text{cell}} + \hat{E}_{\text{junction}} + \hat{E}_{\text{pipes}} + \hat{E}_{\text{spheroid}} + \hat{E}_{\text{boundary}} + \hat{E}_R$ 
		\item Phase 2: $\hat{E}_{\text{cell}} + \hat{E}_{\text{junction}} + \hat{E}_{\text{lumen}} + \hat{E}_{\text{spheroid}} + \hat{E}_{\text{boundary}} + \hat{E}_R$ 
	\end{itemize}
	where $\hat{E}_{\text{lumen}}$ and $\hat{E}_{\text{pipes}}$ correspond to $\hat{E}_{\text{fluid}}$ in the main text, but are separated here for clarity.
	
	During the pre-phase, Phase 1, and Phase 2, we applied the following procedures with key differences in constraints:
	\begin{itemize} 
		\item In pre-phase and Phase 1, we set the junction spring stiffness to zero for one group, either near the boundary $(\alpha, \alpha)$ or near the center $(\beta,\beta)$. 
		\item In Phase 2, we assigned non-zero stiffness to all groups and removed nodes with edge distances exceeding the threshold $(1.1)^4$ after Phases 1–4, accounting for the cumulative swelling ratio. 
		\item We used the energy function $\hat{E}_{\text{pipes}}$ during Phase 1 and switched to $\hat{E}_{\text{lumen}}$ during Phase 2. 
		\item The same swelling target $\hat{S}_R$ for $\hat{E}_R$ was applied throughout all phases. 
	\end{itemize}
	
	The main structural differences between Phase 1 and Phase 2 are the presence of broken springs, variations in stiffness, and the approaches used to handle liquid flow.
	
	To better illustrate the steps involved, we will explain each phase using Fig. \ref{fig-mini_spheroid}. In Fig. \ref{fig-mini_spheroid}(a), the gray-colored cells represent group $\alpha$, while the white-colored cells represent group $\beta$. In this example, we depict the lumen located at the spheroid center, similar to the model presented by \cite{Vasquez2021}. To achieve this, we set the junction spring stiffness to zero for group $(\beta,\beta)$.
	
	After the pre-phase, the absence of spring tension leads to the formation of leaky junctions. During Phase 1, these junctions continue to expand due to liquid flow. Fig. \ref{fig-mini_spheroid}(f) illustrates the initial spring networks, where the connections remain intact. However, after Phase 1, we cut the blue springs because they become too far apart, resulting in neighboring cells no longer adhering to each other. This disconnection models the scenario where the increased liquid content between two cells makes it unlikely for the cell membrane adhesions (springs) to recover their connections. Consequently, a lumen forms and develops as the cells continue to swell, as shown in Fig. \ref{fig-mini_spheroid}(e), with the green-shaded area representing the emerging lumen. In this case, we utilized the connected reservoir (CR) method, resulting in a star-shaped lumen. However, by employing the local lumen (LL) approach, the lumen shape can be made more convex by applying surface tension.
	
	\subsection{Initial Cell Swelling Mechanisms}
\begin{figure}
		\captionsetup{singlelinecheck = false, justification=raggedright}
		\begin{center}
			\begin{tabular}{c c}
				\includegraphics[width=0.22\textwidth]{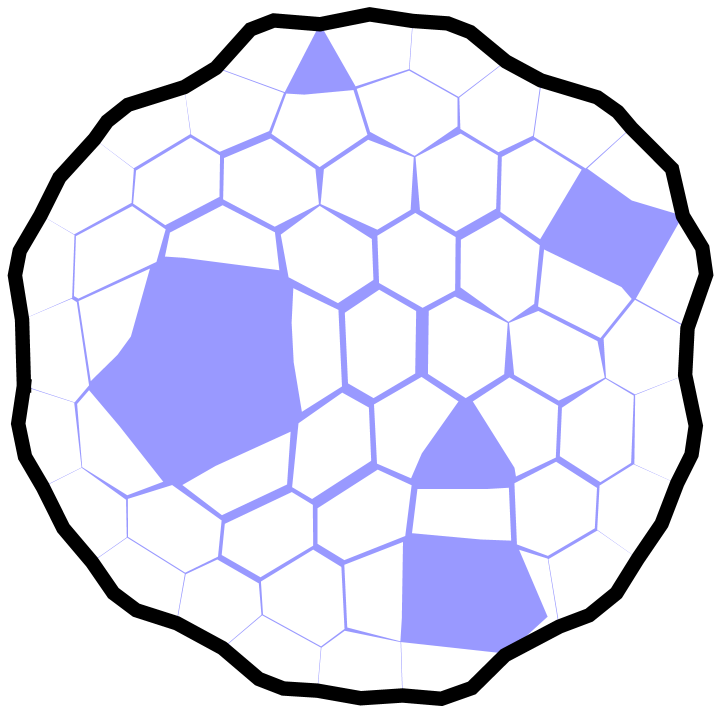}&
				\includegraphics[width=0.22\textwidth]{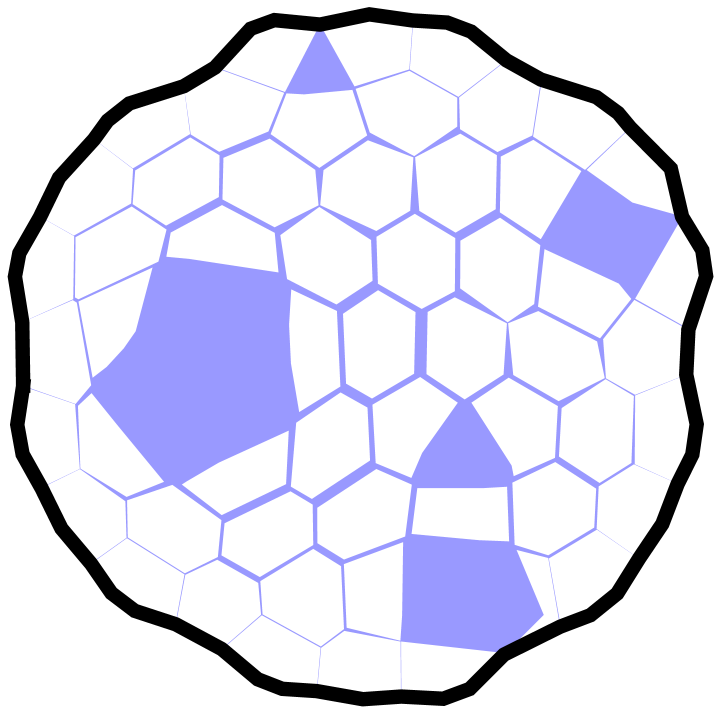}\\
				(a) Inner cells swell fast &(b) Uniform swelling\\
				
				\includegraphics[width=0.22\textwidth]{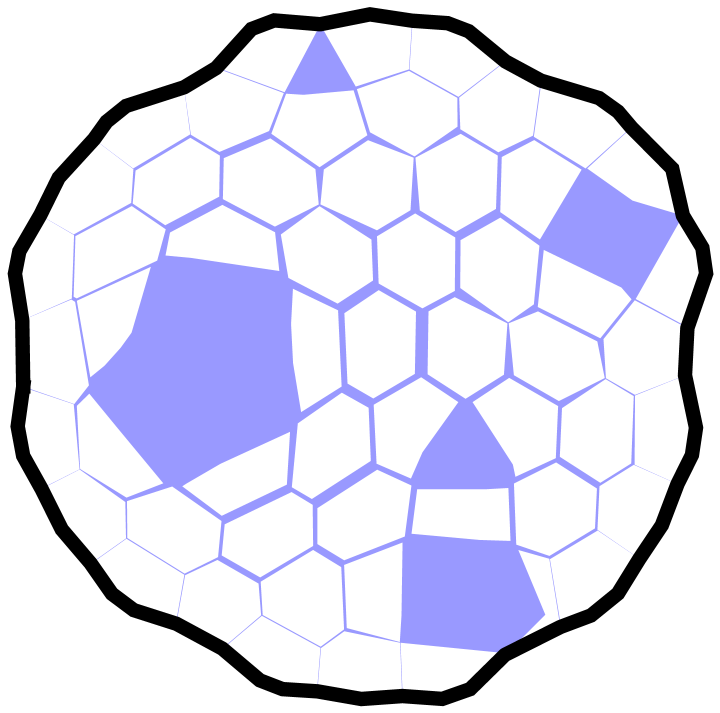}&
				\includegraphics[width=0.22\textwidth]{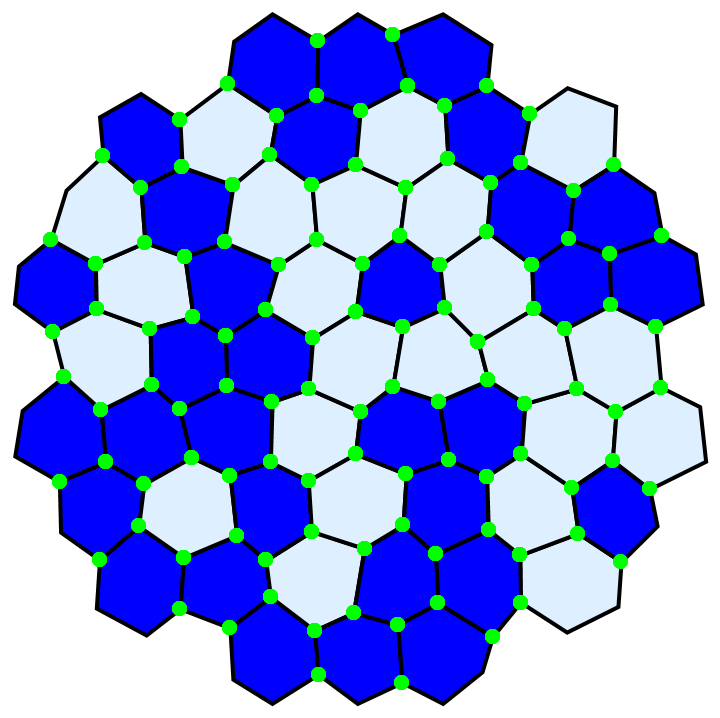}\\
				(c) Boundary cells swell fast &(d) Cell group by color\\
			\end{tabular}
			\caption{{\it Simulation results using different swelling mechanisms:} (a) Inner cells swell approximately 5\% faster. (b) Uniform swelling across all cells. (c) Cells near the boundary swell faster. (d) Cell groups used in the test.}
			\label{fig-result_by_ini}
		\end{center}
	\end{figure}

	For the initial cell swelling, we tested three different methods. The first method represents inner cells swelling approximately 5\% faster than boundary cells (as shown in Equation \ref{eq-cellswelling}). In the second test, we applied uniform swelling where all cells swell equally, with slight disorder introduced by Gaussian noise of 0.1. In this case, there is no relation to distance, meaning $\hat{r}_1 = \hat{r}_2 = 0$. The last case is the opposite: boundary cells swell faster than inner cells.
	
	\begin{equation}
		\hat{v}_i \to \hat{v}_i \cdot \left(1 + \hat{r}_1 - \frac{d(\hat{v}_i, \hat{v}_c)}{\max d(\hat{v}_i, \hat{v}_c)} \cdot \hat{r}_2\right)
		\label{eq-cellswelling}
	\end{equation}
	where $\hat{v}_c$ is the spheroid center, $\hat{r}_1$ and $\hat{r}_2$ represent the swelling ratios for the inner and outer cells, respectively. These different swelling cases are considered as initial inputs only. As expected and shown in Fig. \ref{fig-result_by_ini}, we did not observe significant differences in lumen sizes and locations.
	
	\subsection{Two Different Energy Functions for Pipe Network}
	We compared two formulations for fluid flow energy in Phase 1:
	
	\begin{equation}
		\hat{E}_{\text{pipes}} = -\hat{K}_L \cdot \hat{p} \cdot \hat{A}_{\text{pipes}}
	\end{equation}
	where $\hat{p}$ is constant boundary pressure decreasing toward center.
	
	\begin{equation}
		\hat{E}_{\text{pipes}} = \hat{K}_L \sum \frac{\hat{l}_2 \mathcal{L}(\hat{p})}{\hat{l}_1^2} \hat{A}_{\text{pipes}}
	\end{equation}
	where $\mathcal{L}(\hat{p})$ is discrete Laplacian of non-uniform pressure.
	
\begin{figure}
		\captionsetup{singlelinecheck = false, justification=raggedright}
		\begin{center}
			\begin{tabular}{c c}
				\includegraphics[width=0.22\textwidth]{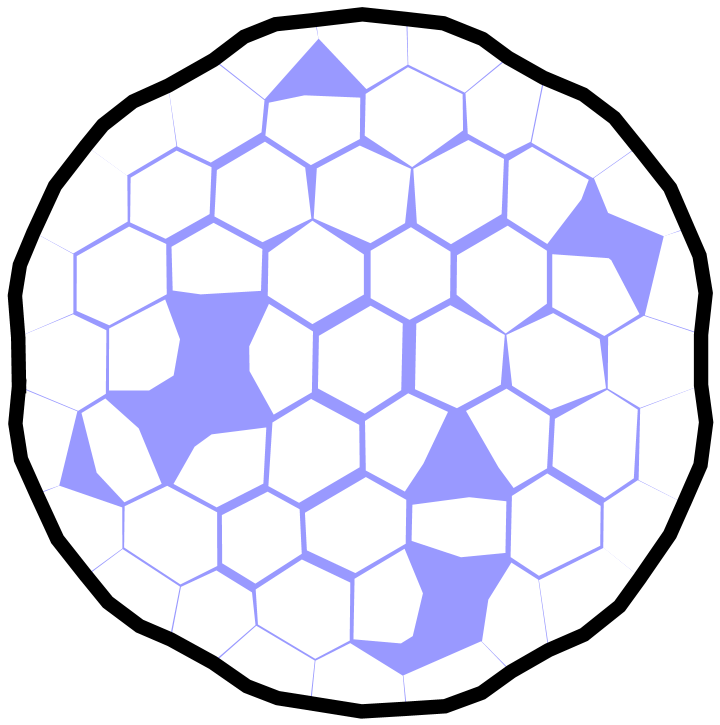}&
				\includegraphics[width=0.22\textwidth]{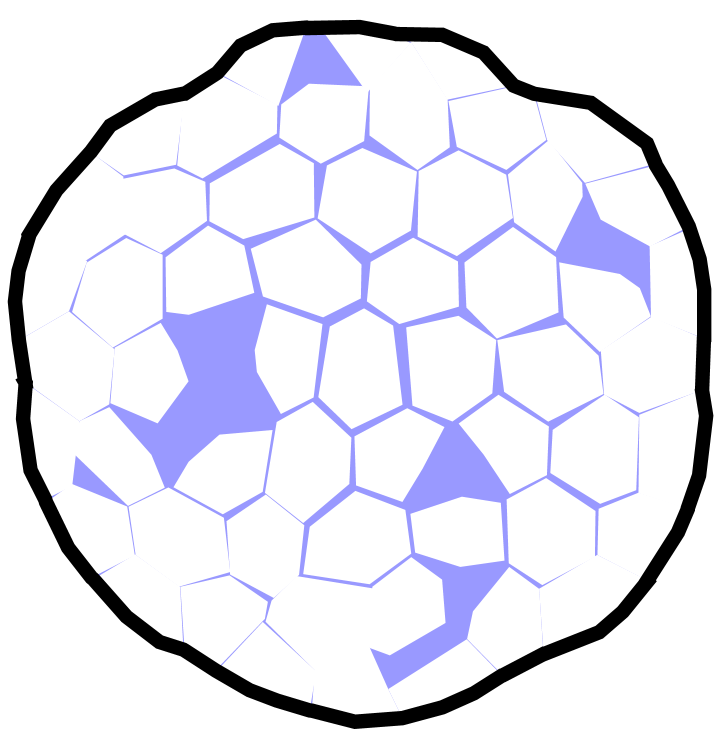}\\
				(a) Phase 1-4 ($\hat{p}\hat{A}$) &(b) Phase 1-3 (Laplacian)\\
			\end{tabular}
			\caption{ {\it Simulation results for different pipe network energy functions:} (a) Phase 1-4, after using simple $\hat{p}\hat{A}$. (b) Phase 1-3: using Laplacian.}
			\label{fig-result_by_pn}
		\end{center}
	\end{figure}

	Fig. \ref{fig-result_by_pn} shows the results comparing different liquid flow energy functions. Although we set smaller coefficients for the Laplacian with the Hagen–Poiseuille equation, we observed faster water flow and quicker lumen growth compared to using the simple $\hat{p}\hat{A}$ function, where $A$ represents the area of the entire pipe network. However, the choice of energy function did not significantly affect the overall shape or location of the lumen. Table \ref{tab-coeff-pre} and \ref{tab-coeff-p1} represent coefficients we have used in the simulation.
	\begin{table}[]
		\centering
		\caption{Coefficients for Pre-Phase}
		\begin{tabular}{c|c}
			Name & coefficients \\
			\hline
			$\hat{S}_A$ (cells with leaky junctions) & 0.9 \\
			$\hat{S}_P$ (cells with leaky junctions) & 0.9\\
			$\hat{S}_A$  & 1.15 \\
			$\hat{S}_P$  & 1.15\\
			$\hat{S}_R$  & 1.1\\
			$\hat{K}_{S1}$ $(\alpha,\alpha)$ &0\\
			$\hat{K}_{S2}$ $(\alpha,\beta)$ &2\\
			$\hat{K}_{S3}$ $(\beta,\beta)$ &2 \\
			$\hat{K}_{R}$ &0.4\\
			\hline
		\end{tabular}
		\label{tab-coeff-pre}
	\end{table}
	\begin{table}[]
		\centering
		\caption{Coefficients for Phase 1}
		\begin{tabular}{c|c}
			Name & coefficients \\
			\hline
			$\hat{S}_A$ (cells with leaky junctions) & 0.9 \\
			$\hat{S}_P$ (cells with leaky junctions) & 0.9\\
			$\hat{S}_A$  & 1.15 \\
			$\hat{S}_P$  & 1.15\\
			$\hat{S}_R$  & 1.1\\
			$\hat{K}_{S1}$ $(\alpha,\alpha)$ &0\\
			$\hat{K}_{S2}$ $(\alpha,\beta)$ &2\\
			$\hat{K}_{S3}$ $(\beta,\beta)$ &2 \\
			$\hat{K}_{B}$ (boundary stiffness) &5\\
			$\hat{K}_L$ ($\hat{p}\hat{A}$) & 1\\
			$\hat{p}$ ($\hat{p}\hat{A}$) &0.01 \\
			$\hat{K}_L$ (Laplacian)& 1\\
			$\hat{p}$ (Laplacian)& 0.001\\
			$\hat{K}_{PS}$ &0.5\\
			$\hat{K}_{R}$ &0.4\\
			\hline
		\end{tabular}
		\label{tab-coeff-p1}
	\end{table}
	
	\subsection{Lumen Shape Functions: CR vs LL}
\begin{figure}
		\centering
		\captionsetup{singlelinecheck = false, justification=raggedright}
		\begin{tabular}{c c}
			\includegraphics[width=0.22\textwidth]{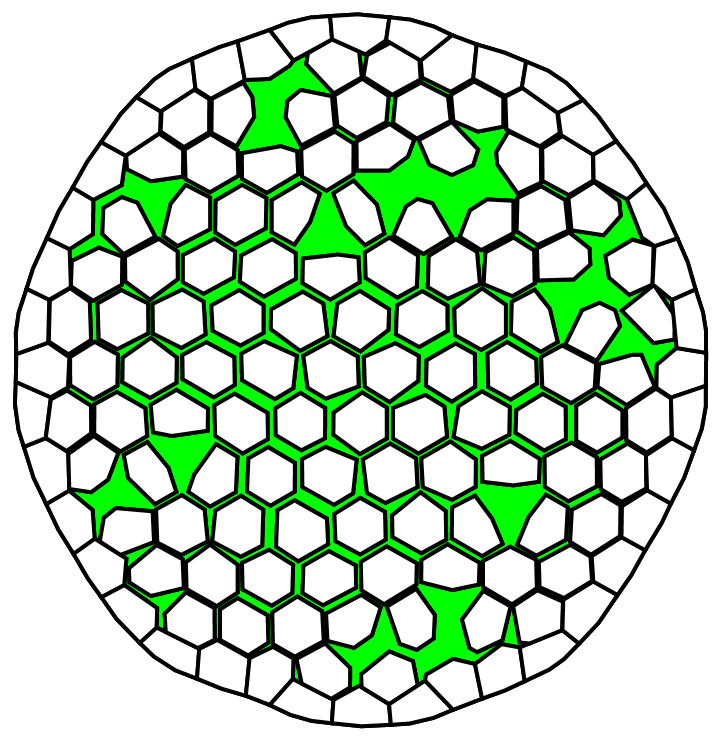}&
			\includegraphics[width=0.22\textwidth]{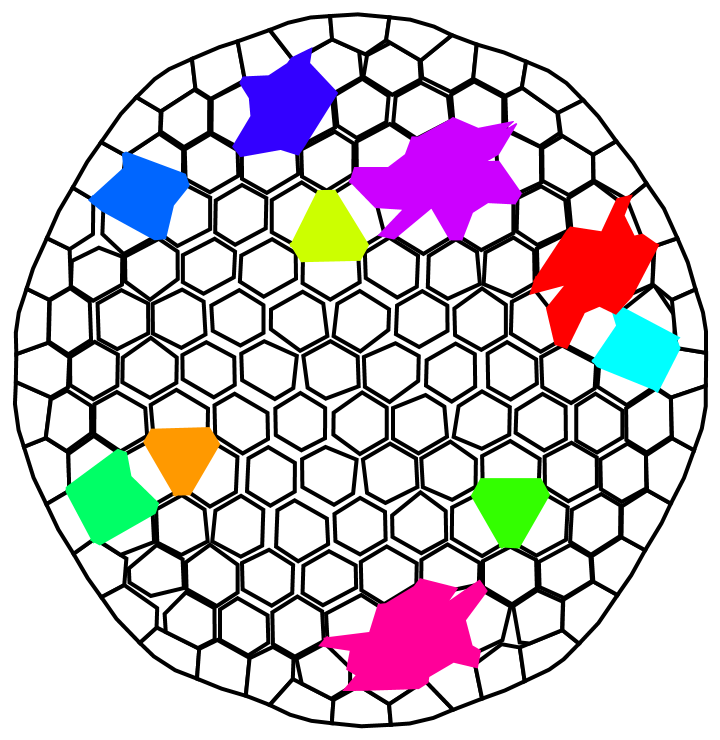}\\
			(a)  Connected Reservoir (CR) &    (b) Local Lumen (LL)\\
		\end{tabular}
		\caption{{\it Visual comparison between two lumen energy functions:} (a) The Connected Reservoir (CR) approach describes the entire gap as a single lumen, represented by shaded green areas. (b) The Local Lumen (LL) approach focuses on individual local reservoirs, depicted as circular paths in different colors.}
		\label{fig-lumen-method}
	\end{figure}

	Lastly, we compared two different methods for calculating lumen shape during Phase 2. 
	\begin{itemize} 
		\item Connected Reservoir Approach (CR): In this approach, we treat the entire gap as a single connected reservoir, similar to a water tank (as depicted in Fig. \ref{fig-lumen-method}(a)). This connected reservoir differs from pipe networks, where pipes generated from adjacent junction edges may leave uncovered areas (e.g., triangular junctions) as they expand. 
		
		\item Local Lumen Approach (LL): During Phase 2, we identify individual local (spherical) lumens and apply pressure based on their respective areas. Although this method may also leave uncovered areas in smaller pipe networks, it models the collection of liquid into local reservoirs more effectively. This approach allows for more precise lumen shaping by incorporating energy functions related to surface tension, as shown in Fig. \ref{fig-lumen-method}(b). 
	\end{itemize}
	
	The corresponding energy functions are as follows:
	\begin{equation}
		\hat{E}_{\text{lumen(CR)}} = -\hat{K}_{L} \cdot \hat{p} \sum \hat{A}_{\text{lumen}}
	\end{equation}
	\begin{equation}
		\hat{E}_{\text{lumen(LL)}} = -\hat{K}_{L} \cdot \hat{p} \sum \hat{A}_{\text{lumen}} + \hat{K}_{PL} \cdot \sum \hat{P}_{\text{lumen}}
	\end{equation}
	
	The first method treats the entire lumen area as a single entity, denoted as $\hat{A}_{\text{lumen}}$. The second method extracts individual lumens and applies the sum of surface tension and lumen energy functions based on their respective areas. We also tested a hybrid approach that incorporates surface tension for individual lumens within the CR framework to observe potential differences. 
	
\begin{figure}
		\captionsetup{singlelinecheck = false, justification=raggedright}
		\begin{center}
			\begin{tabular}{c c}
				\includegraphics[width=0.22\textwidth]{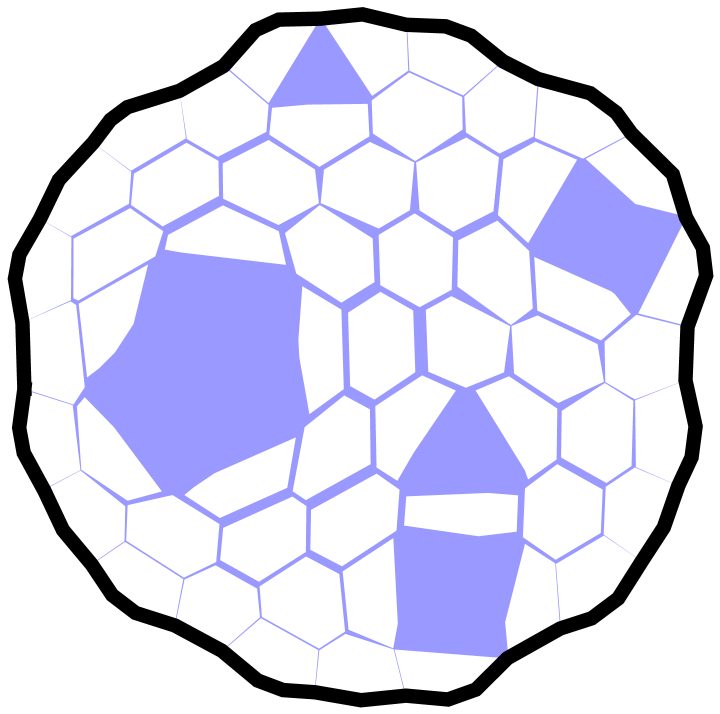}&
				\includegraphics[width=0.22\textwidth]{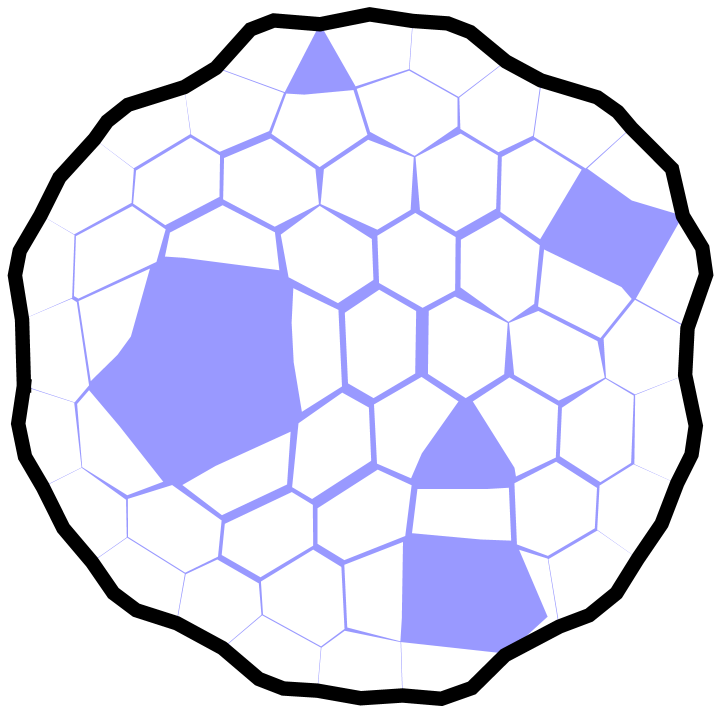}\\
				(a) CR &(b) LL\\
				
				\includegraphics[width=0.22\textwidth]{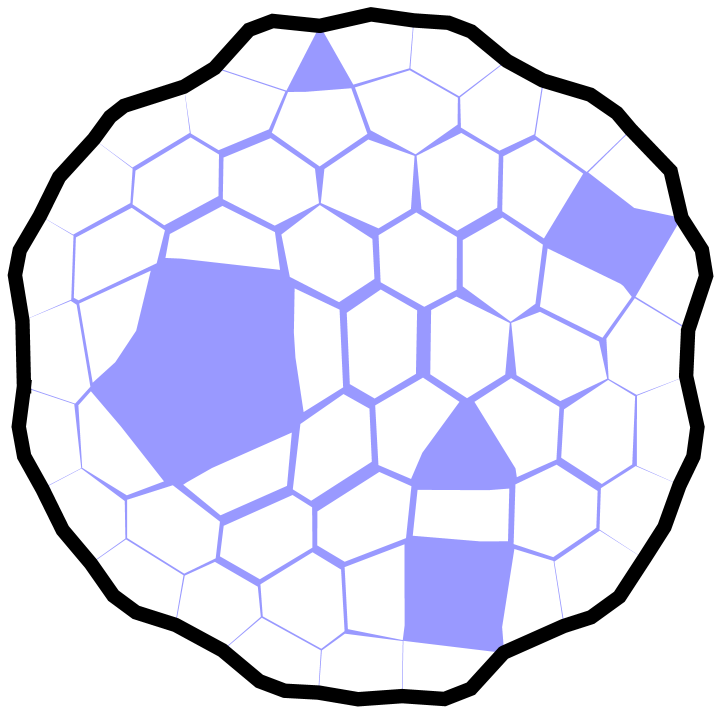}&\\
				(c) CR + surface tension &\\
			\end{tabular}
			\caption{ {\it Simulation results for different lumen energy functions:} (a) Results after phase 2 using the CR energy function. (b) Results after phase 2 using the LL energy function. (c) Results after adding surface tension to the lumen in the CR energy function.}
			\label{fig-result_by_le}
		\end{center}
	\end{figure}

	As shown in Fig. \ref{fig-result_by_le}, applying surface tension to individual lumens results in more convex lumen shapes. However, this effect is not significant, as the spring networks and the overall surface tension of the spheroid already contribute to slight convexity in the lumen shape. Table \ref{tab-coeff-p2} displays the coefficients used in this test.
	
	For multi-lumen scenarios with high surface tension, both methods produce similar results. The LL approach produces slightly more convex individual lumens due to the perimeter term. We used CR for the multi-lumen case to match the concave lumens observed in experiments.
	\begin{table}[]
		\centering
		\caption{Coefficients for Phase 2}
		\begin{tabular}{c|c}
			Name & coefficients \\
			\hline
			$\hat{S}_A$ (cells with leaky junctions) & 0.95 \\
			$\hat{S}_P$ (cells with leaky junctions) & 0.95\\
			$\hat{S}_A$  & 1.05 \\
			$\hat{S}_P$  & 1.05\\
			$\hat{S}_R$  & 1.1\\
			$\hat{K}_{S1}$ $(\alpha,\alpha)$ &5\\
			$\hat{K}_{S2}$ $(\alpha,\beta)$ &5\\
			$\hat{K}_{S3}$ $(\beta,\beta)$ &5 \\
			$\hat{K}_{B}$ (boundary stiffness) &5\\
			$\hat{K}_L$ (LL) & 0.25 \\
			$\hat{p}$ (LL) & 0.1 \\
			$\hat{K}_{PL}$ (LL) & 0.25 \\
			$\hat{K}_L$ (CR)& 0.1\\
			$\hat{K}_{PS}$ &0.5\\
			$\hat{K}_{R}$ &0.4\\
			\hline
		\end{tabular}
		\label{tab-coeff-p2}
	\end{table}

	\subsection{Algorithms}
	
	\begin{algorithm}
		\begin{algorithmic}[1]
			\REQUIRE List of springs $L(\alpha,\alpha)$, where each springs is defined by vertices $(\hat{v}_i, \hat{v}_j, \hat{v}_k)$. Critical length $\hat{l}_{\text{crit}}$.
			\STATE Initialize $L'(\alpha,\alpha) = L(\alpha,\alpha)$
			\FORALL{$(\hat{v}_i, \hat{v}_j, \hat{v}_k) \in L(\alpha,\alpha)$} 
			\IF{$d(\hat{v}_i, \hat{v}_j) + d(\hat{v}_j, \hat{v}_k) + d(\hat{v}_k, \hat{v}_i) > \hat{l}_{\text{crit}}$} 
			\STATE Remove $(\hat{v}_i, \hat{v}_j, \hat{v}_k)$ from $L'(\alpha,\alpha)$
			\ENDIF
			\ENDFOR
			\RETURN $L'(\alpha,\alpha)$
		\end{algorithmic}
		\caption{Algorithm to remove springs from the list}
		\label{algo1}
	\end{algorithm}
	\begin{algorithm}
		\begin{algorithmic}[1]
			\REQUIRE $L(\alpha,\alpha)$ (list of springs), $L'(\alpha,\alpha)$ (modified list after removing springs), $L_\text{pipes}$ (list of pipes, each pipe having four vertices)
			\STATE Initialize $L_\text{lumen} = \emptyset$
			\STATE Set $L_\text{diff} = L(\alpha,\alpha) - (L(\alpha,\alpha) \cap L'(\alpha,\alpha))$ 
			\FORALL{$(\hat{v}_i, \hat{v}_j, \hat{v}_k) \in L_\text{diff}$} 
			\FORALL{$\hat{P}_i \in L_\text{pipes}$} 
			\IF{$\{\hat{v}_i, \hat{v}_j, \hat{v}_k\} \cap \hat{P}_i \neq \emptyset$}
			\STATE Add $\hat{P}_i$ to $L_\text{lumen}$
			\ENDIF
			\ENDFOR
			\STATE Add $(\hat{v}_i, \hat{v}_j, \hat{v}_k)$ to $L_\text{lumen}$
			\ENDFOR
			\RETURN $L_\text{lumen}$
		\end{algorithmic}
		\caption{Algorithm to identify local lumen}
		\label{algo2}
	\end{algorithm}
	\section{Scenario 2 and 3}
	
	\subsection{Relation to existing models of fluid-driven morphogenesis}
	
	The present treatment draws on three existing descriptions, and it is useful to state precisely what is taken from each and what is not.
	
	The closest system is the otic vesicle model of Wang \textit{et al.}~\cite{Wang2026}, which describes a hyaluronan-rich pillar spanning a fluid-filled lumen. There the elevated osmotic pressure of the HA gel arises from Donnan partitioning of mobile ions, Eq.~\eqref{eq:donnan}, with the corresponding osmotic pressure following from the van 't Hoff relation, and water transport across each interface obeying the linear permeability law $J_w = \alpha\left[(\Delta P) - (\Delta\Pi)\right]$. We adopt this physical picture in full: it is the origin of the osmotic drive in our model, and it supplies the justification for the nonlinear dependence of pressure on hyaluronan content, since the excess concentration scales as the square of the fixed charge density rather than linearly with it. What we do not adopt is the geometric reduction. Wang \textit{et al.} treat the system as three lumped compartments whose volumes are projected onto an assumed spherical vesicle and axisymmetric pillar, with the mechanical closure supplied by the Laplace law and Maxwell constitutive relations for the walls. The vertex model replaces this: the extracellular space is resolved as a network of individual compartments, and the shape is an output of energy minimization rather than a prescribed family parametrized by $R_1$, $R_2$ and $h$. This is what allows the present model to address morphologies that are not axisymmetric, including budding and lateral asymmetry.
	
	Sun and Feng~\cite{SunFeng2022} model a spherical lumen enclosed by a thin cell layer, in which solute pumping raises the internal concentration, osmotic pressure drives water influx by Darcy's law, and Hydrostatic fracture of cell--cell contacts intermittently releases the accumulated pressure. From that work we adopt the multiplicative decomposition of the deformation into growth of the stress-free configuration and elastic response about it. As with Wang \textit{et al.}, we do not adopt the lumped transport equations or the shell mechanics. The fracturing--healing switch is likewise omitted, for reasons of timescale discussed below.
	
	Yu \textit{et al.}~\cite{Yu2024} couple a three-dimensional vertex model to a Schnakenberg reaction--diffusion system, in which an activating signal protein and its reactant diffuse between cells, the signal recruits actomyosin, and cellular deformation feeds back to deactivate the signal. From that work we adopt the discretization of the transport problem onto the cell network, that is, diffusion represented as a sum over adjacent compartments weighted by the concentration difference. We do not adopt the reaction kinetics: the autocatalytic terms generate the Turing instability responsible for the patterns reported there, and their inclusion would make the emergent morphology a property of the reaction network rather than of the imposed HAS distribution. Hyaluronan synthesis is enzymatic and does not self-activate, so the linear kinetics retained here are the appropriate reduction. We also do not adopt their active contraction term $\tfrac{1}{2}k_\mathrm{m}m_J A_J^2$, in which the signal recruits actomyosin; in that model osmotic effects are explicitly disregarded, whereas here they are the driving mechanism.
	
	The three descriptions therefore differ in what drives deformation and in how the fluid space is represented. In Wang \textit{et al.} the drive is Donnan osmotic pressure in an HA gel, acting on three lumped compartments; in Sun and Feng it is hydrostatic pressure in a single closed lumen; in Yu \textit{et al.} it is active contraction distributed over cells. Here the drive is osmotic pressure of hyaluronan acting on a spatially resolved extracellular network, entering the energy as the pressure--volume work of Eq.~\eqref{eq:pA}. The compartment areas are not prescribed: they follow from minimization of the total energy, so the flow network adopts whatever configuration balances the osmotic drive against the elasticity of the surrounding cells. Because the pressure varies from compartment to compartment with the local hyaluronan concentration, the deformation is set by the resulting spatial differential rather than by a uniform pressure.
	
	Finally, the direction of coupling differs from Yu \textit{et al.}, which closes the mechanochemical loop through a feedback term in which cell deformation suppresses signalling; it is this feedback that produces the pitchfork and Hopf bifurcations characterizing their phase diagram. The present model does not include such a term: the hyaluronan field sets the cell targets but is itself unaffected by the mechanical state. The morphologies obtained here are therefore not the result of a bifurcation in a feedback loop but of the spatial distribution of synthesis. Introducing a strain dependence into hyaluronan production would close the loop and is the most direct extension of this work.
	
	\subsection{Separation of timescales}
	
	Three timescales are relevant, and the iterative scheme used here sits between them. The fastest processes are mechanical relaxation of the cell network, redistribution of fluid between compartments, and the transient opening and resealing of cell--cell contacts; all are assumed to complete within a single iteration, so that each iteration represents a settled state and the junction configuration entering the minimization is the resealed one. The slowest process is cell division, which acts over a longer interval than the deformations reported here and whose effect enters through the growth of the stress-free area rather than through discrete division events. Between these lies the accumulation of hyaluronan and the associated growth of the stress-free configuration, which is what the iteration index resolves.
	
	This separation is what justifies the quasi-static treatment. Overdamped vertex dynamics is straightforward to implement in the same framework, but integrating the fast relaxation explicitly would resolve transients that do not affect the morphological outcome; the same reasoning has been used to justify quasi-static treatments of epithelial bud growth, where matrix expansion proceeds over hours while actomyosin-driven shape changes relax in minutes~\cite{MUNJAL2021}. The dynamics of the model is therefore carried by the slow variables, the stress-free configuration and the hyaluronan field, with the vertex positions adiabatically slaved to them. In the same sense, resolving the fracturing--healing switch of Sun and Feng explicitly is unnecessary here: that switch modulates transport on a timescale shorter than one iteration, and its cumulative effect is contained in the compartment areas that emerge from the minimization.
	\subsection{Hyaluronan concentration field}
	
	Hyaluronan is synthesized by HAS-expressing cells and released into the
	extracellular space. Its spatial distribution is motivated by diffusion with
	degradation,
	\begin{equation}
		\frac{\partial c}{\partial t}
		=
		D\nabla^2 c-kc+s(\mathbf{r}),
		\label{eq:ha_rd2}
	\end{equation}
	where $c(\mathbf{r})$ is the local hyaluronan concentration, $D$ is an
	effective diffusivity, $k$ is an effective removal rate, and
	$s(\mathbf{r})$ represents local synthesis by HAS-expressing cells. Rather
	than solving Eq.~\eqref{eq:ha_rd2} explicitly at each mechanical iteration,
	we represent the concentration by exponentially screened spatial profiles.
	The corresponding characteristic decay length is motivated by
	$\hat{\lambda}\sim\sqrt{\hat{D}/\hat{k}}$. Thus, $c$ should be interpreted as a reduced
	concentration field that captures the finite spatial range of hyaluronan
	transport.
	
	This construction is motivated by the cell-to-cell reaction--diffusion
	framework of Yu \textit{et al.}, but the nonlinear reaction kinetics are not
	retained. In particular, we use only the idea of a spatially distributed
	concentration field with a finite transport length scale; the concentration
	profiles below are prescribed from the spatial distribution of
	HAS-expressing cells.
	
	\subsubsection{Scenario 2 (budding): steady-state field on the compartment network}
	
	In Scenario 2 the concentration field is obtained by solving Eq.~\eqref{eq:ha_rd} at steady state directly on the extracellular network, rather than by prescribing a profile. Discretizing the diffusion operator on the compartment adjacency graph, in which two compartments are neighbors when they share an edge, the stationary condition becomes the linear system
	\begin{equation}
		\left( \hat{k}\,\mathbf{I} + \hat{D}\,\mathbf{L} \right)\mathbf{\hat{c}} = \mathbf{\hat{s}} ,
		\label{eq:ha_steady}
	\end{equation}
	where $\mathbf{L}$ is the graph Laplacian of the network, with $L_{\alpha\alpha}$ the number of neighbors of compartment $\alpha$ and $L_{\alpha\beta} = -1$ for adjacent pairs, and $\mathbf{\hat{s}}$ has entries $\hat{s}_0$ in compartments adjacent to HAS-expressing cells and zero elsewhere. Equation~\eqref{eq:ha_steady} is sparse and is solved at each iteration on the current, deformed configuration, so that the profile and the geometry remain mutually consistent. The simulations use $\hat{D} = 0.5$, $\hat{k} = 0.05$ and $\hat{s}_0 = 1$, giving a decay length $\hat{\lambda} = \sqrt{\hat{D}/\hat{k}} \simeq 3.2$ in units of the compartment spacing. The resulting field is normalized by its maximum before entering the mechanical coupling, so that the amplitude of the osmotic drive is set by the coefficient of the pressure term alone.
	
	Because the discrete Laplacian is isotropic, the field carries no directional preference of its own; its spatial structure follows from the arrangement of the source compartments and from the geometry of the network on which it is solved.
	
	\subsubsection{Scenario 3 (lateral asymmetry): superposition from localized sources}
	
	In Scenario 3, HAS expression is localized to a set $\mathcal{S}$ of cells and the field is represented directly by superposing the screened kernel that solves Eq.~\eqref{eq:ha_rd2} for an isolated point source,
	\begin{equation}
		\hat{c}_\alpha = \hat{c}_A \sum_{\mu\in\mathcal{S}} \exp\left( -\frac{\left|\mathbf{\hat{r}}_\alpha-\mathbf{\hat{r}}_\mu\right|}{\hat{\lambda}} \right),
		\label{eq:ha_superposition2}
	\end{equation}
	where $\mathbf{\hat{r}}_\alpha$ is the centroid of extracellular compartment $\alpha$, $\mathbf{\hat{r}}_\mu$ is the centroid of HAS-expressing cell $\hat{\mu}$, $\hat{c}_A$ sets the amplitude and $\hat{\lambda}$ is the decay length. The simulations use $\hat{\lambda} = 30$ and $\hat{c}_A = 0.5$; unlike Scenario 2 the field is not normalized, so that its magnitude as well as its shape is carried into the mechanical coupling.
	
	The kernel in Eq.~\eqref{eq:ha_superposition2} is isotropic, so the directional structure of the field arises from the spatial arrangement of the HAS-expressing cells rather than from an anisotropic transport law. Localizing the sources to one lateral region produces concentration gradients both across and along the tissue, whereas a source distributed uniformly along an entire boundary would produce an approximately one-dimensional profile by symmetry.
	
	\subsubsection{Decay length and the origin of the mechanical differential}
	
	The two scenarios use decay lengths differing by an order of magnitude, of roughly three and thirty compartment spacings, because the mechanical differential acts on different scales in each case. Budding requires the concentration to fall appreciably within the thickness of the cell layer, so that compartments on either side of the HAS-expressing row experience different osmotic pressures; lateral asymmetry instead requires a gradient spanning the width of the tissue, so that the two sides swell by different amounts. The decay length is therefore not a single material constant in these simulations but the parameter that selects which of the two differentials dominates.
	
	Two distinct routes of asymmetry are present in the model. The first is the concentration field, which sets the osmotic pressure acting on the extracellular compartments. The second is the set of target areas and perimeters assigned to the cells, which are prescribed as fixed swelling ratios drawn from experimental measurement and from reported values. The latter are imposed rather than computed, for two reasons. First, the targets are what render the vertex energy bounded below: without them the area and perimeter terms lose their reference state, the minimization is underconstrained, and the network collapses to a point or to arbitrary polygons. Second, using measured swelling ratios keeps the cellular deformation anchored to observation rather than to a further modelling assumption. In principle the same Donnan argument that gives the osmotic pressure of the extracellular gel, Eq.~\eqref{eq:donnan}, would also determine the swelling of the cells from the local concentration, and the target ratios could then be derived rather than assigned; we have not done so here. In either case the form of the constitutive law is unchanged: the area and perimeter terms retain their standard isotropic form, representing bulk and shape elasticity of the cell, with no directional term added.
	
	\subsection{Simulation Details}\label{app-units}
	We first state the energy contributions in dimensional form (simulation units). Lengths are measured in $\mu$m$^*$ and stresses in Pa$^*$, so that every term carries the energy unit Pa$^*\!\cdot\!\mu$m$^{*3}$:
	\begin{align}
		E_{\text{cell}} &= \frac{1}{2}\sum_{\alpha} \left[ K_A (A_\alpha - A_{0\alpha})^2 + K_P (P_\alpha - P_{0\alpha})^2 \right], \\
		E_{\text{junction}} &= \sum_{\langle \alpha\beta \rangle} K_S \, d_{\alpha\beta}^2, \\
		E_{\text{fluid}} &= -\sum_{\alpha} K_\Pi \, \Pi_\alpha A^f_\alpha, \\
		E_{\text{spheroid}} &= K_{PS} \, P_{\text{spheroid}}, \\
		E_{\text{boundary}} &= \sum_{i,j \in \partial\Omega} K_B \, d(v_i, v_j)^2, \\
		E_{\text{tension}} &= K_t L_t .
	\end{align}
	The corresponding coefficient units are $[K_A] = \mathrm{Pa}^*/\mu\mathrm{m}^*$, $[K_P] = [K_S] = [K_{PS}] = [K_B] = [K_R] = \mathrm{Pa}^*\!\cdot\!\mu\mathrm{m}^*$, and $[K_t]  = \mathrm{Pa}^*\!\cdot\!\mu\mathrm{m}^{*2}$. The osmotic pressure $\Pi_\alpha$ is a true pressure, $[\Pi] = \mathrm{Pa}^*$, and the coefficient $K_\Pi$, with $[K_\Pi] = \mu\mathrm{m}^*$, converts the pressure--area product into the common energy unit; we set $K_\Pi = 1$~$\mu$m$^*$ throughout, so that the fluid term is numerically equal to $\Pi_\alpha A^f_\alpha$.
	
	In the simulations we work with the dimensionless form of these expressions, obtained by dividing by the elastic energy scale $K_A A_0^2$, measuring areas in units of the reference cell area $A_0$, and lengths in units of $\sqrt{A_0}$, so that $\hat{E} = E/(K_A A_0^2)$, $\hat{A} = A/A_0$, $\hat{P} = P/\sqrt{A_0}$, $\hat{d} = d/\sqrt{A_0}$ and $\hat{L} = L/\sqrt{A_0}$. Hatted symbols denote these normalized quantities throughout, and the coefficients reported below are the resulting pure numbers. The energy contributions become
	\begin{align}
		\hat{E}_{\text{cell}} &= \tfrac{1}{2}\sum_{\alpha} \left[ (\hat{A}_\alpha - \hat{A}_{0\alpha})^2 + \hat{K}_P (\hat{P}_\alpha - \hat{P}_{0\alpha})^2 \right], \\
		\hat{E}_{\text{junction}} &= \sum_{\langle \alpha\beta \rangle} \hat{K}_S \, \hat{d}_{\alpha\beta}^2, \\
		\hat{E}_{\text{fluid}} &= -\sum_{\alpha} \hat{\Pi}_\alpha \hat{A}^f_\alpha, \\
		\hat{E}_{\text{spheroid}} &= \hat{K}_{PS} \, \hat{P}_{\text{spheroid}}^2, \\
		\hat{E}_{\text{boundary}} &= \sum_{i,j \in \partial\Omega} \hat{K}_B \, \hat{d}(v_i, v_j)^2, \\
		\hat{E}_{\text{tension}} &= \hat{K}_t \hat{L}_t .
	\end{align}
	where $\hat{A}_0$ represents the dimensionless swelling target and $\hat{K}_P = 1$ if not specified.
	
	Junction stiffnesses are assigned based on measured values in the literature. Laser ablation of epithelial junctions gives vertex recoil velocities of approximately $1.5$, $3.5$ and $10.5$~$\mu$m/s for wild-type, claudin-knockout and zonula-occludens-knockout conditions respectively \cite{Mukenhirn2024}. Because a faster recoil reflects a weaker junction with greater paracellular leakage, we invert these ratios to obtain relative stiffnesses, so that the wild-type condition is the stiffest and the zonula-occludens knockout the most compliant. Normalizing the stiffest case to $5$~Pa$^*$ places the three conditions at approximately $5.0$, $2.2$ and $0.7$~Pa$^*$, and we therefore sample $\tfrac{1}{2}{K}_S$ over the range $0$--$5$~Pa$^*$, with $0$~Pa$^*$ representing fully disrupted junctions.
	
	An independent estimate confirms that this range spans physiological values. Forces of order $100$--$120$~nN transmitted across contact areas of order $50$~$\mu$m$^2$ correspond to junction stresses of $2$--$2.4$~kPa \cite{Liu2010,Maruthamuthu2011}. In two dimensions the force is carried along a line rather than distributed over an area, which introduces a geometric factor of $25$ and places these stresses at $1.25$--$1.5$~Pa$^*$, near the middle of our sampled range. Values of $1$--$2$~Pa$^*$ thus correspond to normal junction function, while $3$--$5$~Pa$^*$ represents the tight-junction state associated with barrier formation. 
	
	Because the stiffnesses are relative, the correspondence does not require the simulated tissue to match experimental dimensions: spheroids in Fig.~\ref{fig-result_by_groups} contain of order $200$ cells with mean area $\sim$$1$~$\mu$m$^{2*}$, whereas the experiments involve of order $10^4$ cells with mean area $\sim$$50$~$\mu$m$^2$. Because the cells swell over time, the energy function is normalized on the initial configuration rather than the instantaneous one. The dimensionless stiffness $\hat{K}_S = K_S/(K_A A_0)$ therefore spans $0$ to $10$ in Scenarios 1--3, where the regular hexagonal cell area is set to $A_0 = 1$ and a small perturbation is applied to the vertex positions.
	
	\begin{figure}
		\captionsetup{singlelinecheck = false, justification=raggedright}
		\begin{center}
			\begin{tabular}{c c}
				\includegraphics[width=0.23\textwidth]{figs/Asymmetry03_tension5to1.png}&
				\includegraphics[width=0.23\textwidth]{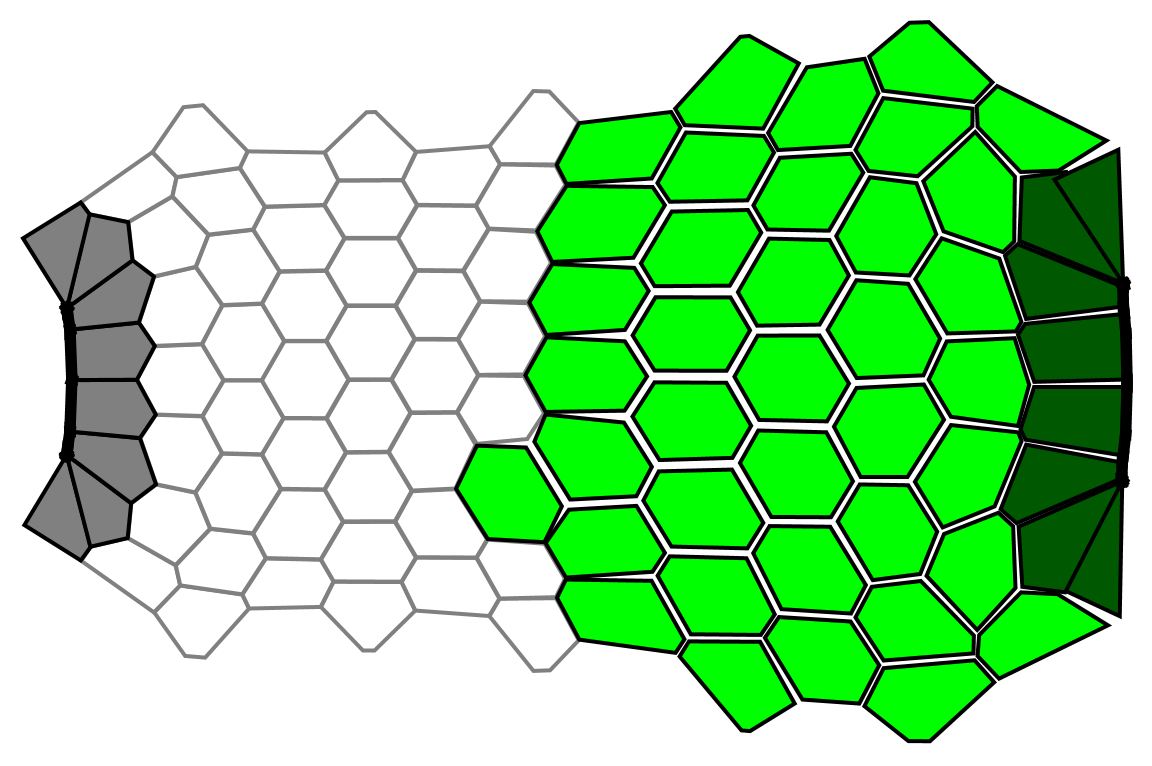}  \\
				(a)  & (b) \\
				\includegraphics[width=0.23\textwidth]{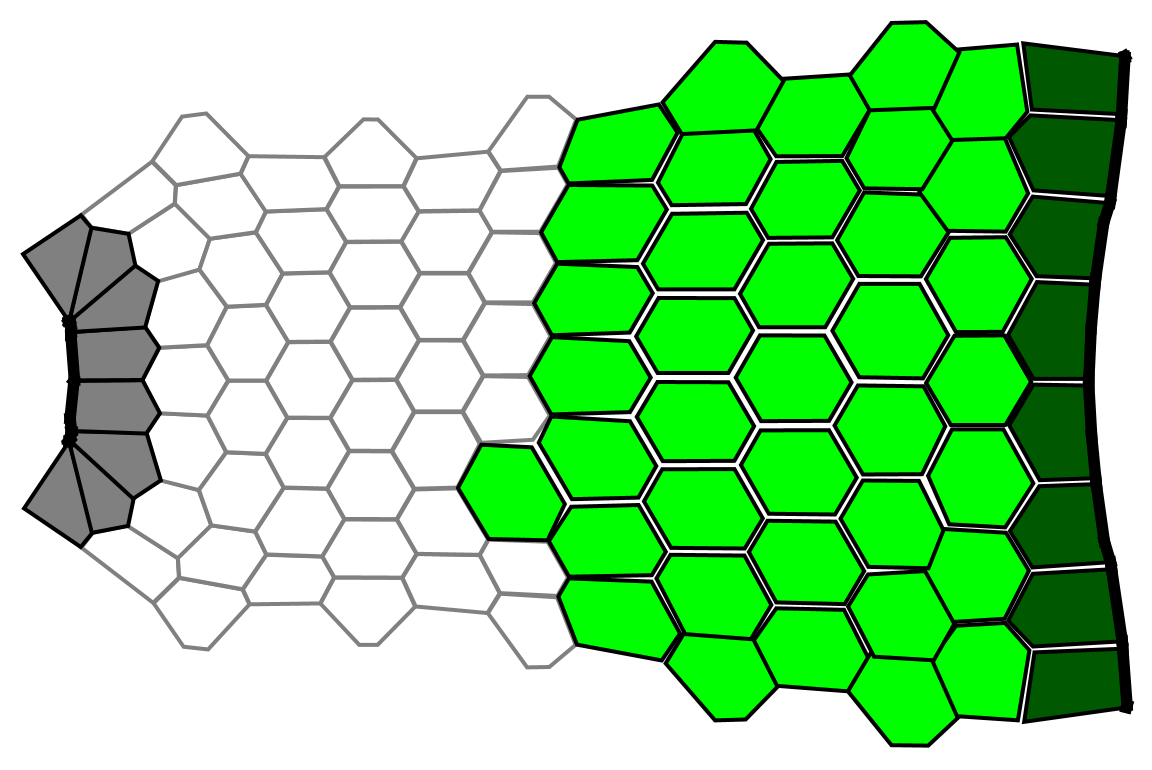} & \includegraphics[width=0.23\textwidth]{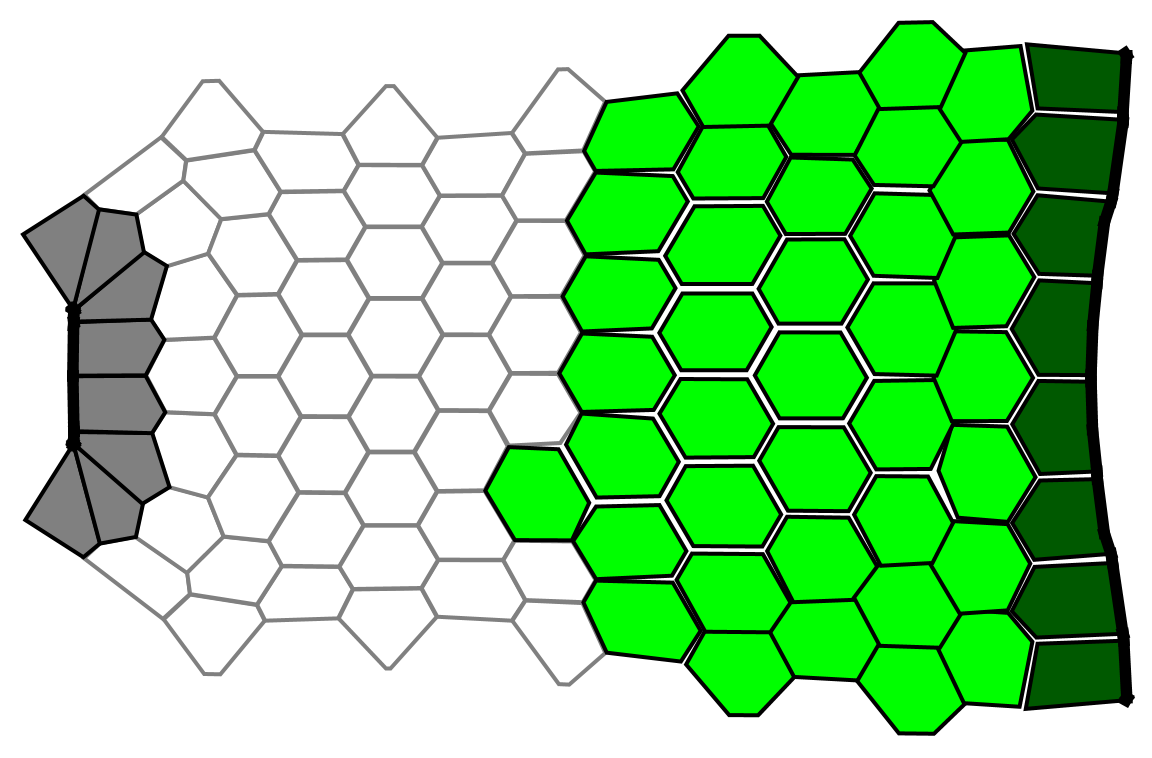} \\
				(c)&(d)\\
				\includegraphics[width=0.23\textwidth]{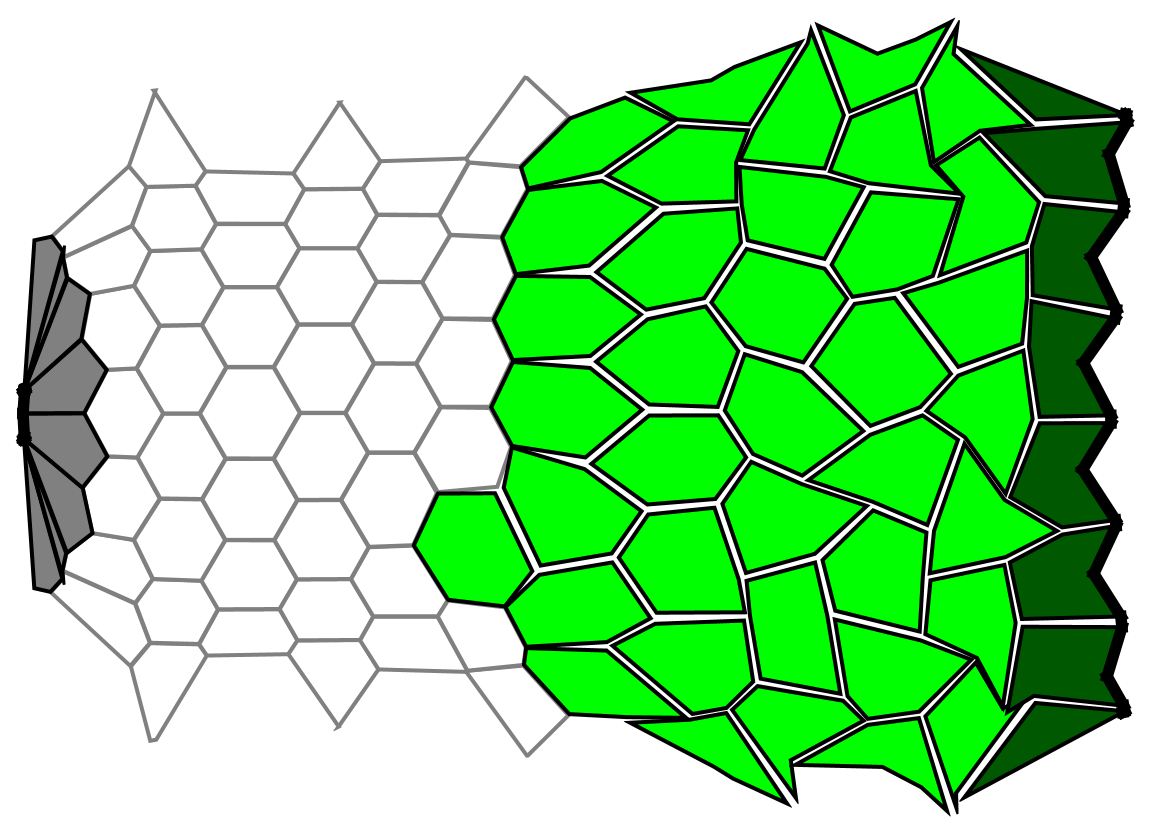} & \includegraphics[width=0.23\textwidth]{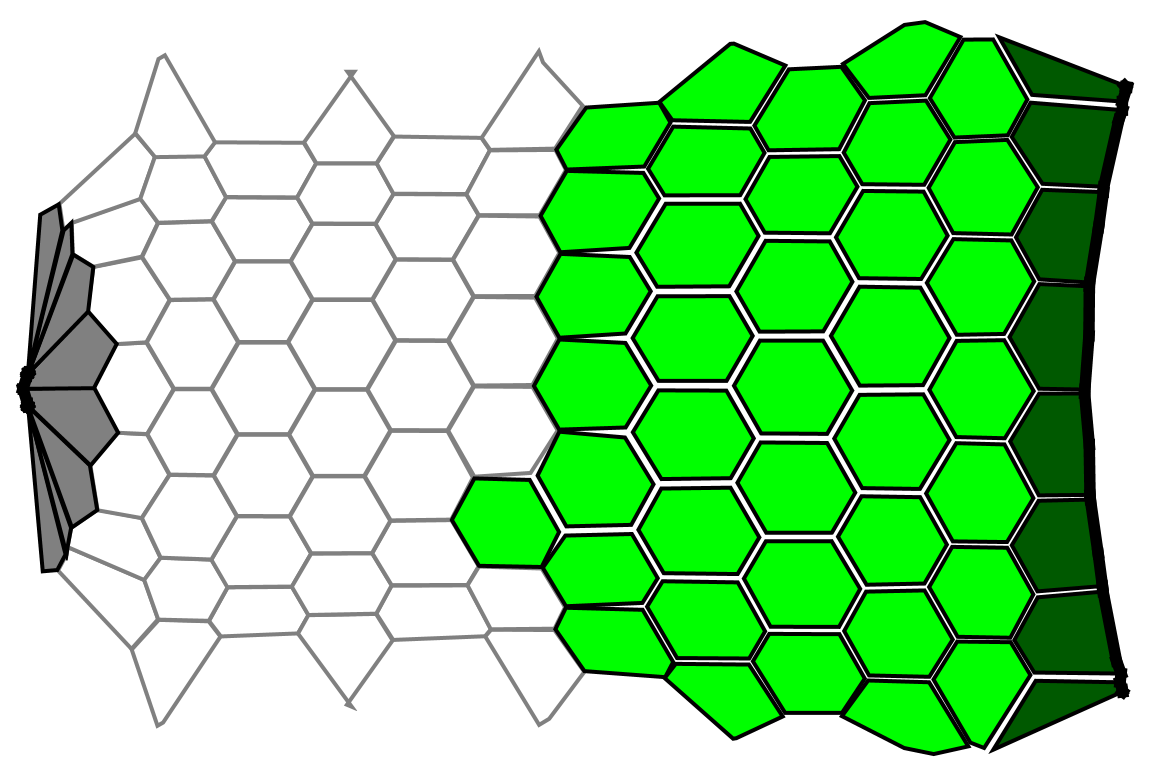} \\
				(e)&(f)\\ 
				\includegraphics[width=0.23\textwidth]{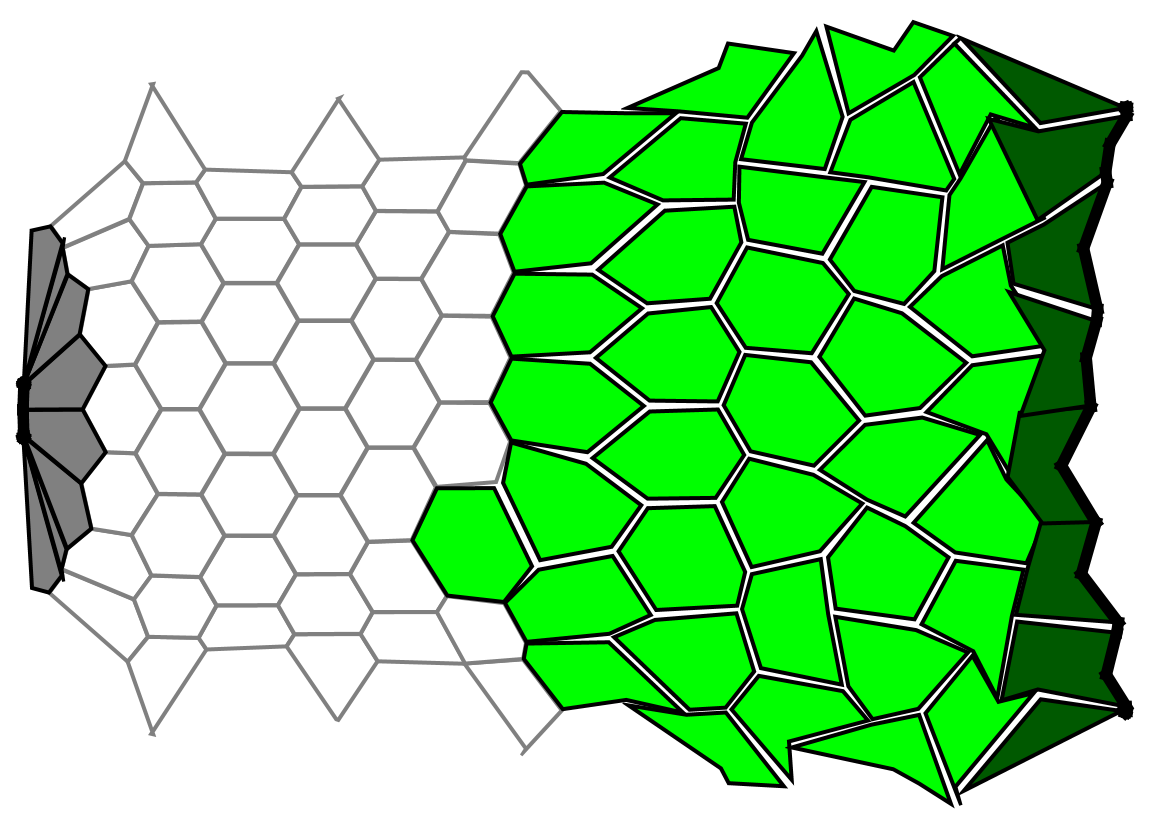} & \includegraphics[width=0.23\textwidth]{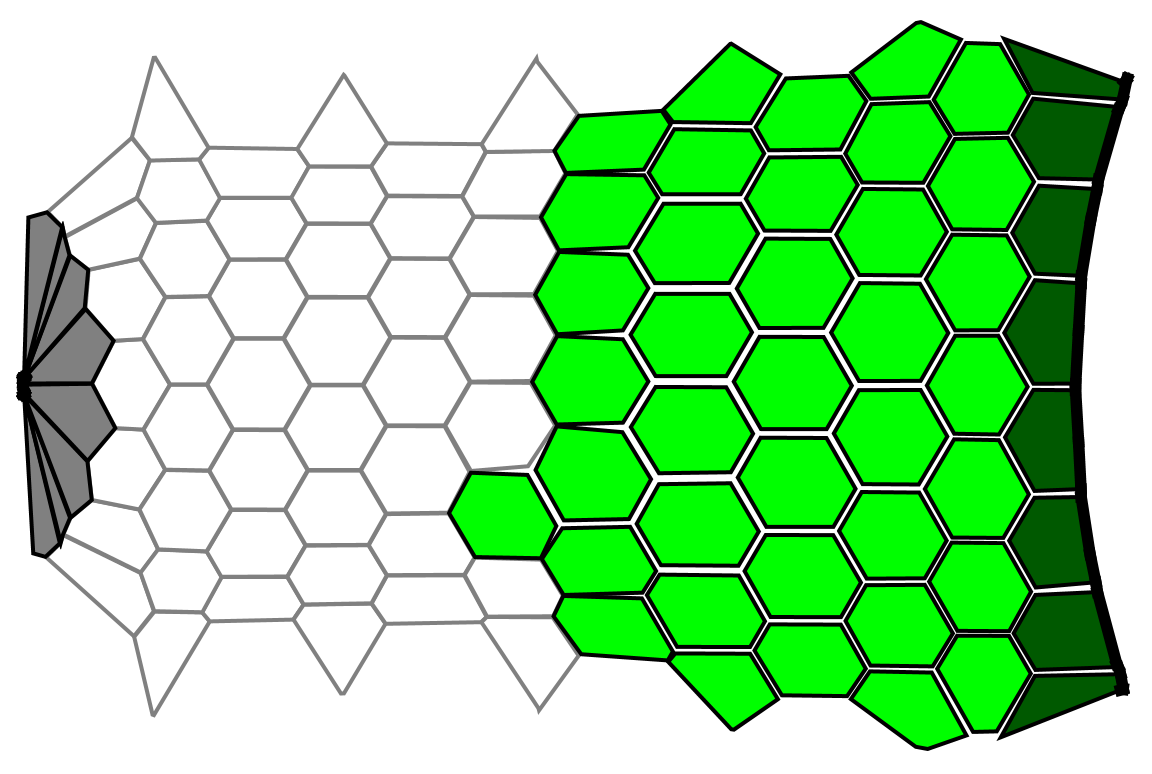} \\
				(g)&(h)\\
			\end{tabular}
			\caption{{\it Comparison of results based on different parameters.}
				(a)~Reference, the same as Fig.~\ref{fig-result_by_asymmetry}(b). (b)~With the same surface tension on the left and right sides. (c)~As (a), except that all cells are included in the pipe network. (d)~As (a), with two modifications: the target area of the left-hand cells is held at its reference value, and all cells are included in the pipe network. (e)~Simulation result using target swelling ratios assigned from the rescaled pressure field, $R_i \in [0.99, 1.01]$, with the shape index forced by $s = 1.01$ (right side only). (f)~As (e), with $s = 1$, so that the shape index is not forced. (g)~Same condition as (e), without the pipe network. (h)~As (e), with almost no cell swelling, $R_i \in [0.9999, 1.0001]$, and the pipe network retained.}\label{fig-result_by_asymmetry2}
		\end{center}
	\end{figure}
	The swelling ratio in Scenario 1 was determined from the experimental measurements. A value close to that used in Scenario 3 was adopted, but reduced from $1.04$ to $1.02$ in the absence of direct experimental data, reflecting the shorter duration of this scenario. In Scenario 2 a larger value of $1.05$ was used, consistent with the larger deformation. On the basis of rough timescales reported in the reference, Scenarios 2 and 3 correspond to approximately one to two days or less, while Scenario 4 corresponds to about a week, giving an estimated simulation timescale of approximately 2--20 hours per iteration.
	
	Next, we discuss variants of Scenario 3 in which the asymmetric surface tension is made symmetric and the cells are included in the pipe network, the latter because the contribution of osmotic swelling is otherwise difficult to separate. The morphologies in Fig.~\ref{fig-result_by_asymmetry2}(b)--(f) differ in detail from the reference case, but the lateral asymmetry is unchanged in each, indicating that none of these modifications alters the mechanism. Tables~\ref{tab-coeff-s2} and~\ref{tab-coeff-s3} list the coefficients used in the simulations. For Fig.~\ref{fig-result_by_asymmetry2}(e), instead of assigning arbitrary targets, we assumed that swelling is proportional to pressure: each cell's average pressure was rescaled to the range $R_{i-1} \in [0.99, 1.01]$, computed from the positions at iteration $i-1$, and this relative ratio was used to assign the target cell swelling ratio, where $s_i$ represents the shape index change for the cells on the right side, with $s = 1.01$ in (e) and $s = 1$ in (f) (on the left side $s$ is fixed at $1$). In place of the equation above, we used
	\begin{align}
		\hat{E}_{\text{cell},i} = \tfrac{1}{2}\sum_{\alpha} &\big [ (\hat{A}_\alpha - R_{i-1} \cdot\hat{A}_{(i-1)\alpha})^2 + \\ \notag
		&\hat{K}_P (\hat{P}_\alpha - (s)^i \sqrt{R_{i-1}} \cdot \hat{P}_{(i-1)\alpha})^2 \big]
	\end{align}
	where $i-1$ denotes the cell area and perimeter at the previous iteration. The target is assigned by multiplying the previous area by $R_{i-1}$ and the previous perimeter by $\sqrt{R_{i-1}}$, so that each cell either contracts or swells. Panel (f) more closely resembles (a)--(d) than (e) does, in which the shape index was forced, showing that the pressure distribution can be used to guide both cell swelling and the pipe network size. We also tested removing the fluid energy from (e), shown in (g); the result is almost equivalent to (f), and to the almost-no-swelling case (h). This does not isolate the two contributions as long as cell morphologies are not well characterized. If we ignore intercellular gaps, then cell swelling guided by pressure gradients can by itself characterize the overall morphological symmetry.
	\begin{table}[]
		\centering
		\caption{Coefficients for Scenario 2 (budding)}
		\begin{tabular}{c|c}
			Name & coefficients \\
			\hline
			$\hat{S}_A$ (HAS-treated) & 1 \\
			$\hat{S}_P$ (HAS-treated) & 1.05 \\
			$\hat{S}_A$ (untreated) & 1 \\
			$\hat{S}_P$ (untreated) & 1 \\
			$\hat{K}_{S1}$ (non-HAS to HAS junctions) & 5 \\
			$\hat{K}_{S2}$ (all other junctions) & 1 \\
			$\hat{K}_{R}$ (boundary cells) & 10 \\
			$\hat{K}_L$ ($\hat{p}\hat{A}$) & 0.5 \\
			$\hat{p}$ ($\hat{p}\hat{A}$) & 0.1 \\
			$\hat{\sigma}$ (arc 1) & 0.5 \\
			$\hat{\sigma}$ (arc 2) & 0.5 \\
			$\hat{D}$ (diffusivity) & 0.5 \\
			$\hat{k}$ (removal rate) & 0.05 \\
			$\hat{s}_0$ (source strength) & 1.0 \\
			$\hat{\lambda} = \sqrt{\hat{D}/\hat{k}}$ & 3.16 \\
			\hline
		\end{tabular}
		\label{tab-coeff-s2}
	\end{table}
	
	\begin{table}[]
		\centering
		\caption{Coefficients for Scenario 3 (lateral asymmetry)}
		\begin{tabular}{c|c}
			Name & coefficients \\
			\hline
			$\hat{S}_A$ (left) & 0.99 \\
			$\hat{S}_P$ (left) & 0.99 \\
			$\hat{S}_A$ (right) & 1.02 \\
			$\hat{S}_P$ (right) & $\sqrt{(1.01)(1.02)}$ \\
			$\hat{K}_{S1}$ (right junctions) & 1 \\
			$\hat{K}_{S2}$ (center junctions) & 5 \\
			$\hat{K}_{S3}$ (left junctions) & 10 \\
			$\hat{K}_L$ ($\hat{p}\hat{A}$) & 0.5 \\
			$\hat{p}$ ($\hat{p}\hat{A}$) & 0.1 \\
			$\hat{\sigma}$ (right boundary) & 0.05 \\
			$\hat{\sigma}$ (left boundary) & 0.5 \\
			$\hat{\lambda}$ & 30 \\
			$\hat{c}_A$ (amplitude) & 0.5 \\
			\hline
		\end{tabular}
		\label{tab-coeff-s3}
	\end{table}
\end{document}